\documentclass[twocolumn]{aastex63}
\usepackage{textcomp}
\usepackage{txfonts}
\usepackage{graphicx}
\usepackage{lipsum}
\usepackage{longtable}
\usepackage{threeparttablex}
\usepackage{booktabs}

\newcommand{\tess}{\emph{TESS}}
\newcommand{\gaia}{\emph{Gaia}}
\newcommand{\kepler}{\emph{Kepler}}

\newcommand{\sname}{TOI-421}
\newcommand{\planetb}{TOI-421\,b}
\newcommand{\planetc}{TOI-421\,c}
\newcommand{\pyaneti}{\href{https://github.com/oscaribv/pyaneti}{\texttt{pyaneti}}}

\newcommand{\logrhk}{$\rm log\,R^{\prime}_\mathrm{HK}$}

\newcommand{\kms}{\,km\,s$^{-1}$} 
\newcommand{\ms}{\,m\,s$^{-1}$} 

\newcommand{\mstar}{M$_{\star}$}
\newcommand{\rstar}{R$_{\star}$}

\newcommand{\rsun}{$R_{\odot}$}

\newcommand{\vsini}{$v$\,sin\,$i_\star$}

\newcommand{\vmic}{$v_{\rm mic}$}
\newcommand{\vmac}{$v_{\rm mac}$}

\newcommand{\teff}{$T_{\rm eff}$}
\newcommand{\logg}{log\,{\it g$_\star$}}
\newcommand{\feh}{[Fe/H]}

\newcommand{\Tzeroc}[1][days]   {$8440.13162_{-0.00068}^{+0.00070}$~#1} 
\newcommand{\Pc}[1][days]  {$16.06819\pm0.00035$~#1} 
\newcommand{\esinc}[1][ ]  {$0.348_{-0.086}^{+0.065}$~#1} 
\newcommand{\ecosc}[1][ ]  {$-0.164_{-0.078}^{+0.084}$~#1} 
\newcommand{\bc}[1][ ]  {$0.738_{-0.032}^{+0.029}$~#1} 
\newcommand{\dentrheec}[1][${\rm g^{1/3}\,cm^{-1}}$]  {$1.93\pm0.13$~#1} 
\newcommand{\rrc}[1][ ] {$0.0542_{-0.0010}^{+0.0011}$~#1} 
\newcommand{\kc}[1][${\rm m\,s^{-1}}$] {$4.66\pm0.29$~#1} 
\newcommand{\mpc}[1][$M_{\oplus}$] {$16.42_{-1.04}^{+1.06}$~#1} 
\newcommand{\rpc}[1][$R_{\oplus}$] {$5.09_{-0.15}^{+0.16}$~#1} 
 
\newcommand{\ec}[1][ ] {$0.152\pm0.042$~#1} 
\newcommand{\wc}[1][deg] {$114.7_{-13.3}^{+15.6}$~#1} 
\newcommand{\ic}[1][deg] {$88.353_{-0.084}^{+0.078}$~#1} 
 
\newcommand{\ac}[1][AU]  {$0.1189\pm0.0039$~#1}

\newcommand{\insolationc}[1][${\rm F_{\oplus}}$] {$34.32_{-2.16}^{+2.34}$~#1}

\newcommand{\Teqc}[1][K] {$673.6_{-10.9}^{+11.2}$~#1} 
\newcommand{\ttotc}[1][hours] {$2.71_{-0.038}^{+0.043}$~#1}

\newcommand{\denpc}[1][${\rm g\,cm^{-3}}$] {$0.685_{-0.072}^{+0.080}$~#1}

\newcommand{\Tzerob}[1][days] {$8441.2847_{-0.0018}^{+0.0020}$~#1} 
\newcommand{\Pb}[1][days]   {$5.19672\pm0.00049$~#1} 
\newcommand{\esinb}[1][ ]   {$0.27_{-0.21}^{+0.15}$~#1} 
\newcommand{\ecosb}[1][ ]   {$-0.268_{-0.090}^{+0.117}$~#1} 
\newcommand{\bb}[1][ ]   {$0.933_{-0.024}^{+0.016}$~#1} 
\newcommand{\rrb}[1][ ]   {$0.0285_{-0.0018}^{+0.0019}$~#1} 
\newcommand{\kb}[1][${\rm m\,s^{-1}}$]   {$2.97\pm0.27$~#1} 
\newcommand{\mpb}[1][$M_{\oplus}$] {$7.17\pm0.66$~#1} 
\newcommand{\rpb}[1][$R_{\oplus}$] {$2.68_{-0.18}^{+0.19}$~#1} 
 
\newcommand{\eb}[1][ ]  {$0.163_{-0.071}^{+0.082}$~#1} 
\newcommand{\wb}[1][deg]   {$128.9_{-27.2}^{+24.9}$~#1} 
\newcommand{\ib}[1][deg]   {$85.68_{-0.46}^{+0.36}$~#1} 
 
\newcommand{\ab}[1][AU]   {$0.0560\pm0.0018$~#1}

\newcommand{\insolationb}[1][${\rm F_{\oplus}}$] {$154.57_{-9.73}^{+10.53}$~#1}

\newcommand{\Teqb}[1][K]   {$981.4_{-15.8}^{+16.3}$~#1} 
\newcommand{\ttotb}[1][hours] {$1.107_{-0.063}^{+0.065}$~#1}

\newcommand{\denpb}[1][${\rm g\,cm^{-3}}$] {$2.05_{-0.41}^{+0.52}$~#1}

\newcommand{\Tzerod}[1][days] {$8430.33_{-4.98}^{+4.77}$~#1} 
\newcommand{\Pd}[1][days] {$43.24_{-0.55}^{+0.57}$~#1}

\newcommand{\kd}[1][${\rm m\,s^{-1}}$]   {$2.36_{-0.3}^{+0.3}$~#1}
\newcommand{\qone}[1][]   {$0.269_{-0.083}^{+0.121}$~#1} 
\newcommand{\qtwo}[1][]   {$0.65_{-0.35}^{+0.24}$~#1}

\newcommand{\qoneground}[1][]   {$0.164_{-0.097}^{+0.101}$~#1} 
\newcommand{\qtwoground}[1][]   {$0.348_{-0.100}^{+0.101}$~#1}

\newcommand{\HARPS}[1][${\rm km\,s^{-1}}$] {$79.54382_{-0.00025}^{+0.00024}$~#1} 
\newcommand{\FIES}[1][${\rm km\,s^{-1}}$]  {$-0.0169_{-0.0015}^{+0.0015}$~#1} 
\newcommand{\PFS}[1][${\rm km\,s^{-1}}$]   {$0.00222_{-0.00093}^{+0.00091}$~#1} 
\newcommand{\HIRES}[1][${\rm km\,s^{-1}}$] {$-0.00096_{-0.00043}^{+0.00041}$~#1} 
\newcommand{\jHARPS}[1][${\rm m\,s^{-1}}$] {$1.88_{-0.18}^{ + 0.20}$~#1} 
\newcommand{\jFIES}[1][${\rm m\,s^{-1}}$]  {$0.85_{-0.65}^{ + 1.41}$~#1} 
\newcommand{\jPFS}[1][${\rm m\,s^{-1}}$]   {$2.44_{-0.64}^{ + 1.00}$~#1} 
\newcommand{\jHIRES}[1][${\rm m\,s^{-1}}$] {$2.11_{-0.34}^{ + 0.40}$~#1}

\received{}
\revised{}
\accepted{}
\submitjournal{ApJ}

\shorttitle{The transiting multi-planet system TOI-421}
\shortauthors{Carleo et al.}
\graphicspath{{./}{figures/}}

\begin{document}

\title{The multi-planet system TOI-421 \footnote{Based on observations made with ESO Telescopes at the La Silla Observatory under programs ID 1102.C-0923, 0103.C-0874, 0103.C-0759, 0103.C-0442, and 60.A-970.
Based on observations obtained with the Nordic Optical Telescope (NOT), operated on the island of La Palma jointly by Denmark, Finland, Iceland, Norway, and Sweden, in the Spanish Observatorio del Roque de los Muchachos (ORM) of the Instituto de Astrof\'isica de Canarias (IAC).
This paper includes data gathered with the 6.5 meter Magellan Telescopes located at Las Campanas Observatory, Chile.
This work makes use of observations from the LCOGT network.} 
\\ A warm Neptune and a super puffy mini-Neptune transiting a G9 V star in a visual binary }

\correspondingauthor{Ilaria Carleo}
\email{icarleo@wesleyan.edu}

\author[0000-0002-0810-3747]{Ilaria Carleo}
\affiliation{Astronomy Department and Van Vleck Observatory, Wesleyan University, Middletown, CT 06459, USA}
\affiliation{INAF -- Osservatorio Astronomico di Padova, Vicolo dell'Osservatorio 5, I-35122, Padova, Italy}

\author[0000-0001-8627-9628]{Davide Gandolfi}
\affiliation{Dipartimento di Fisica, Universit\`a degli Studi di Torino, via Pietro Giuria 1, I-10125, Torino, Italy}

\author[0000-0003-0563-0493]{Oscar Barrag\'an}
\affiliation{Sub-department of Astrophysics, Department of Physics, University of Oxford, Oxford, OX1 3RH, UK}

\author[0000-0002-4881-3620]{John~H.~Livingston}
\affiliation{Department of Astronomy, University of Tokyo, 7-3-1 Hongo, Bunkyo-ku, Tokyo 113-0033, Japan}

\author[0000-0003-1257-5146]{Carina~M.~Persson}
\affiliation{Department of Space, Earth and Environment, Chalmers University of Technology, Onsala Space Observatory, 439 92 Onsala, Sweden}

\author{Kristine~W.\,F.~Lam}
\affiliation{Center for Astronomy and Astrophysics, TU Berlin, Hardenbergstr. 36, 10623 Berlin, Germany}

\author{Aline Vidotto}
\affiliation{School of Physics, Trinity College Dublin, The University of Dublin College Green, Dublin 2, Dublin, Ireland}

\author[0000-0003-2527-1598]{Michael B. Lund}
\affiliation{NASA Exoplanet Science Institute/Caltech-IPAC, MC 314-6, 1200 E California Blvd, Pasadena, CA 91125, USA}

\author{Carolina Villarreal D’Angelo}
\affiliation{School of Physics, Trinity College Dublin, The University of Dublin College Green, Dublin 2, Dublin, Ireland}

\author[0000-0001-6588-9574]{Karen A.\ Collins}
\affiliation{Center for Astrophysics \textbar \ Harvard \& Smithsonian, 60 Garden Street, Cambridge, MA 02138, USA}

% ...

\begin{abstract}


We report the discovery of a warm Neptune and a hot sub-Neptune transiting \sname\ (BD-14\,1137, TIC 94986319), a bright ($V$=9.9) G9 dwarf star in a visual binary system observed by the \tess\ space mission in Sectors 5 and 6. We performed ground-based follow-up observations -- comprised of LCOGT transit photometry, NIRC2 adaptive optics imaging, and FIES, CORALIE, HARPS, HIRES, and PFS high-precision Doppler measurements -- and confirmed the planetary nature of the 16-day transiting candidate announced by the \tess\ team. We discovered an additional radial velocity signal with a period of 5 days induced by the presence of a second planet in the system, which we also found to transit its host star. We found that the inner mini-Neptune, \planetb, has an orbital period of P$_\mathrm{b}$=\Pb, a mass of M$_\mathrm{b}$\,=\,\mpb\ and a radius of R$_\mathrm{b}$\,=\,\rpb, whereas the outer warm Neptune, \planetc, has a period of P$_\mathrm{c}$=\Pc, a mass of M$_\mathrm{c}$\,=\,\mpc, a radius of R$_\mathrm{c}$\,=\,\rpc\ and a density of $\rho_\mathrm{c}$=\denpc. With its characteristics the inner planet ($\rho_\mathrm{b}$=\denpb) is placed in the intriguing class of the super-puffy mini-Neptunes. \planetb\ and \planetc\ are found to be well suitable for atmospheric characterization. Our atmospheric simulations predict significant Ly-$\alpha$ transit absorption, due to strong hydrogen escape in both planets, and the presence of detectable CH$_4$ in the atmosphere of \planetc\ if equilibrium chemistry is assumed. 


\end{abstract}

\keywords{Exoplanet astronomy: Exoplanet systems --- High
resolution spectroscopy --- stars: fundamental parameters --- 
techniques: radial velocities, spectroscopic, photometric}


\section{Introduction}
\label{sec:intro}
The \tess\ (\textit{Transiting Exoplanet Survey Satellite}, \citealt{Rickeretal2014}) mission's primary scientific driver is to measure masses for transiting planets smaller than 4~$R_\oplus$ around bright stars, in order to explore the transition from sub-Neptunes (with extended envelopes) to rocky planets (with compact atmospheres), that occurs at about 1.8~$R_\oplus$.
The \textit{Kepler} mission revealed that small planets (especially in the super-Earth and sub-Neptune regime), in compact coplanar multi-planet systems, are very common (\citealt{lathametal2011, Lissaueretal2011, Lissaueretal2014, Roweetal2014}). While most of the \kepler\ stars are distant and faint, making radial velocity (RV) follow-up very difficult, the \tess\ mission is focused on the nearest and brightest stars, so that an intensive follow-up, as well as atmospheric characterization, can more easily be achieved. Those prospects are rather important, for example for future space-based observations with \textit{JWST}. 

Since July 2018 \tess\ has been scanning the sky and performing a photometric search for planets transiting bright stars. In its primary mission, this survey will cover 26 Sectors, each of them monitored for $\sim$27\,days, with candidate alerts released almost every month. \tess\ is expected to detect $\sim$10,000 transiting exoplanets (\citealt{Sullivanetal2015,Barclayetal2018,Huangetal2018a}). More than 1,000 planet candidates have been revealed, with dozens of confirmed planets so far \citep[e.g.][]{Espositoetal2019, Brahmetal2019, Lendletal2020, diazetal2020}, some of which are multi-planet systems \citep[e.g,][]{Gandolfietal2018, Huangetal2018b, Dragomiretal2019, Gunteretal2019, Quinnetal2019,Gandolfietal2019}. 

Multi-planet systems are prime targets for testing planetary formation and evolution theories. Orbiting the same star, they offer an opportunity to simplify the assumptions of initial conditions and compare planets with different sizes and compositions in the same system. Such systems are also interesting for transmission spectroscopy, which allows us to characterize planetary atmospheres and compare them at different levels of incident stellar flux. 

In the present paper we report on the discovery and characterization of a sub-Neptune (\sname\ b) and a Neptune (\sname\ c) transiting the bright star BD-14\,1137 (\sname, TIC 94986319), observed by \tess\ in Sectors 5 and 6. The work presented here is part of the RV follow-up project carried out by the {\tt KESPRINT} collaboration\footnote{\url{www.kesprint.science}.} \citep[e.g.,][]{Grziwa2016b,VanEylenetal2016, Gandolfi2017,Nowak2017,Barragan2018,Perssonetal2019,Korth2019}, which aims to confirm and characterize planet candidates from the \textit{K2} and \tess\ space missions. 

This paper is organized as follows. In Section \ref{sec:obs}, we describe the observations carried out with different space- and ground-based facilities. These include \tess\ photometry, AO imaging, ground-based photometry, and high-resolution spectroscopy. In Section \ref{sec:stellar} we derive the stellar parameters of \sname. Section~\ref{sec:tessphot} reports the \tess\ photometry analysis with the detection of the transit signals. Section~\ref{sec:rvfreq} presents the HARPS frequency analysis to confirm the planetary nature of the transiting companions and investigate additional signals associated to the stellar activity. Sections \ref{sec:rv} and \ref{sec:joint} present a preliminary RV modeling and a joint analysis of \tess\ light curve and RV data, respectively.
We discuss the results in Section \ref{sec:disc}, including the simulations of possible atmospheric signals in different wavelength bands and of the atmospheric evolution of both planets. Finally, we draw our conclusion in Section \ref{sec:concl}. 

\begin{table}
\centering
\caption{Main identifiers, equatorial coordinates, proper motion, parallax, optical and infrared magnitudes, and fundamental parameters of \sname.}
\label{tab:stellar}
\begin{tabular}{lrr}
\hline
Parameter & Value & Source \\
\hline
\multicolumn{3}{l}{\it Main identifiers}  \\
\noalign{\smallskip}
\multicolumn{2}{l}{TIC}{94986319} & ExoFOP\footnote{\url{https://exofop.ipac.caltech.edu/}} \\
\multicolumn{2}{l}{BD}{-14 1137}  & ExoFOP \\
\multicolumn{2}{l}{TYC}{5344-01206-1} & ExoFOP \\
\multicolumn{2}{l}{2MASS}{J05272482-1416370}  & ExoFOP \\
\multicolumn{2}{l}{\gaia\ DR2}{2984582227215748864}  & \gaia\ DR2\footnote{\cite{GaiaDR2}} \\
\hline
\multicolumn{3}{l}{\it Equatorial coordinates, parallax, and proper motion}  \\
\noalign{\smallskip}
R.A. (J2000.0)	&    05$^\mathrm{h}$27$^\mathrm{m}$24.83$^\mathrm{s}$	& \gaia\ DR2 \\
Dec. (J2000.0)	& $-$14$\degr$16$\arcmin$37.05$\arcsec$	                & \gaia\ DR2 \\
$\pi$ (mas) 	& $13.3407\pm0.0361$                                    & \gaia\ DR2 \\
$\mu_\alpha$ (mas\,yr$^{-1}$) 	& $-35.687 \pm 0.046$		& \gaia\ DR2 \\
$\mu_\delta$ (mas\,yr$^{-1}$) 	& $ 50.450 \pm 0.064$		& \gaia\ DR2 \\
\hline
\multicolumn{3}{l}{\it Optical and near-infrared photometry} \\
\noalign{\smallskip}
$\tess$              & $9.2711\pm0.006$     & TIC v8\footnote{\cite{Stassun2018}}         \\
\noalign{\smallskip}
$G$				 & $9.7778\pm0.0002$	& \gaia\ DR2 \\
$B_\mathrm{p}$   & $10.2034\pm0.0012$   & \gaia\ DR2 \\
$R_\mathrm{p}$   & $9.2265\pm 0.0012$   & \gaia\ DR2 \\
\noalign{\smallskip}
$B$              & $10.735 \pm 0.076$          & TIC v8 \\
$V$              & $ 9.931 \pm 0.006$          & TIC v8 \\
\noalign{\smallskip}
$J$ 			&  $8.547\pm0.020$      & 2MASS\footnote{\cite{cutri2003}} \\
$H$				&  $8.219\pm0.033$      & 2MASS \\
$Ks$			&  $8.071\pm0.018$      & 2MASS \\
\noalign{\smallskip}
$W1$			&  $8.058\pm0.023$      & All{\it WISE}\footnote{\cite{cutri2013}} \\
$W2$			&  $8.110\pm0.020$      & All{\it WISE} \\
$W3$             & $8.060\pm0.021$      & All{\it WISE} \\
$W4$             & $7.809\pm0.168$      & All{\it WISE} \\
\hline
\multicolumn{3}{l}{\it Fundamental parameters}   \\
\vsini\ (\kms)      & $1.8 \pm 1.0$       & This work \\
\teff\ (K) & $5325^{+78}_{-58}$ & This work \\
\logg\ (cgs) & $4.486^{+0.025}_{-0.018}$ & This work \\
\feh\ (dex) & $-0.02 \pm 0.05$ & This work \\
\mstar\ ($\mathrm{M_\odot}$) & $0.852^{+0.029}_{-0.021}$ & This work \\
\rstar\ ($\mathrm{R_\odot}$) & $0.871 \pm 0.012$  & This work \\
Age (Gyr) & $9.4^{+2.4}_{-3.1}$ & This work \\
Distance (pc) & $74.94 \pm 0.58$ & This work \\
$\mathrm{A_V}$ (mag) & $0.11^{+0.12}_{-0.08}$ & This work \\
\hline
\end{tabular}
\end{table}

\section{Observations} \label{sec:obs}

\subsection{\tess\ photometry}\label{sec:2.1tess}

\sname\ (TIC\,94986319) -- whose identifiers, coordinates, proper motion, optical and infrared magnitudes, and fundamental parameters are listed in Table~\ref{tab:stellar} -- was monitored in Sectors 5 and 6 of the \tess\ mission between 15 November 2018 and 07 January 2019. In Sector 5, the target was imaged on CCD 3 of camera 2, and on CCD 4 of camera 2 in Sector 6. A total of 34,622 photometric data points were collected, each with an exposure time of 2 minutes. The target light curves were processed by the Science Processing Operations Center \citep[SPOC;][]{Jenkins2016} data reduction pipeline. The Simple Aperture Photometry (SAP) was used in the SPOC pipeline to produce the light curves (\citealt{Twickenetal2010,morrisetal2017}), and the Presearch Data Conditioning (PDCSAP) algorithm was used to remove known instrumental systematics \citep{Smith2012,Stumpe2012,Stumpeetal2014}. For the transit detection and analysis presented in this work (Sec. \ref{sec:tessphot}), we downloaded \sname's PDCSAP light curves, which are publicly available at the Mikulski Archive for Space Telescopes (MAST) web-page\footnote{\url{https://archive.stsci.edu/tess/}.}.

\subsection{Ground-based Photometry}

\subsubsection{LCOGT}
We observed \sname\ in Pan-STARSS Y-band on UT 2019-02-05 using the South Africa Astronomical Observatory node of the Las Cumbres Observatory Global Telescope (LCOGT) 1-m network \citep{Brown:2013}. We used the {\tt TESS Transit Finder}, which is a customized version of the {\tt Tapir} software package \citep{Jensen:2013}, to schedule our transit observation. The 1-m telescopes are equipped with $4096\times4096$ LCO SINISTRO cameras having an image scale of 0.389$\arcsec$ pixel$^{-1}$ resulting in a $26\arcmin\times26\arcmin$ field of view. The images were calibrated using the standard LCOGT BANZAI pipeline \citep{McCully:2018} and the photometric data were extracted using the {\tt AstroImageJ} ({\tt AIJ}) software package \citep{Collins:2017}. The images have a mean stellar PSF FWHM of $1.78\arcsec$. The optimum target star photometric aperture used to extract the data for the analyses in this work had a radius of 15 pixels ($5.8\arcsec$). 
Furthermore, the transit was detected in apertures as small as 5 pixels ($1.9\arcsec$) with higher model residuals. 

\subsubsection{WASP-South}
\label{Sec:WASP-South}

The field of \sname\ was also observed by WASP-South each year from 2008 to 2015, covering a span of $\sim$120 nights per year.  WASP-South was an array of 8 cameras located in Sutherland, South Africa, being the Southern station of the WASP project \citep{2006PASP..118.1407P}. Until 2012 it used 200-mm, f/1.8 lenses, observing fields with a typical 10-min cadence, and accumulated 14\,800 photometric data points on \sname. It then switched to 85-mm, f/1.2 lenses using an SDSS-$r$ filter, and accumulated another 77\,000 observations of \sname. We did not find any significant periodicity. 

\subsection{AO Imaging}

High-resolution adaptive optics (AO) imaging observations of \sname\ were made with NIRC2 on Keck\,II (\url{http://www2.keck.hawaii.edu/inst/nirc2/}) on 2019 Mar 25 UT. Weather was dry and stable but clouds affected the observations throughout the night. Extinction due to clouds was estimated to be between 1.2 - 3 magnitudes. \sname\ was observed at an airmass of 1.5. Observations were made in natural guide star, narrow camera (0.009942 \arcsec/pixel) mode, and used the full 1024\arcsec\,$\times$\,1024\arcsec\ FOV. A standard 3-point dither pattern was used to avoid the noisy lower left quadrant of the detector. Each pointing was done with a 3\arcsec\ nod to find any off-axis bright sources. Observations were made in the $K$-band for a total integration time of 180 seconds once all pointings were co-added. The AO observations of \sname\ resulted in a spatial resolution of 0.053\arcsec\ (FWHM) in the $K$-band (Fig. \ref{fig:ao_contrast}).

\subsection{Ground-based Spectroscopy} 
We carried out spectroscopic follow-up observations of \sname\ using different facilities -- as described in the sub-sections below -- to spectroscopically confirm the planetary nature of the transit signals detected in the \tess\ light curve and determine the masses of the two transiting planets.


\subsubsection{FIES}

We started the radial velocity follow-up of \sname\ in February 2019 using the FIbre-fed Echell\'e Spectrograph \citep[FIES;][]{Frandsen1999,Telting2014} mounted at the 2.56\,m Nordic Optical Telescope (NOT) of Roque de los Muchachos Observatory (La Palma, Spain). FIES is mounted inside an insulated building where the temperature is kept constant within 0.02\,°C and fed with octagonal fibres to improve the radial velocity (RV) stability of the spectrograph \citep{Stuermer2018}. We employed the intermediate resolution fibre, which provides a resolving power of R\,=\,45\,000 over the wavelength range 3660\,-\,9275\,\AA, and acquired 10 spectra between 2 February and 13 March 2019. We used the same observing strategy as in \citet{Buchhave2010} and \citet{Gandolfi2013}, i.e., we split the observations in three sub-exposures to remove cosmic ray hits and bracketed the three exposures with long-exposed (T$_\mathrm{exp}$\,$\approx$\,80\,sec) ThAr spectra to trace the instrument drift and improve the wavelength solution. We reduced the data using IRAF and IDL standard procedures and extracted relative RV measurements by cross-correlating the observed Echell\'e spectra with the first epoch spectrum (Table~\ref{Table:FIES}). 

\subsubsection{CORALIE}
We also took 7 high-resolution spectra ($R$\,$\approx$\,60\,000) of \sname\ using the CORALIE Echell\'e spectrograph on the Swiss 1.2 m Euler telescope at La Silla Observatories, Chile \citep{CORALIE}. 
We extracted the radial velocity measurements by cross-correlating the CORALIE Echell\'e spectra with a binary G2 mask \citep{Pepe2002a}. The Doppler measurements show no significant RV variation at a level of $\sim$8\,\ms\ and exclude that \sname\ is an eclipsing binary mimicking planetary transits.

\subsubsection{HARPS}

We acquired 105 high-resolution ($R$\,=\,115\,000) spectra  of \sname\ using the High Accuracy Radial velocity Planet Searcher \citep[HARPS][]{Mayor2003} spectrograph mounted at the ESO-3.6\,m telescope of La Silla Observatory, Chile. Installed in a pressure- and temperature-controlled enclosure and fed with octagonal fibres, HARPS has demonstrated a long-term precision at the 1\,\ms\ level and below \citep{Lovis2006}. Our HARPS observations were performed over two observing seasons between February 2019 and January 2020, as part of the observing programs 1102.C-0923, 0103.C-0874, 0103.C-0759, 0103.C-0442, and 60.A-9709. We used the second fibre of the instrument to monitor the sky background and set the exposure time to 900-2100\,sec depending on sky conditions and constraints of the observing schedule. We reduced the data using the dedicated HARPS Data Reduction Software (DRS) and computed the RVs by cross-correlating the \'Echelle spectra with a G2 numerical mask \citep{Baranne1996,Pepe2002a,Lovis2007}. We also used the DRS to extract the line profile asymmetry indicators, namely, the full width at half maximum (FWHM) and the bisector inverse slope (BIS) of the cross-correlation function (CCF), and the Ca\,{\sc ii} H\,\&\,K lines activity indicator\footnote{We adopted a B$-$V color index of 0.710, as listed in the APASS catalog \citep{Henden2015}.} \logrhk. We report the HARPS RV measurements and their uncertainties, along with the FWHM, BIS, \logrhk, exposure time, and signal-to-noise (S/N) ratio per pixel at 550 nm in Table~\ref{Table:HARPS}.

\subsubsection{HIRES}

Between 17 September 2019 and 3 March 2020, we obtained 28 spectra ($R$\,=\,55\,000) of \sname\ over 27 nights using the High Resolution \'Echelle Spectrometer \citep[HIRES;][]{Vogt1994} on the 10-m Keck-I telescope. The spectra were collected using an iodine cell for wavelength reference. The median exposure time was 680\,s, which allowed us to achieve a S/N ratio of 200 per reduced pixel at 5500\,\AA. We also obtained a high S/N ratio template spectrum without the iodine cell, which was used as input for the standard forward-modeling procedures of the California Planet Search \citep{Howard2010}. We report the RVs and uncertainties based on the weighted mean and weighted error in the mean of the $\sim$700 individual spectral chunks in Table~\ref{Table:HIRES}.

\subsubsection{PFS}

\sname\ was also selected as a high priority target by the WINE (Warm gIaNts with tEss) collaboration, which aims at systematically characterizing long period transiting giant and Neptune-size planets from the \tess\ mission \citep[see, e.g.,][]{Brahm2019,Jordan2019}. In this context, we monitored \sname\ with the Planet Finder Spectrograph \citep[PFS; ][]{craneetal2006,craneetal2008,Craneetal2010} mounted on the 6.5\,m Magellan II Clay Telescope at Las Campanas Observatory (LCO) in Chile. The observations were performed in 9 different nights, between 18 February and 11 October 2019. For these observations we used the 0.3$\arcsec\times$2.5$\arcsec$ slit, which delivers high-resolution spectra with a resolving power of $R$\,=\,130\,000. The observing strategy consisted of obtaining two consecutive 1200-sec spectra per night of \sname\ to increase the total S/N ratio per epoch and also to average out the stellar and instrumental jitter. These observations were performed with the use of an iodine cell for determining the radial velocity of the star. The PFS data were processed with a custom IDL pipeline that is capable of delivering RVs with a precision less than 1 \ms\ \citep{Butler1996}. Additionally, 3 consecutive 1200-sec iodine-free exposures of \sname\ were obtained to construct a stellar spectral template for computing the RVs. The PFS RVs are listed in Table~\ref{Table:PFS}.

\section{Stellar parameters}
\label{sec:stellar}

\begin{figure*}[htbp!]
    \centering
	\includegraphics[width=2\columnwidth]{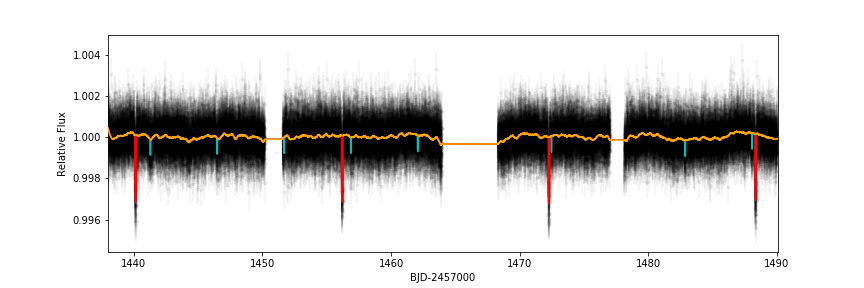}
    \caption{\tess\ PDCSAP light curve of \sname\ is denoted by the black points. The orange line shows the variability filter applied by the transit search algorithm. The detected transits of the 16.1 d planet and the 5.2 d planet are shown in red and cyan, respectively.}
    \label{fig:PDCSAP_lightcurve}
\end{figure*}

The fundamental stellar parameters of \sname\ were independently determined by three different methods -- one based on spectral synthesis, one on template matching, and one using different sets of isochrones, as described in the paragraphs below. For this purpose, we utilized the co-added HARPS spectrum, which has a S/N ratio of $\sim$700 per pixel at 5500\,\AA.\\

\emph{Method 1}. We derived the effective temperature $T_\mathrm{eff}$, surface gravity \logg, iron abundance [Fe/H], and projected rotational velocity \vsini\ with the spectral analysis package Spectroscopy Made Easy \citep[{\tt{SME}}, version 5.2.2;][]{vp96, pv2017}. We used  the {\tt{ATLAS12}}  model spectra \citep{Kurucz2013} and the line lists from the Vienna Atomic Line Database\footnote{{\url{http://vald.astro.uu.se}}.} \citep{Ryabchikova2015} to model the co-added HARPS spectrum. Following the modelling procedure detailed in \citet{2017A&A...604A..16F}  and \citet{2018A&A...618A..33P}, we measured $T_\mathrm{eff}$ from the wings of H-$\alpha$ line, and \logg\ from the line wings of the Ca and Mg triplets around 6100~\AA~and 5100~\AA, respectively. \vsini\ and [Fe/H] were derived from narrow unblended iron lines between 6000 and 6600~\AA. The micro- and macro-turbulent velocities, \vmic\ and \vmac, were kept fixed using the \citet{Bruntt2010b} and \citet{Doyle2014} calibration equations for Sun-like stars to 0.9~\kms\ and 2.5~\kms, respectively. The final best fitting model spectrum was checked using the Na doublet at 5888~\AA~and 5895~\AA. We found $T_\mathrm{eff}$\,=\,5194\,$\pm$\,60~K, \mbox{$\log g_\star\,=\,4.45\,\pm\,0.05$\,(cgs)}, [Fe/H]\,=\,$-0.04$\,$\pm$\,0.06~(dex), and  \mbox{\vsini\,=\,1.8\,$\pm$\,1.0~\kms}. The uncertainties are internal error bars that do not account for the systematic uncertainties of the atmospheric models.

We measured the visual interstellar extinction ($A_\mathrm{V}$) along the line of sight to \sname\ following the method described in \citet{Gandolfi2008}. Briefly, we built the spectral energy distribution (SED) of the star from the broadband photometry listed in Table~\ref{tab:stellar}, and fitted the SED using synthetic magnitudes computed from a low-resolution BT-NextGen model spectrum \citep{Allard2012} with the same spectroscopic parameter as the star. We adopted the \citet{Cardelli1998}'s extinction law and assumed a total-to-selective extinction ratio of R$_\mathrm{V}$\,=\,(A$_\mathrm{V})/\mathrm{E(B-V)}$\,=\,3.1. We found that the interstellar extinction along the line of sight is consistent with zero ($A_\mathrm{V}$\,=\,0.03\,$\pm$\,0.03), as expected given the relatively short distance to the star ($\sim$75\,pc).

In order to compute the stellar mass, radius, and age we employed the Bayesian {\tt{PARAM\,1.3}}\footnote{\url{http://stev.oapd.inaf.it/cgi-bin/param_1.3}.} model tool tracks \citep{daSilva2006} with the  {\tt{PARSEC}} isochrones   \citep{2012MNRAS.427..127B}. Input parameters are \teff~and \feh~from {\tt{SME}}, the apparent visual magnitude, and the parallax (Table~\ref{tab:stellar}). We added 0.1 mas in quadrature to the parallax uncertainty of 0.0361~mas to account for systematic errors of \gaia’s astrometry \citep{2018A&A...616A...9L}. The resulting stellar parameters are $M_\star\,=\,0.86\,\pm\,0.02~M_{\odot}$, $R_\star\,=\,0.86\,\pm\,0.02~R_{\odot}$, $\log g_\star\,=\,4.48\,\pm\,0.02$ (cgs), and an age of $9.2\,\pm\,2.3$~Gyr. We note that the derived \logg\ is in very good agreement with the spectroscopic value.

\emph{Method 2}. We used {\tt SpecMatch-Emp} \citep{Yee2017} to analyze the co-added HARPS spectrum via comparison with a spectroscopic library of well-characterized stars, enabling empirical estimates of the effective temperature, stellar radius, and photospheric iron content. Following the procedure described in \citet{Hirano2018} to adapt the HARPS spectrum for use with {\tt SpecMatch-Emp}, we obtained \teff\,=\,5337\,$\pm$\,110\,K, \rstar\,=\,0.972\,$\pm$\,0.100\,\rsun, and \feh\,=\,0.05\,$\pm$\,0.09\,dex. 

\emph{Method 3}. We also computed a set of uniformly inferred stellar parameters using {\tt isochrones} \citep{2015ascl.soft03010M} and \texttt{MIST} \citep{2016ApJ...823..102C} to fit 2MASS $JHKs$ photometry \citep{2006AJ....131.1163S} and \gaia\ data release 2 (DR2) parallax \citep{2016A&A...595A...1G, 2018A&A...616A...1G}, adding 0.1\,mas of uncertainty in quadrature to account for systematics \citep{2018A&A...616A...9L}. We assumed priors on \teff\ and \feh\ based on the spectroscopic results from {\tt SME}, using a more conservative prior width of 100 K for \teff\ to account for systematic errors, and sampled the posterior using {\sc MultiNest} \citep{2013arXiv1306.2144F}. We obtained the following parameter estimates: \teff\,=\,$5325^{+78}_{-58}$~K,
\logg\,=\,$4.486^{+0.025}_{-0.018}$, \feh\,=\,$-0.02 \pm 0.05$\,dex, \mstar\,=\,$0.852^{+0.029}_{-0.021}$\,$\mathrm{M_\odot}$, \rstar\,=\,$0.871 \pm 0.012$\,$\mathrm{R_\odot}$, age\,=\,$9.4^{+2.4}_{-3.1}$\,Gyr, distance\,=\,$74.94\,\pm\, 0.58$\,pc, and $\mathrm{A_V}$\,=\,$0.11^{+0.12}_{-0.08}$.

We note that the stellar parameter estimates obtained using these three independent methods agree well within 1-1.5\,$\sigma$, and serve to ensure the accuracy of our results.

Since the derived parameters we find using the \emph{Method 3} agree more with those found in \emph{Method 2}, we adopted the results of \emph{Method 2} and list the final adopted parameter estimates in Table~\ref{tab:stellar}. The reason for doing so, is that since \emph{Method 1} rely on the profile of the Balmer $H\alpha$ in order to determine the \teff, and this method become less accurate for cooler temperatures. The quoted projected rotational velocity \vsini\ has been, however, derived following \emph{Method 1}, since \emph{Method 2} does not provide it. The spectroscopic parameters of \sname\ translate into a spectral type and luminosity class of G9\,V \citep{Pecaut2013}.\\

We also looked for solar-like oscillations \citep{garciaballot2019} using an optimized aperture for asteroseismology (Gonz\'alez-Cuesta et al. in prep.) for both sectors. While the probability of detection computed following \cite{Schofieldetal2019} is very low for this star, we applied the A2Z pipeline \citep{Mathuretal2010} on the concatenated light curves of sectors 5 and 6. Some excess of power and periodicity of the modes (also known as the large frequency separation) was found around 2000\,$\mu$Hz and 3000\,$\mu$Hz but with very low confidence level. The expected frequency of maximum power being around 3500\,$\mu$Hz and the signal-to-noise ratio being still very low, the seismic analysis is not conclusive.

\section{\tess\ photometric analysis and planet detection}
\label{sec:tessphot}

The detection of a 16-day transit signal was issued by the TESS Science Office QLP pipeline in Sector 5, and subsequently identified in the SPOC pipeline \citep{Twicken2018} in the Sector 6 data set. The SPOC Data Validation difference image centroid offsets for \sname\ in the multi-sector run indicated that the source of the transit signature was within 1.4\arcsec\ of the proper motion corrected location of the target star. The detection was then released as a planetary candidate via the TOI releases portal\footnote{\url{https://tess.mit.edu/toi-releases/}.} on 08 February 2019. We independently performed transit searches on the PDCSAP light curve using the \texttt{DST} algorithm \citep{Cabrera2012}. The variability in the light curve was filtered using the Savitzky-Golay method \citep{Savitzky1964,Press2002}, and a transit model described by a parabolic function was used for transit searches. The algorithm recovered the detection of \planetc\ where the transit signal has a period of 16.069\,$\pm$\,0.002~d, a transit depth of 2735.90\,$\pm$\,76.04~ppm and a duration of 2.72\,$\pm$\,0.05~hours. 

The detection of a 5.2-day signal in the follow-up RV data (see Section \ref{sec:rvfreq}), prompted further analysis of the \tess\ light curve. The \texttt{DST} algorithm further detected a transit signal with a period of 5.197\,$\pm$\,0.003~d, a depth of 654.21\,$\pm$\,88.70~ppm and a duration of 1.23\,$\pm$\,0.14 hours. The detection of both transit signals were also confirmed with the software package \texttt{EXOTRANS} \citep{Grziwa2016}, which applies the Box least-squares algorithm \citep[BLS;][]{Kovacs2002} for transit searches. Figure~\ref{fig:PDCSAP_lightcurve} shows the PDCSAP light curve of \sname\ along with the detection of the two planets.

\section{Contamination from possible stellar companions}
\label{sec:contamination}

\begin{figure}[!t]
\resizebox{\hsize}{!}{\includegraphics[trim={0 0 0 0},clip]{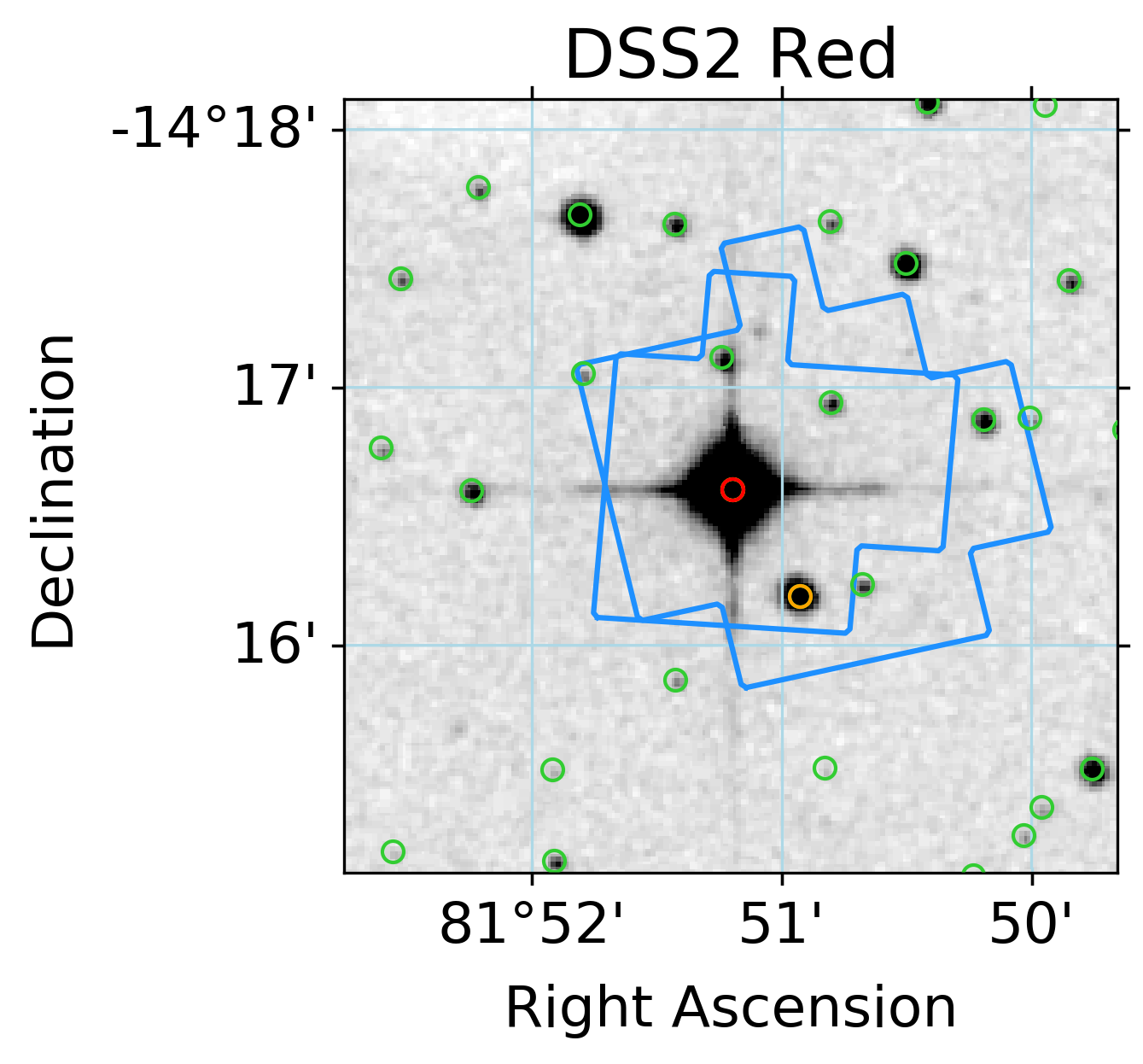}}
\caption{3\,\arcmin\,$\times$\,3\,\arcmin\ DSS2 (Red filter) image with the Sectors 5 and 6 SPOC photometric apertures overplotted in blue. Colored circles denote the positions of \gaia\ DR2 sources within 2\arcmin\ of \sname; the red circle is \sname\ (2984582227215748864), the orange circle is a likely bound M dwarf companion (2984582227215748224), and other sources are in green. We computed the contamination from the companion and other stars contributing flux to the aperture, but it is insignificant at 1.8$\pm$0.4\% (and consistent with the TIC contamination ratio of 0.024605, see Section~\ref{sec:obs}).}
\label{fig:tessaperture}
\end{figure}

In order to check if the 5.2 and 16-day transit signals could arise from another source and assess the dilution level of \sname, we visually inspected archival images and compared the positions of \gaia\ DR2 \citep{GaiaDR2} sources with the SPOC photometric apertures from Sectors 5 and 6. We used the coordinates of \sname\ from the \tess\ Input Catalog\footnote{Available at \url{https://mast.stsci.edu/portal/Mashup/Clients/Mast/Portal.html}.} \citep[][]{Stassun2018} to retrieve \gaia\ DR2 sources and a 3\,\arcmin\,$\times$\,3\,\arcmin\ image from DSS2\footnote{\url{https://skyview.gsfc.nasa.gov/current/cgi/titlepage.pl}.}. Following the procedures described in \citet{Gandolfietal2019}, we computed a photometric dilution level of 1.8\,$\pm$\,0.4\% for \sname, which is a little smaller than the SPOC contamination ratios of 0.028 and 0.024 in Sectors 5 and 6, respectively. This small difference is most likely due to the fact that the SPOC target star is affected by an artefact in the 2MASS catalog caused by a diffraction spike from the 2MASS telescope secondary spider. In our specific case this creates a TICv8 neighbor $6.9\arcsec$ North of the target star, which does not exist. We note that our dilution calculation is not affected by this issue because it is based on \gaia\ DR2.

We discovered that the nearby fainter star, spatially located at $\sim$29.4\arcsec\ NW of \sname\ (\gaia\ ID 2984582227215748224, $\Delta$G\,=\,4.8, indicated by an orange circle in Figure~\ref{fig:tessaperture}), has a parallax and \gaia\ G-band extinction that are consistent within the error bars with those of \sname, and that two stars have similar proper motions. We concluded that the pair forms very likely a visual binary and \sname\ is the primary component. According to \gaia\ DR2 effective temperature (\teff\,=\,3676$_{-385}^{+376}$\,K), the secondary component is an M dwarf. The angular separation and parallax imply a separation between the two stars of about 2200\,AU.

\begin{figure}[!t]
   \centering
\includegraphics[width=1.0\linewidth, angle=180]{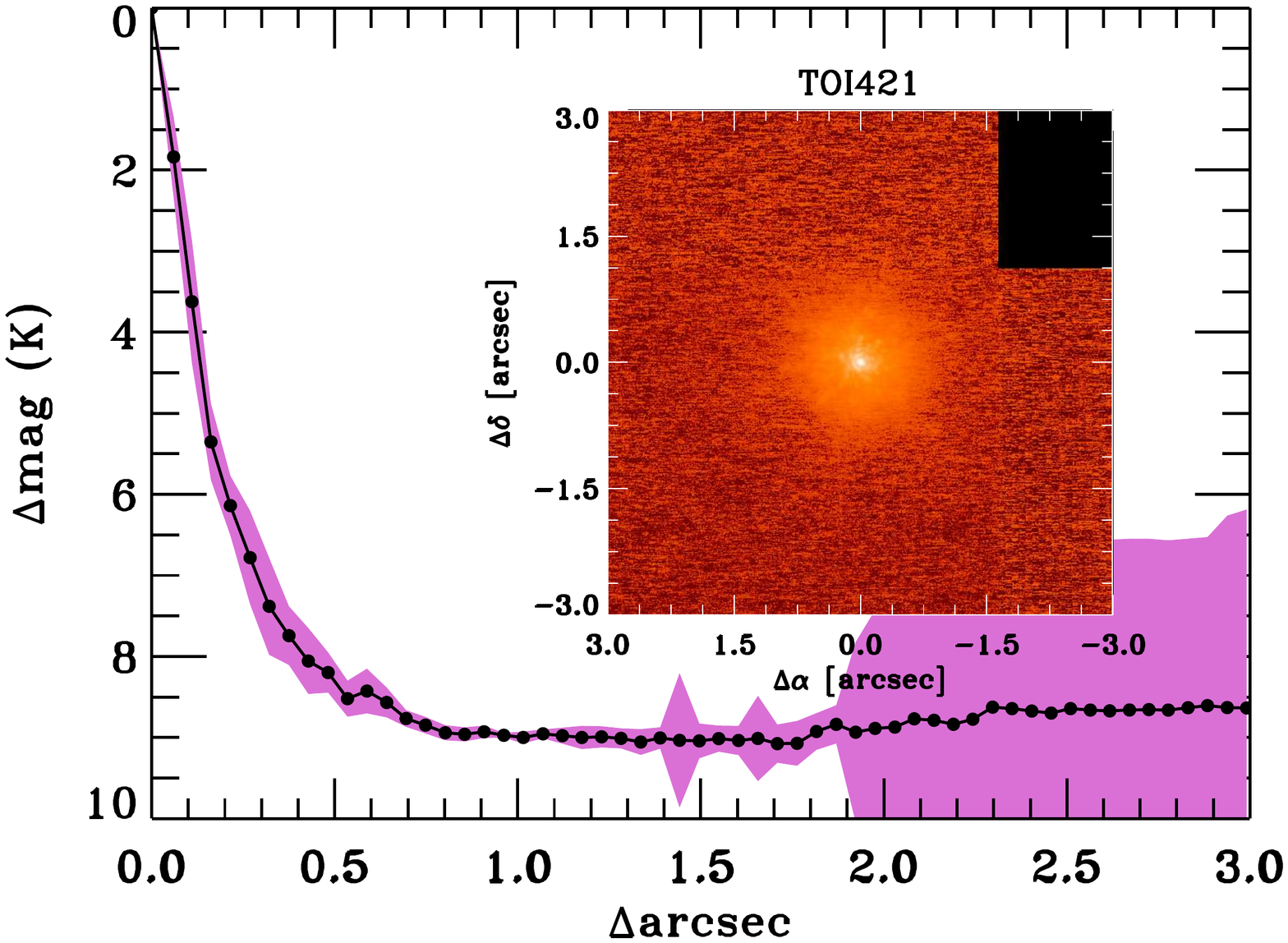}
\caption{Companion sensitivity for the Keck adaptive optics imaging.  The black points represent the 5$\sigma$ limits and are separated in steps of 1 FWHM ($\sim 0.053$\arcsec); the purple represents the azimuthal dispersion (1$\sigma$) of the contrast determinations (see text). The inset image is of the primary target showing no additional companions to within 3\arcsec\ of the target.}
\label{fig:ao_contrast}
\end{figure}

The Keck AO observations show no additional stellar companions were detected to within a resolution $\sim$0.053$\arcsec$ FWHM (Figure \ref{fig:ao_contrast}). The sensitivities of the final combined AO image were determined by injecting simulated sources azimuthally around the primary target every $45^\circ $ at separations of integer multiples of the central source's FWHM \citep{Furlanetal2017}. The brightness of each injected source was scaled until standard aperture photometry detected it with $5\sigma $ significance. The resulting brightness of the injected sources relative to the target set the contrast limits at that injection location. The final $5\sigma $ limit at each separation was determined from the average of all of the determined limits at that separation; the uncertainty on the 5$\sigma$ limit was set by the rms dispersion of the azimuthal slices at a given radial distance. The sensitivity curve is shown in Figure \ref{fig:ao_contrast} along with an inset image zoomed to primary target showing no other companion stars.

We also fitted the FIES, HARPS, HIRES, and PFS measurements using the same RV models presented in Table~\ref{tab:models} (Sect.~\ref{sec:rv}) including an RV linear trend to account for the presence of a potential outer companion. We found an acceleration consistent with zero within less than 1-$\sigma$.

\section{Frequency analysis of the HARPS data and stellar activity}\label{sec:rvfreq}

\begin{figure}[!ht]
\resizebox{\hsize}{!}{\includegraphics[trim={0 0 0 0},clip]{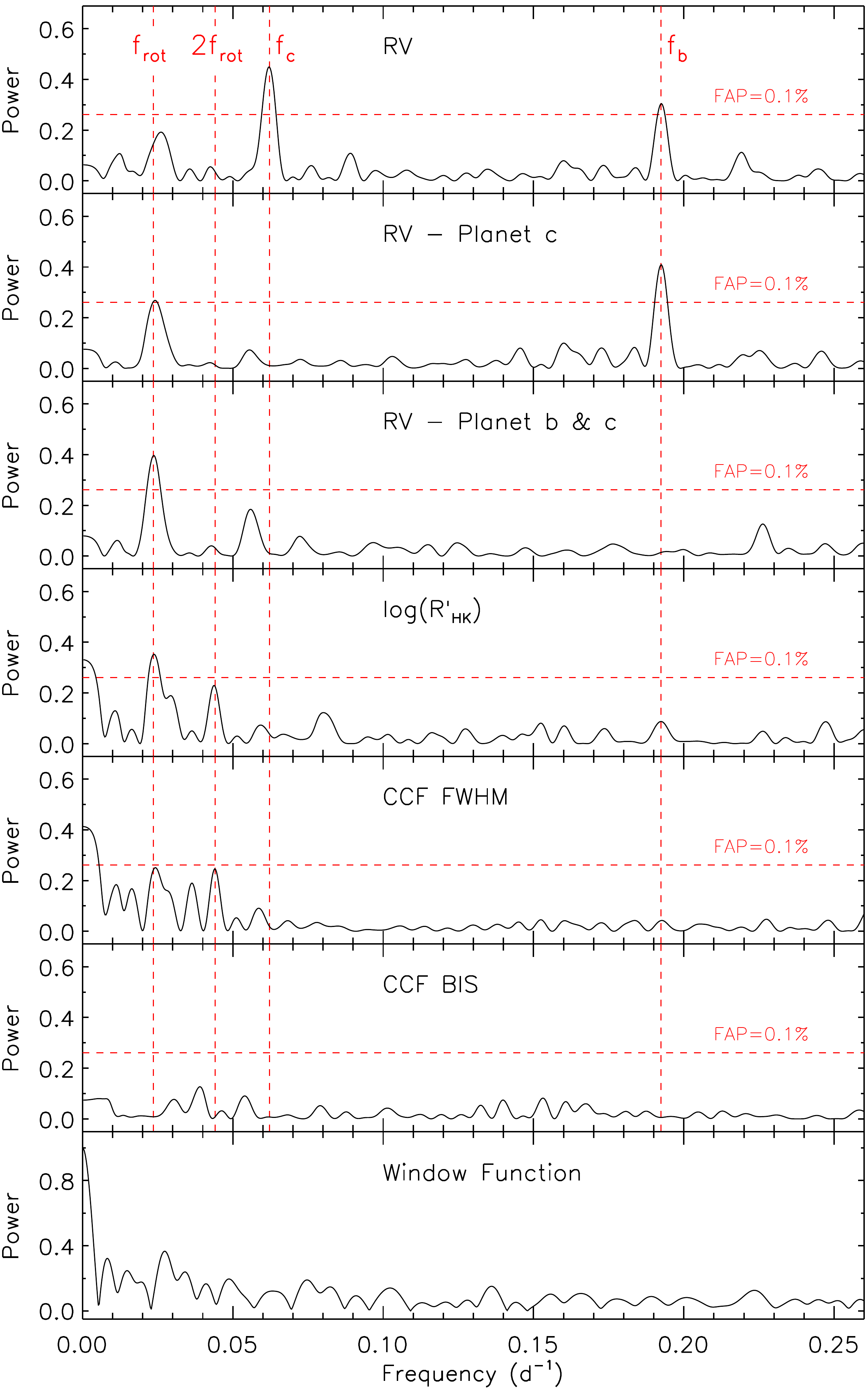}}
\caption{Generalized Lomb-Scargle periodogram of the 98 HARPS measurements acquired between August 2019 and January 2020 (upper panel) and RV residuals, following the subtraction of the Doppler signals of planet c (second panel), and planets b \& c (third panel). The periodogram of the Ca {\sc ii} H \& K lines activity indicator \logrhk, of the FWHM and BIS, and of the window function are shown in the last four panels. The horizontal dashed lines mark the false alarm probability at 0.1\,\%. The orbital frequencies of planet b ($f_\mathrm{b}$\,$\approx$\,0.193\,d$^{-1}$) and c ($f_\mathrm{c}$\,$\approx$\,0.062\,d$^{-1}$), as well as the stellar rotation frequency ($f_\mathrm{rot}$\,$\approx$\,0.024\,d$^{-1}$) and its first harmonic (2$f_\mathrm{rot}$\,$\approx$\,\,0.047\,d$^{-1}$) are marked with vertical dashed lines.} 
\label{fig:GLS_Periodograms}
\end{figure}

In order to search for the Doppler reflex motion induced by the 16-day transiting planet candidate and unveil the presence of possible additional signals induced by other orbiting planets and/or stellar activity, we performed a frequency analysis of the RV measurements, as well as of the Ca {\sc ii} activity index (\logrhk) and CCF asymmetry indicators (FWHM and BIS). To this aim, we used only the HARPS measurements, as they form the largest homogeneous data-set among our spectroscopic data, and analyzed only the 98 HARPS measurements collected between August 2019 and January 2020, to avoid the presence of spurious peaks associated to the 1-year sampling in the power spectrum of the HARPS time series.

The generalized Lomb-Scargle periodogram \citep{Zechmeister2009} displays a significant peak at the frequency of the transit signal reported by the \tess\ QLP pipeline  (f$_\mathrm{c}$\,$\approx$\,0.062\,d$^{-1}$, P$_\mathrm{c}$\,$\approx$\,16.1\,d). Following the bootstrap method \citep{Murdoch1993,Hatzes2016}, we assessed its false alarm probability (FAP) by computing the GLS periodogram of $10^5$ mock time-series obtained by randomly shuffling the Doppler measurements and their uncertainties, while keeping the time-stamps fixed. We found that the peak at f$_\mathrm{c}$ has a FAP\,$\ll 0.1$~\%. We note that the periodogram of the Ca {\sc ii} activity index \logrhk, as well as those of the CCF asymmetry indicators (FWHM and BIS) do not show any significant peak at f$_\mathrm{c}$ (Fig.~\ref{fig:GLS_Periodograms}, fourth, fifth and sixth panels), providing strong evidence that this Doppler signal is due to the stellar reflex motion induced by the transiting planet \sname\,c detected in the \tess\ light curve.

The periodogram of the 98 HARPS RVs shows also a significant peak (FAP$<$0.1\,\%) at f$_\mathrm{b}$\,$\approx$\,0.193\,d$^{-1}$ (P$_\mathrm{b}$\,=\,5.2\,d) whose power increases once the Doppler signal of \sname\,c is removed\footnote{We removed the Doppler signal of \sname\,c from the HARPS RVs by fitting a circular model, fixing period and time of first transit to the \tess\ ephemeris, while allowing for the systemic velocity and RV semi-amplitude to vary.} (Fig.~\ref{fig:GLS_Periodograms}; first and second panel). This second peak has no counterpart in the periodograms of \logrhk, FWHM, and BIS, indicating that it is not due to stellar activity. A re-analysis of the \tess\ light curve unveils the presence of a transit signal at 5.2~d, as described in Section~\ref{sec:tessphot}, confirming that the Doppler signal discovered in the RV time series is associated to a second transiting planet (\sname\,b). 

The periodogram of the RV residuals following the subtraction of the Doppler signal induced by \sname\,c (Fig.~\ref{fig:GLS_Periodograms}; second panel) shows also a second significant peak\footnote{We estimated the uncertainty from the width of the peak by fitting a Gaussian function.} (FAP$<$0.1\,\%) at f$_\mathrm{rot}$\,=\,0.024\,$\pm$\,0.003\,d$^{-1}$, corresponding to a period of P$_\mathrm{rot}$\,=\,$42\,^{+6}_{-5}$~d and an RV semi-amplitude variation of $\sim$2.4\,\ms, whose power becomes stronger once the Doppler signal of \sname\,b is also removed (third panel). This peak is also significantly detected in the GLS periodogram of \logrhk\ (FAP\,$<$\,0.1\,\%; fourth panel). It is also found in the periodogram of the FWHM (fifth panel), although with a higher false alarm probability (FAP\,$\approx$\,0.2\,\%). This suggests that the rotation period of the star is close to $\sim$42\,days and that the third Doppler signal at f$_\mathrm{rot}$ is induced by intrinsic stellar variability associated with the presence of active regions rotating on and off the visible stellar disk. We note that the periodograms of \logrhk\ and FWHM (Figure~\ref{fig:GLS_Periodograms}, fourth and fifth panels) display also a peak at the first harmonic of the rotation period (2f$_\mathrm{rot}$), which is likely due to the presence of active regions at opposite stellar longitudes. 

\begin{figure}
   \centering
\includegraphics[width=0.9\linewidth]{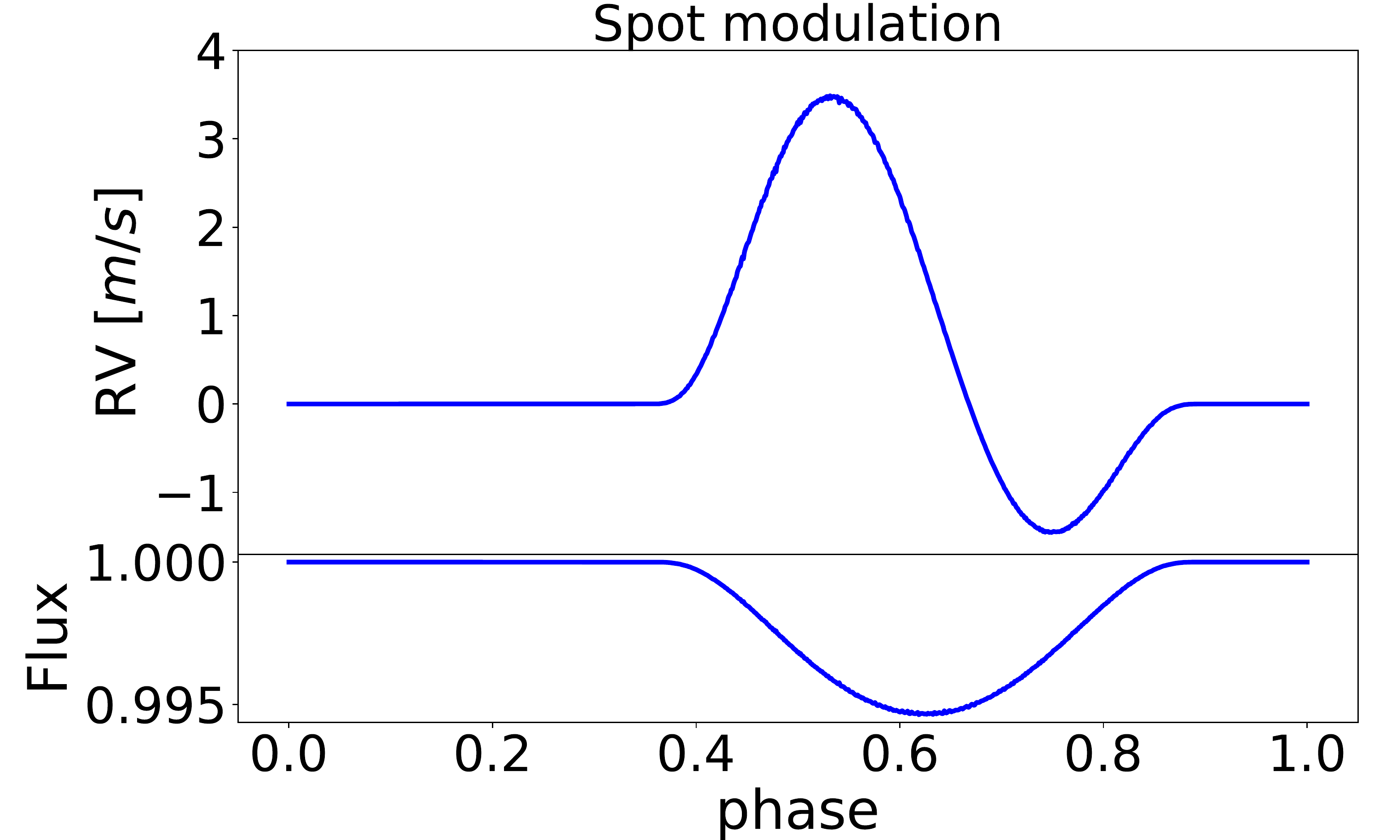}
\includegraphics[width=0.9\linewidth]{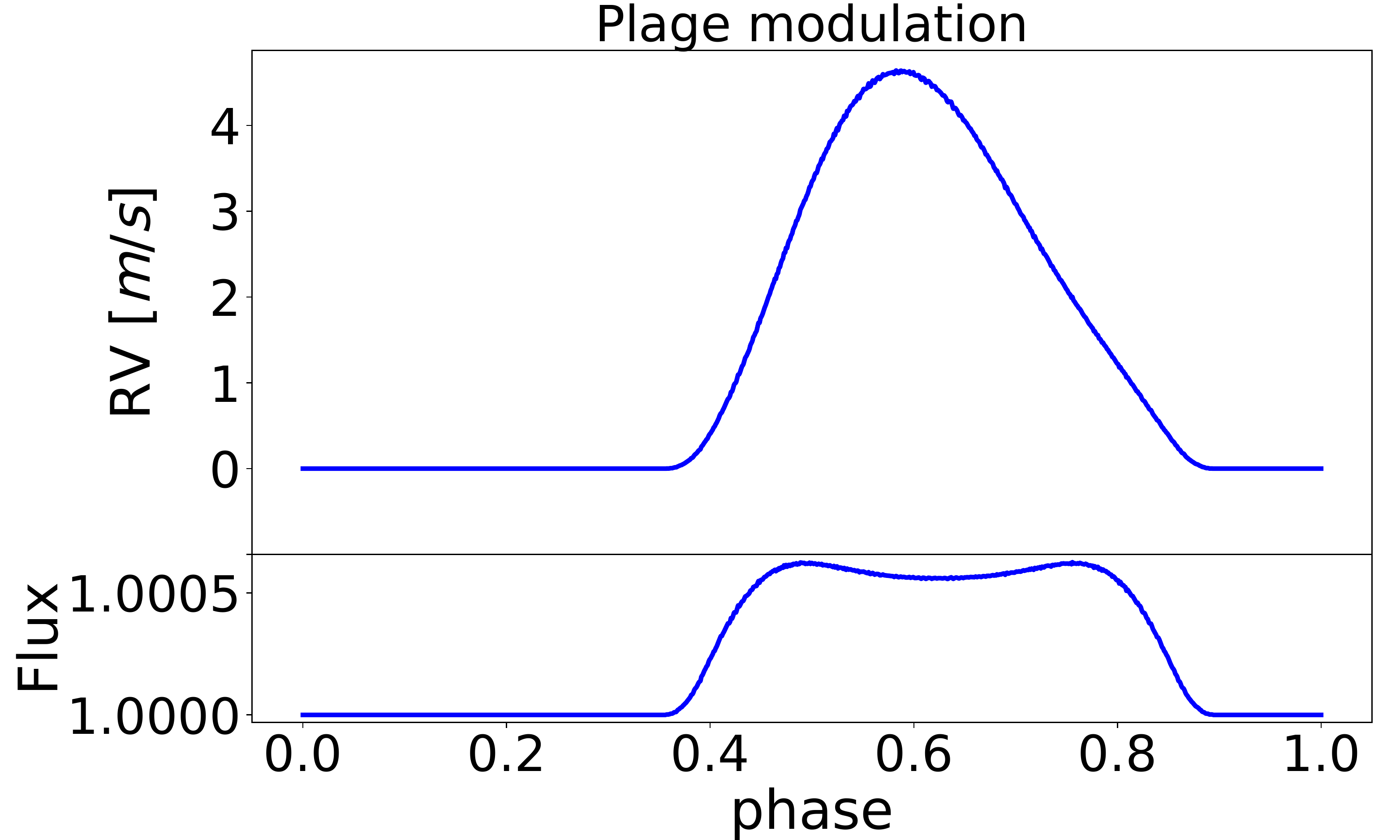}
\caption{Simulated photometric and spectroscopic effects induced by activity regions with radius 0.12~$R_\star$ and located at a stellar latitude of $30^{\circ}$. The upper panel shows the results for a spot with $\Delta T = 663K$ with respect of the quiet photosphere of \sname. The lower panel shows the same effects for a plage with $\Delta T = 251K$.}
\label{spot_plage}
\end{figure}

In order to further investigate the nature of the $\sim$42-day signal, we searched the WASP-South data (Section~\ref{Sec:WASP-South}) for any rotational modulation using the methods from \citet{2011PASP..123..547M}. We found no significant periodicity, with a 95\%-confidence upper limit on the amplitude of 1 mmag. We did find a significant periodicity compatible with the lunar cycle, but this was seen only in the data from the 85-mm lenses, which are more vulnerable to moonlight, and not in the 200-mm data; furthermore, in the 85-mm data the same signal was also seen in adjacent field stars, so we concluded that it is not intrinsic of \sname.

With a mean \logrhk\ of  $-$4.93\,$\pm$\,0.04 (Table~\ref{Table:HARPS}), \sname\ is a relatively quiet star. The lack of significant rotational modulation in the WASP-South light curve might be explained if the spot-induced variability of \sname\ is too low to produce a photometric signal with an amplitude higher than the WASP-South photometric precision. We further investigated this scenario with the code \texttt{SOAP2.0} \citep{Dumusque}, which simulates stellar activity using a fine grid to model the photosphere of a spotted rotating star. For each grid cell, \texttt{SOAP2.0} simulates the local CCF using as a reference the solar HARPS CCF, and accounts for the contribution of spots and plages using the HARPS CCF of a solar active region. The adoption of of the solar CCF is an advantage for our test, because \sname\ is a G9 star, meaning that it is more likely to behave as our Sun. For a given set of stellar parameters and spots/plages distribution, size, and temperature, \texttt{SOAP2.0} can estimate the photometric and Doppler signals induced by active regions, accounting for the inhibition of the convective blue-shift and limb darkening/brightening effects.

For \sname\ we used the effective temperature \teff\ and stellar radius $R_\star$ listed in Table~\ref{tab:stellar}, and assumed a rotation period of $P_\mathrm{rot}$\,=\,42~days. To account for the wavelength range covered by the HARPS spectra, we adopted the linear and quadratic limb darkening coefficients of the Sun \citep[$q_1$\,=\,0.29 and $q_2$\,=\,0.34;][]{Claret1}. Assuming a simplified model with one single spot, we found that the $\sim$2.4\,\ms\ RV semi-amplitude variation induced by stellar activity could be accounted for by a typical sunspot, with a temperature contrast  with respect to the quiet photosphere of $\Delta T$\,=\,663~K \citep{Meunier}, a radius of 0.10\,$R_\star$ and placed at a latitude of $30^{\circ}$, which is the average active latitude for the Sun \citep{Donati09, Strassmeier09}. The corresponding photometric variation would have an amplitude of 5000~ppm, equivalent to $\sim$ 5\,mmag, which is higher than the 1~mmag upper limit of the WASP-South time-series of \sname. We performed a similar test by replacing the spot with a plage. We assumed a temperature contrast of $\Delta T$\,=\,251~K \citep[typical plage contrast for the Sun;][]{Meunier} and a radius of 0.12 $R_\star$, while we kept latitude identical to that of the spot ($30^{\circ}$). While the resulting RV semi-amplitude is $\sim$2.4\,\ms\, the photometric variation is 500~ppm, which corresponds to $\sim$0.5~mmag, i.e., lower than the 1 mmag upper limit on the amplitude observed in the WASP-South photometry. In Figure \ref{spot_plage}, we show the results of our simulations. The upper panel displays the effect of the simulated spot on both the RV and the photometric signal, whereas the bottom panel shows the same for a plage. The photometric effect induced by a plage is thus one order of magnitude smaller than that caused by a spot. These results agree with those reported in \citet{Dumusque} and, more recently, by \citet{Shapiroetal2016} and \citet{2019ApJ...874..107M}

\begin{figure}[t]
   \centering
\includegraphics[width=0.9\linewidth]{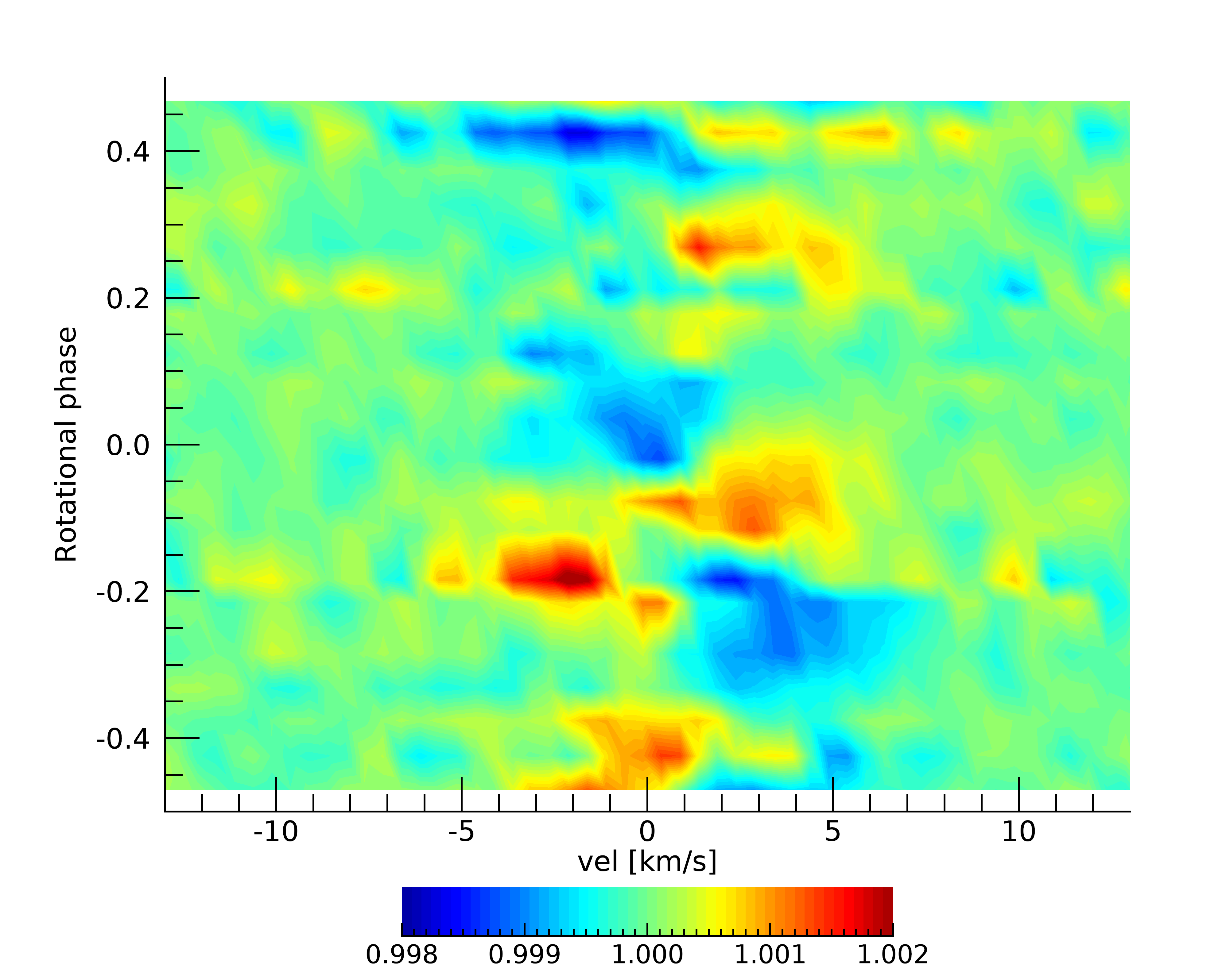}
\caption{Contour map of the CCF residuals of \sname\ versus radial velocity and rotational phase. The color bar indicates relative CCF amplitude with respect to the mean CCF.}
\label{fig:CCFresiduals}
\end{figure}

To understand if the Doppler signal of \sname\ is spot- or plage-dominated, we produced an average contour map of the HARPS CCF residuals (after the division for the average CCF), plotted versus radial velocity and stellar rotation phase. The results are displayed in Figure~\ref{fig:CCFresiduals}. Positive deviations (i.e., cool star spots) are shown in red, while negative deviations (hot spots) are shown in blue. These can account for the RV variation due to stellar activity, if we consider their associated perturbation to be $\Delta RV$\,$\simeq$\,2 $\times$\,FWHM\,$\times$\,$\Delta I$\,$\times$\,$f$\,$\simeq$\,14\,$\times$\,$f$, 
where $\Delta I$\,$\sim$\,0.004 is the intensity range and $f$\,$\le$\,1 the filling factor \citep{carleo2020}. The contours show that the activity of the star is dominated by plages, though some spots are also evident.

We conclude that the activity-induced Doppler signal of \sname\ can be explained by plages that would also account for the non detection of any photometric variability in the WASP-South time-series. Alternatively, the star was photometrically quieter at the time of the WASP-South observations (2008-2015) and more active during our spectroscopic follow-up (2019-2020). This is corroborated by the fact that the contour map of the HARPS CCF shows also the presence of spots.

We also analyzed the \tess\ data to look for surface rotation modulation in the light curve due to the passage of spots or active regions on the visible disk. We did the analysis with two different light curves. The first one is the one described in Section \ref{sec:2.1tess} and the second one is the one optimized for asteroseismology following Gonz\'alez-Cuesta et al., in prep. The optimized aperture was obtained by selecting larger and larger apertures with different thresholds in the flux starting from the SPOC aperture (the smaller one) up to the larger one with a threshold of 10e-/s with increments of 10e-/s. In the resultant light curve only points with a quality flag equal to zero have been retained. Missing points have been interpolated using inpainting techniques as in \cite{garciaetal2014a}. The light curve has also been corrected from outliers and stitched together following \cite{garciaetal2011}. The optimization is done by comparing the signal measured at the expected region for the modes (around 3500\,$\mu$Hz, Section 3) and the high-frequency noise above 2000\,$\mu$Hz. The best aperture found was the one with a threshold of 80\,e-/s for sector 5 and a threshold of 10\,e-/s for sector 6.
For both light curves we removed the transits and concatenated them. Then we applied our rotation pipeline following \cite{garciaetal2014b}, \cite{Mathuretal2014}, and \cite{Santosetal2019}. Our methodology consists of performing a time-frequency analysis with wavelets, computing the auto-correlation function (ACF) and a composite spectrum (CS) that is a combination of the first 2 methods (see \citealt{Ceillieretal2017}). We found a signal at 45 $\pm$ 3.54 days in the three methods. The heights of the ACF and CS are above the thresholds defined in \cite{Ceillieretal2017} to reliably select a rotation period. However, we usually require that we observe 3 rotation periods to have more reliability on the rotation period. With only 58 days of observations we cannot not fulfill this criteria. However, having this period obtained with these 3 different methods and using 2 different processing of the light curves (in terms of apertures) suggests that this period could be from stellar origin and it is independent on the processing of the light curve or the aperture selected. Because the HARPS RV analysis also finds a rotation period of $\sim$42 days, it gives more weight on it being real and the analysis of \tess\ data complements the spectroscopic analysis. Indeed with the \tess\ photometry alone we could not be confident enough about the period found as it could still be a harmonic of a longer periodicity or still something of instrumental origin.

\section{RV modeling}\label{sec:rv}

Motivated by the results of our frequency analysis, we performed a series of fits to the RV data to enable model selection and obtain system parameter estimates. Specifically we used \texttt{RadVel} \footnote{\url{https://github.com/California-Planet-Search/radvel}.} \citep{2018PASP..130d4504F} to test six different two-planet models: circular orbits (``2c''), eccentric orbits (``2e''), circular orbits with a Gaussian Process (GP) noise model (``2cGP''), eccentric orbits with a GP noise model (``2eGP''), circular orbits with an additional sine curve for the stellar activity (``2cS''), and eccentric orbits with an additional sine curve for the stellar activity (``2eS''). We used a GP model with a quasi-periodic kernel, which has been used extensively in the literature to model stellar RV signals (see, e.g., \citealt{2014MNRAS.443.2517H,2015ApJ...808..127G,2017AJ....154..226D}); to avoid over-fitting we imposed wide Gaussian priors on the hyper-parameters loosely informed by our frequency analysis. 

\begin{table}
\centering
\caption{RV model selection. \label{tab:models}}
\begin{tabular}{lccccccr}
\hline
Model &     AICc & BIC & $N_\mathrm{free}$ & $N_\mathrm{data}$ &    RMS$^a$ & $\ln{\mathcal{L}}^b$ \\
\hline
2eS &   587.83 &   633.88 &     19 &    123 &   2.25 &  -260.96 \\
2cS &   591.48 &   629.18 &     15 &    123 &   2.38 &  -268.33 \\
2eGP &   596.78 &   644.79 &     20 &    123 &   1.57 &  -267.24 \\
2cGP &   601.06 &   640.92 &     16 &    123 &   1.69 &  -275.19 \\
2c &   630.25 &   661.16 &     12 &    123 &   2.78 &  -282.70 \\
2e &   634.49 &   674.36 &     16 &    123 &   2.71 &  -279.66 \\
\hline
\multicolumn{7}{l}{$^a$ Root mean square of the data minus the model.}\\
\multicolumn{7}{l}{$^b$ Log-likelihood of the data given the model.}\\
\end{tabular}
\end{table}

To compare the quality of these models, we computed both the commonly used Bayesian Information Criterion (BIC) and the Akaike Information Criterion (AICc; corrected for small sample sizes), which is a second-order estimator of information loss. The results of these fits are presented in Table~\ref{tab:models}, sorted in ascending order of AICc (best to worst). The 2eS model is strongly favored over the other models by the AICc, suggesting that the orbits of the two transiting planets are significantly eccentric and that the stellar activity is reasonably well described by a sinusoid. We note that the BIC presents a slight preference for the 2cS model, but the AICc has been suggested to have practical advantages over the BIC \citep{Burnham2014}. We performed a full MCMC exploration of the parameter space of the 2eS model using \texttt{RadVel}, yielding semi-amplitudes of $K_{b} = 4.33^{+0.37}_{-0.35}$ and $K_{c} = 3.05^{+0.35}_{-0.34}$ m\,s$^{-1}$ for the planetary components, and $2.52\pm 0.36$ m\,s$^{-1}$ for the stellar component; the eccentricities are significant at the $\sim$2$\sigma$ level and constrained to $e_{b} = 0.147^{+0.069}_{-0.065}$ and $e_{c} = 0.171^{+0.087}_{-0.086}$.

\section{Joint analysis}\label{sec:joint}

\begin{figure*}[ht!]
    \centering
    \includegraphics[width=0.98\textwidth]{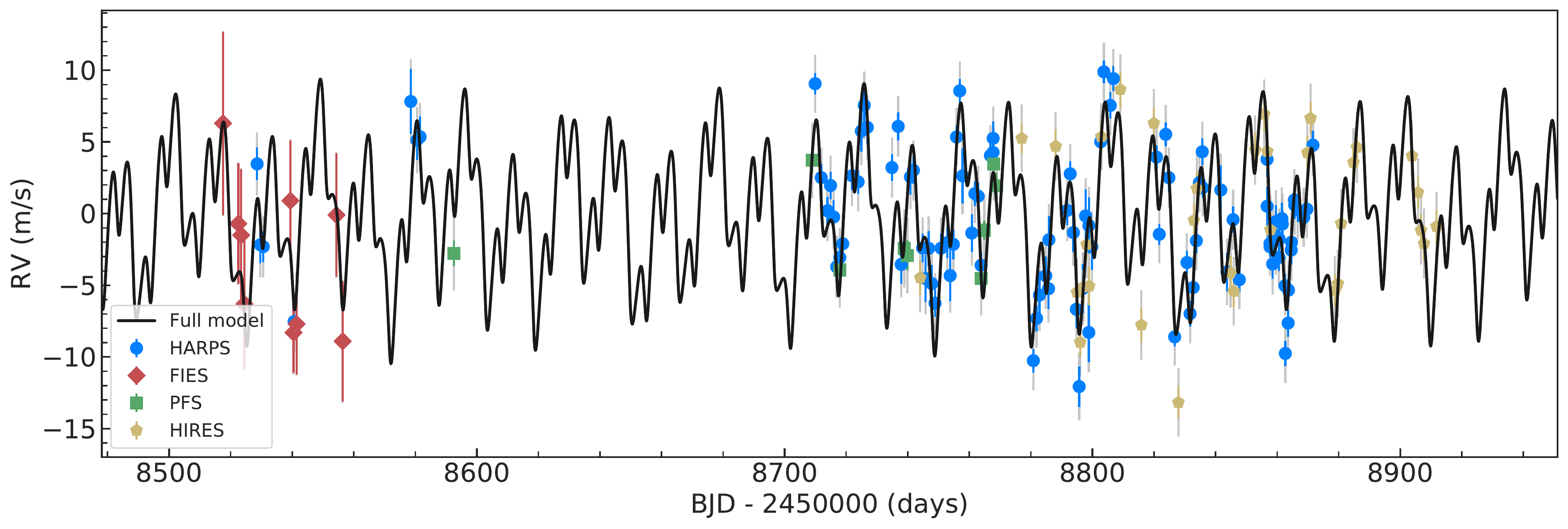}\\
    \caption{RV time-series. HARPS (blue circles), FIES (red diamonds), PFS (green squares), and HIRES (yellow pentagons) data are shown following the subtraction of the each inferred offset. The inferred full model (i.e. two planet signals plus the activity induced signal) is presented as solid continuous line. Grey error bars account for the inferred jitter for each instrument.}
    \label{fig:timeseries}
\end{figure*}

\begin{figure}[ht!]
    \centering
    \includegraphics[width=0.47\textwidth]{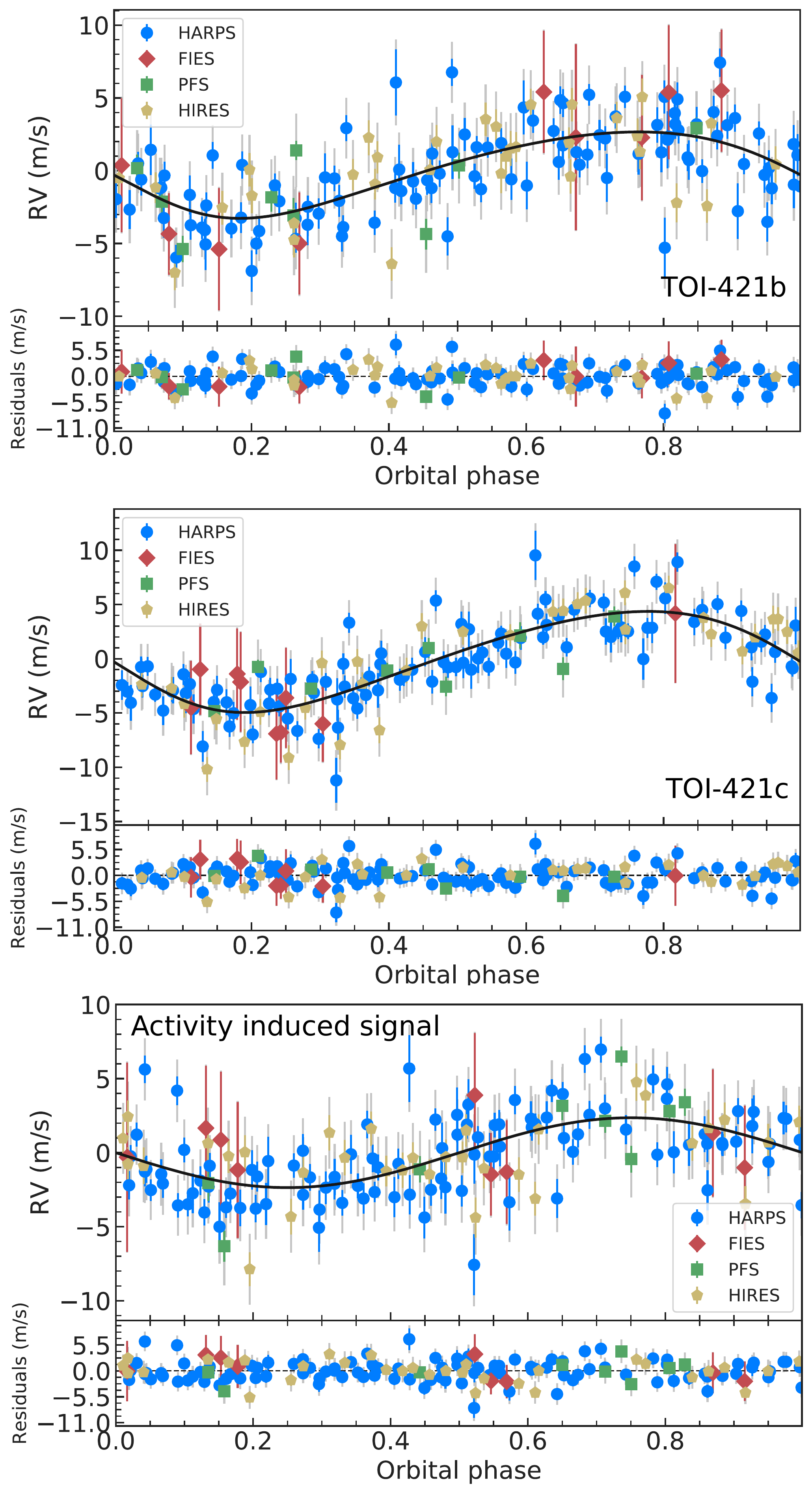}
    \caption{Phase-folded RV plots with residuals for \sname b (top panel), \sname c (middle panel), and activity induced signal (lower panel). HARPS (blue circles), FIES (red diamonds), PFS (green squares), and HIRES (yellow pentagons) data are shown following the subtraction of the each inferred offset and the other signals. Black solid line shows the inferred model for each case. Grey error bars account for the inferred jitter for each instrument. }
    \label{fig:rvs}
\end{figure}

\begin{figure}
    \centering
    \includegraphics[width=0.47\textwidth]{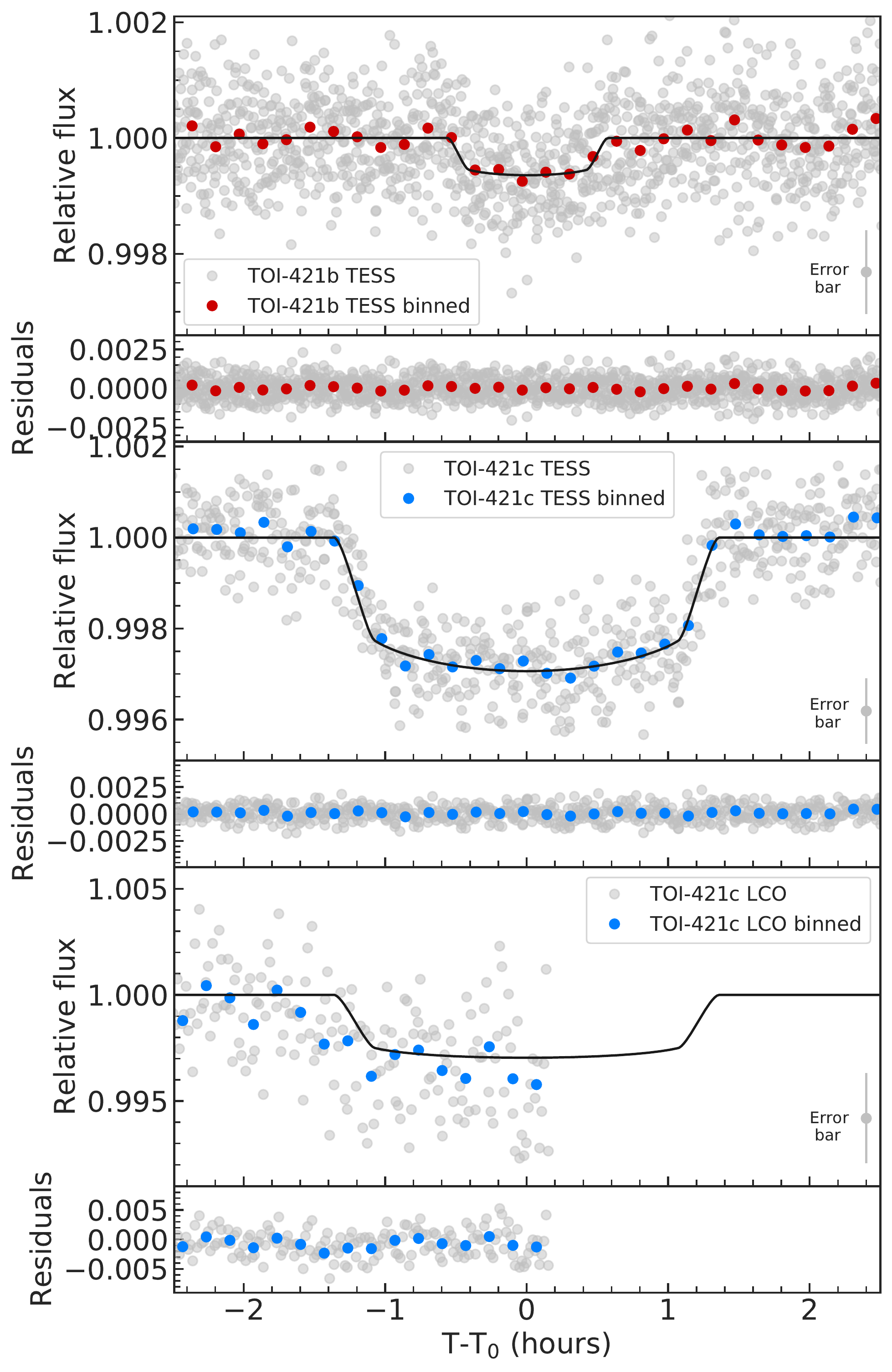}\\
    \caption{Light curves around the transit with residuals of \sname b (upper and middle panels refer to \tess\ and LCOGT, respectively) and \sname c (\tess, lower panel) with inferred transit model over-plotted. Data are shown in the nominal short-cadence mode and binned to 10 min. 
    Typical error bar for nominal data are shown at the bottom right for each panel.}
    \label{fig:transits}
\end{figure}

We performed a global analysis of our RV and transit data with the open-source code \pyaneti\ \citep{pyaneti}. Briefly, \pyaneti\ creates posterior distributions of the fitted parameters using a Markov chain Monte Carlo (MCMC) sampling approach. 

Transits are modelled following a \citet{Mandel2002} quadratic limb darkening model, implemented with the parametrization suggested by \citet{Kipping2013}. A preliminary fit of the light curve shows that the limb-darkening coefficients are not well constrained for the LCOGT data (we note that the limb-darkening coefficients for the \tess\ data are constrained by data themself). Therefore, we used the code \texttt{LDTk} \citep{Husser2013,Parviainen2015} to estimate the limb darkening coefficients corresponding to a star with the stellar parameters presented in Table~\ref{tab:stellar}, and the Pan-STARSS Y-band ($940 - 1060$ nm). We used a Gaussian prior with mean given by the \texttt{LDTk} estimation and a conservative standard deviation of 0.1.
We also note that we sampled for the stellar density $\rho_\star$, and recover $a/R_\star$ for each planet using Kepler's third law \citep[for more details see, e.g.,][]{Winn2010}. We implemented a Gaussian prior on the stellar density using the stellar mass and radius provided in Section~\ref{sec:stellar} to constrain better the orbit eccentricities \citep[see e.g., ][]{vaneylen2015}. 
For the transit analysis, we did not include the potential contamination of nearby stars because the expected effect is smaller than the white noise of the light curve (see Sect.~\ref{sec:contamination}).
We note that the combination of this transit analysis, together with the RV data, provides a stronger constrain on the orbital eccentricities. The final analysis supports the conclusions of Section~\ref{sec:rv} (See Table~\ref{tab:parstoi421}).

The RV model was chosen following the results presented in Section~\ref{sec:rv}. We used two Keplerian orbits to model the Doppler reflex motion induced by the two transiting planets, and an extra sinusoidal to take into account the activity signal induced by the star at its rotation period.

We used 500 Markov chains to explore the parameter space. We stopped the sampling once all chains converged \citep[following] [we define convergence when $R < 1.02$ for all parameters]{Gelman1992}. The posterior distribution were created using the last 2500 converged iterations and the 500 chains, leading to a posterior of 1,250,000 points per parameter.

Details on the fitted parameters, adopted priors, and parameter estimates are given in Table~\ref{tab:parstoi421}. Inferred parameters are defined as the median and 68\% region of the credible interval of the posterior distributions for each fitted parameter. Figure~\ref{fig:timeseries} shows the RV model time-series. The phase-folded RV and transit plots are shown in Figures \ref{fig:rvs} and \ref{fig:transits}, respectively.
We note that there is an apparent shift of $\sim 20$ min on the LCO transit in Fig. \ref{fig:transits}. Given that the expected TTVs in the system are smaller (see Section \ref{sec:ttvs}), it is likely that this effect is caused by systematics in the ground-based data.

\begin{table*}
  \footnotesize
  \caption{\sname\ parameters. \label{tab:parstoi421}}  
  \centering
  \begin{tabular}{lcc}
  \hline
  Parameter & Prior$^{(\mathrm{a})}$ & Value$^{(\mathrm{b})}$  \\
  \hline
  \multicolumn{3}{l}{\emph{ \bf Model Parameters for  \sname b}} \\
    Orbital period $P_{\mathrm{orb}}$ (days)  & $\mathcal{U}[ 5.1917 , 5.2017]$   & \Pb[] \\
    Transit epoch $T_0$ (BJD - 2,450,000)  & $\mathcal{U}[8441.2335 , 8441.3335 ]$  & \Tzerob[]  \\
    $\sqrt{e} \sin \omega_\star$ &  $\mathcal{U}(-1,1)$ & \esinb \\
    $\sqrt{e} \cos \omega_\star$  &  $\mathcal{U}(-1,1)$ & \ecosb \\
    Scaled planetary radius $R_\mathrm{p}/R_{\star}$ &  $\mathcal{U}[0,0.1]$ & \rrb[]  \\
    Impact parameter, $b$ &  $\mathcal{U}[0,1.1]$  & \bb[] \\
    Radial velocity semi-amplitude variation $K$ (m s$^{-1}$) &  $\mathcal{U}[0,50]$ & \kb[] \\
    \multicolumn{3}{l}{\emph{ \bf Model Parameters for  \sname c}} \\
    Orbital period $P_{\mathrm{orb}}$ (days)  &  $\mathcal{U}[16.0642 , 16.0741]$ & \Pc[] \\
    Transit epoch $T_0$ (BJD - 2,450,000)  & $\mathcal{U}[  8440.0804 , 8440.1804 ]$  & \Tzeroc[]  \\
    $\sqrt{e} \sin \omega_\star$ &  $\mathcal{U}(-1,1)$ & \esinc \\
    $\sqrt{e} \cos \omega_\star$  &  $\mathcal{U}(-1,1)$ & \ecosc \\
    Scaled planetary radius $R_\mathrm{p}/R_{\star}$ &  $\mathcal{U}[0,0.1]$ & \rrc[]  \\
    Impact parameter, $b$ &  $\mathcal{U}[0,1.1]$  & \bc[] \\
    Radial velocity semi-amplitude variation $K$ (m s$^{-1}$) &  $\mathcal{U}[0,50]$ & \kc[] \\
    \multicolumn{3}{l}{\emph{ \bf Model parameters of activity induced RV sinusoidal signal}} \\
     Period $P_{\mathrm{rot}}$ (days)  &  $\mathcal{U}[ 35 , 50]$ & \Pd[] \\
    Epoch $T_0$ (BJD - 2,450,000)  & $\mathcal{U}[8412.2835 , 8452.2835 ]$ & \Tzerod[]  \\
    Radial velocity semi-amplitude variation $K$ (m s$^{-1}$) &  $\mathcal{U}[0,50]$ & \kd[] \\
    \multicolumn{3}{l}{\emph{ \bf Other system parameters}} \\
    Stellar density $\rho_\star$ &  $\mathcal{N}[1.91,0.14]$ & \dentrheec[]  \\
    Systemic velocity $\gamma$ (\kms)$^{(\mathrm{c})}$ & $\mathcal{U}[ 79.0318 , 80.0537]$ & \HARPS[] \\
    Instrumental systemic velocity $\gamma$ (\kms)$^{(\mathrm{c})}$ & $\mathcal{U}[ -0.5, 0.5]$ & \FIES[] \\
    Instrumental systemic velocity $\gamma$ (\kms)$^{(\mathrm{c})}$ & $\mathcal{U}[ -0.5, 0.5]$ & \PFS[] \\
    Instrumental systemic velocity $\gamma$ (\kms)$^{(\mathrm{c})}$  & $\mathcal{U}[  -0.5, 0.5 ]$ & \HIRES[]  \\
    RV jitter term $\sigma_{\rm HARPS}$ (\ms) & $\mathcal{U}[0,100]$ & \jHARPS[] \\
    RV jitter term $\sigma_{\rm FIES}$ (\ms) & $\mathcal{U}[0,100]$ & \jFIES[] \\
    RV jitter term $\sigma_{\rm PFS}$ (\ms) & $\mathcal{U}[0,100]$ & \jPFS[] \\
    RV jitter term $\sigma_{\rm HIRES}$  (\ms) & $\mathcal{U}[0,100]$ & \jHIRES[] \\
    Limb darkening $q_1$ \tess\  & $\mathcal{U}[0,1]$ & \qone \\
    Limb darkening $q_2$ \tess\ & $\mathcal{U}[0,1]$ & \qtwo \\
    Limb darkening $q_1$ Pan-STARSS Y-band & $\mathcal{N}[0.24,0.1]$ & \qoneground \\
    Limb darkening $q_2$ Pan-STARSS Y-band & $\mathcal{N}[0.36,0.1]$ & \qtwoground \\
    \noalign{\smallskip}
    \hline
    \multicolumn{3}{l}{\textbf{Derived parameters for \sname b}} \\
    Planet mass ($M_{\oplus}$)  & $\cdots$ & \mpb[] \\
    Planet radius ($R_{\oplus}$)  & $\cdots$ & \rpb[] \\
    Planet density (${\rm g\,cm^{-3}}$) & $\cdots$ & \denpb[] \\
    semi-major axis $a$ (AU)  & $\cdots$ & \ab[] \\
    $e$  & $\cdots$ & \eb  \\
    $\omega_\star $ (deg)  &  $\cdots$ & \wb[]  \\
    Orbital inclination $i$ (deg)  & $\cdots$ & \ib[] \\
    Transit duration (hours) & $\cdots$ & \ttotb[] \\
     Equilibrium temperature$^{(\mathrm{d})}$ $T_{\rm eq}$ ($K$)   & $\cdots$ & \Teqb[]  \\
    Insolation $F_{\rm p}$ ($F_{\oplus}$)   & $\cdots$ & \insolationb[] \\
     \multicolumn{3}{l}{\textbf{Derived parameters for \sname c}} \\
    Planet mass ($M_{\oplus}$)  & $\cdots$ & \mpc[] \\
    Planet radius ($R_{\oplus}$)  & $\cdots$ & \rpc[] \\
    Planet density (${\rm g\,cm^{-3}}$) & $\cdots$ & \denpc[] \\
    semi-major axis $a$ (AU)  & $\cdots$ & \ac[] \\
    $e$  & $\cdots$ & \ec  \\
    $\omega_\star $ (deg)  &  $\cdots$ & \wc[]  \\
    Orbital inclination $i$ (deg)  & $\cdots$ & \ic[] \\
    Transit duration (hours) & $\cdots$ & \ttotc[] \\
     Equilibrium temperature$^{(\mathrm{d})}$ $T_{\rm eq}$ ($K$)   & $\cdots$ & \Teqc[]  \\
    Insolation $F_{\rm p}$ ($F_{\oplus}$)   & $\cdots$ & \insolationc[] \\
   \noalign{\smallskip}
   \hline
   \noalign{\smallskip}
  \end{tabular}
  \footnotesize
  ~\\
  \emph{Note} -- $^{(\mathrm{a})}$ $\mathcal{U}[a,b]$ refers to uniform priors between $a$ and $b$, $\mathcal{N}[a,b]$ to Gaussian priors with median $a$ and standard deviation $b$, and $\mathcal{F}[a]$ to a fixed value $a$.\\  
  $^{(\mathrm{b})}$ Parameter estimates and corresponding uncertainties are defined as the median and 68.3\% credible interval of the posterior distributions.\\
  $^{(\mathrm{c})}$ HARPS RVs are absolute measurements: the HARPS $\gamma$ velocity corresponds to systemic velocity. FIES, PFS, and HIRES RVs are relative measurements.\\
  $^{(\mathrm{d})}$ Assuming albedo equal to zero.
\end{table*}

\begin{figure}
    \centering
    \includegraphics[width=0.49\textwidth]{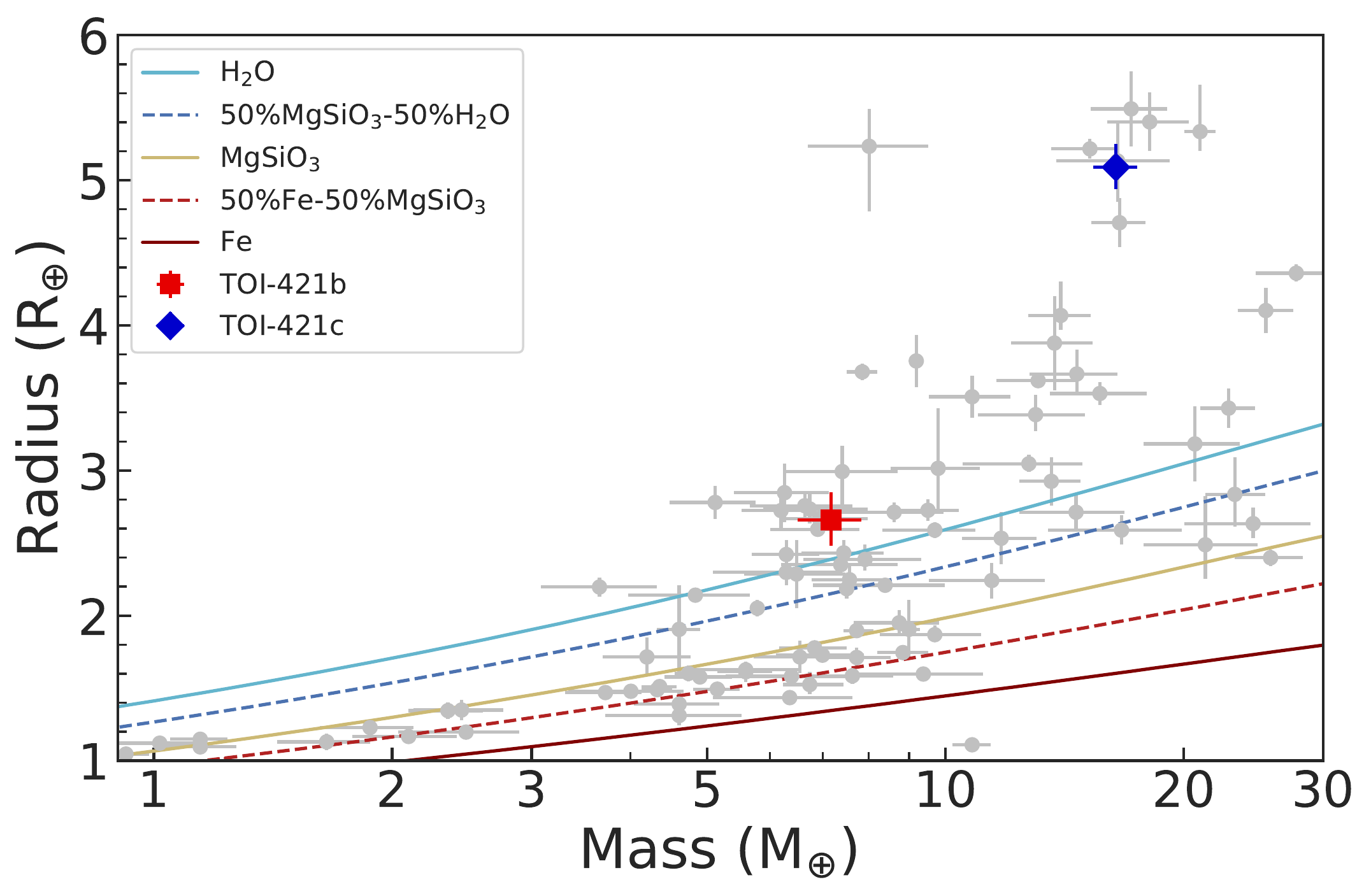}
    \caption{Mass-radius  diagram  for planets  with  mass  and  radius measurement precision  better  than  20\%  \citep[gray points, from  the  TEPCat database;][]{tepcat}. \sname b (red square) and  \sname (blue diamond) are shown for comparison. \citet[][]{Zeng2016}'s theoretical composition models  are shown using different lines and colours.}
    \label{fig:massradius}
\end{figure}{}

\section{Discussion} \label{sec:disc}
The innermost planet, \planetb, ($P_\mathrm{b}$\,=\,5.2\,days) has a mass of $M_\mathrm{b}$=\mpb\ and a radius of $R_\mathrm{b}$=\rpb, yielding a density of $\rho_\mathrm{b}$=\denpb. The outer transiting planet, \planetc, ($P_\mathrm{c}$\,$\approx$\,16.1~days) has a mass of $M_\mathrm{c}$\,=\,\mpc\ and a radius of $R_\mathrm{c}$\,=\,\rpc\, resulting in a mean density of $\rho_\mathrm{c}$\,=\,\denpc.  Figure~\ref{fig:massradius} shows the position of \planetb\ and c in the mass-radius diagram along with the sample of small planets ($R_\mathrm{p}$\,$\leqslant$\,6\,$R_{\oplus}$) whose masses and radii have been measured with a precision better than 20\%. Given their positions with respect to theoretical mass-radius relations, both planets are expected to host an atmosphere dominated by light elements, namely H and He. We performed a series of simulations -- including hydrogen escape rate (Section~\ref{sec:hyescape}), planetary atmospheric evolution (Section~\ref{sec:evolution}), and retrievals of the transmission spectrum (Section~\ref{sect:Retrievals}) -- which indicate that \planetb\ and c are very intriguing planets for atmospheric characterization. Finally, we compute the Helium 10830 \AA\ simulation following the approach in \cite{oklopcicetal2018} finding no significant Helium absorption ($\lesssim 0.5$\% at line center) for either of the two planets.



\subsection{Hydrogen escape}\label{sec:hyescape}



Atmospheric escape in close-in planets takes place when the  high-energy (X-ray$+$EUV; hereafter XUV) stellar photons photoionize and heat up the planetary upper atmospheres (e.g., \citealt{2009ApJ...693...23M}). 
\planetb's and c's close distances to the star suggest that their atmospheres may be significantly heated by the stellar high-energy emission and hence are undergoing escape. Therefore, these two planets are very appealing objects for studying the effects of mass loss. Interestingly, two other objects with similar bulk densities, that is K2-18\,b ($M_p=8.63M_\oplus$, $R_p=2.6R_\oplus$, $\rho_p=2.67\,$g\,cm$^{-3}$, \citealt{Bennekeetal2019}) and GJ3470\,b ($M_p=13.9M_\oplus$, $R_p=4.6R_\oplus$, $\rho_p=0.80\,$g\,cm$^{-3}$, \citealt{Awiphanetal2016}), have shown the presence of atmospheric escape through the detection of Ly-$\alpha$ absorption during transit \citep{Bourrier2018, dosSantos2020}. 

Here, we use the 1D hydrodynamic atmospheric escape model presented in \citet{2019MNRAS.490.3760A} to predict the behaviour of the planetary atmospheres under the influence of the photoionizing flux of the host star. With this we can infer the current properties of the planetary upper atmospheres, including the mass-loss rate. For the output of this 1D model we study the atmospheric signatures of \planetb\ and \planetc\ of the neutral hydrogen Ly-$\alpha$ and H-$\alpha$ lines during transit. Our model uses as input the XUV stellar luminosity, which was derived considering the median \logrhk\ value ($-$4.93\,$\pm$\,0.04) from Table~\ref{Table:HARPS}, converting it into a Ca\,II\,H\&K chromospheric emission flux using the equations listed in \citet{fossati2017a} and then by converting this emission in XUV flux using the scaling relations of \citet{linsky2013} and \citet{linsky2014}. We find an XUV flux at a distance of 1\,AU of $F_{\rm XUV}\,=\,$23.12\,erg\,cm$^{-2}$\,s$^{-1}$, corresponding to an XUV luminosity of $L_{\rm XUV} = 6.5 \times 10^{28} {\rm erg ~s^{-1}}$. This implies XUV fluxes of $F_{\rm XUV}\,=\,$1654.8\,erg\,cm$^2$\,s$^{-1}$ and $F_{\rm XUV}\,=\,$7452.0\,erg\,cm$^2$\,s$^{-1}$ at the distance of planet b and c, respectively. 

\begin{figure}[ht]
   	\centering
	\includegraphics[width=0.30\textwidth]{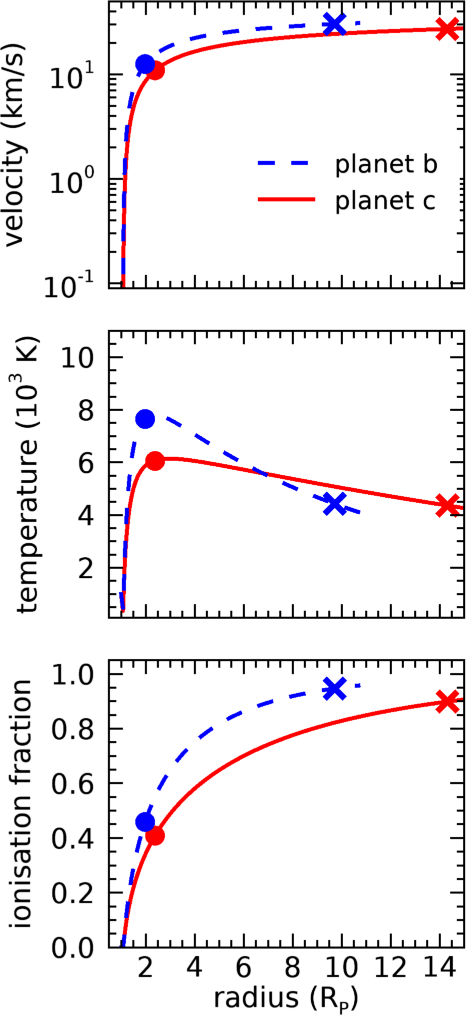}
	\caption{Profiles of hydrodynamic hydrogen escape for planet b (blue) and c (red). From top to bottom: radial velocity of the escaping atmosphere, temperature and ionisation fraction. The dot and the cross indicate the sonic point (when planetary material reaches sound speed) and Roche lobe boundary, respectively.}
	\label{fig.profile}
\end{figure}

Figure \ref{fig.profile} summarizes some key properties of the escaping atmosphere for both planets: radial component of the velocity (top), temperature (middle) and ionisation fraction (bottom). For planet b (inner planet), we derive an atmospheric escape rate of $4.5 \times 10^{10}$g s$^{-1}$. We calculate the Roche lobe distance to be at $9.7\,R_p$ and, at this distance, the speed of the escaping material is about 30~km s$^{-1}$. Planet c has a similar escape rate of $4.4 \times 10^{10}$g s$^{-1}$, with material reaching a speed of 28~km s$^{-1}$ at the Roche lobe distance ($14.3\,R_p$). The reason for comparable escape is that, although planet c receives an XUV flux that is 4.5 times smaller than planet b due to its larger orbital distance, it has a lower surface gravity ($g_c \simeq 24\%$ of Jupiter's gravity versus $g_b\simeq  38\%$ for planet b). It is more difficult for low-gravity planets to hold on to their atmospheres, thus the lower gravity of planet c compensates for its lower incident XUV flux, reaching comparable escape to planet b. 


\begin{figure}[ht]
   	\centering
	\includegraphics[width=0.495\textwidth]{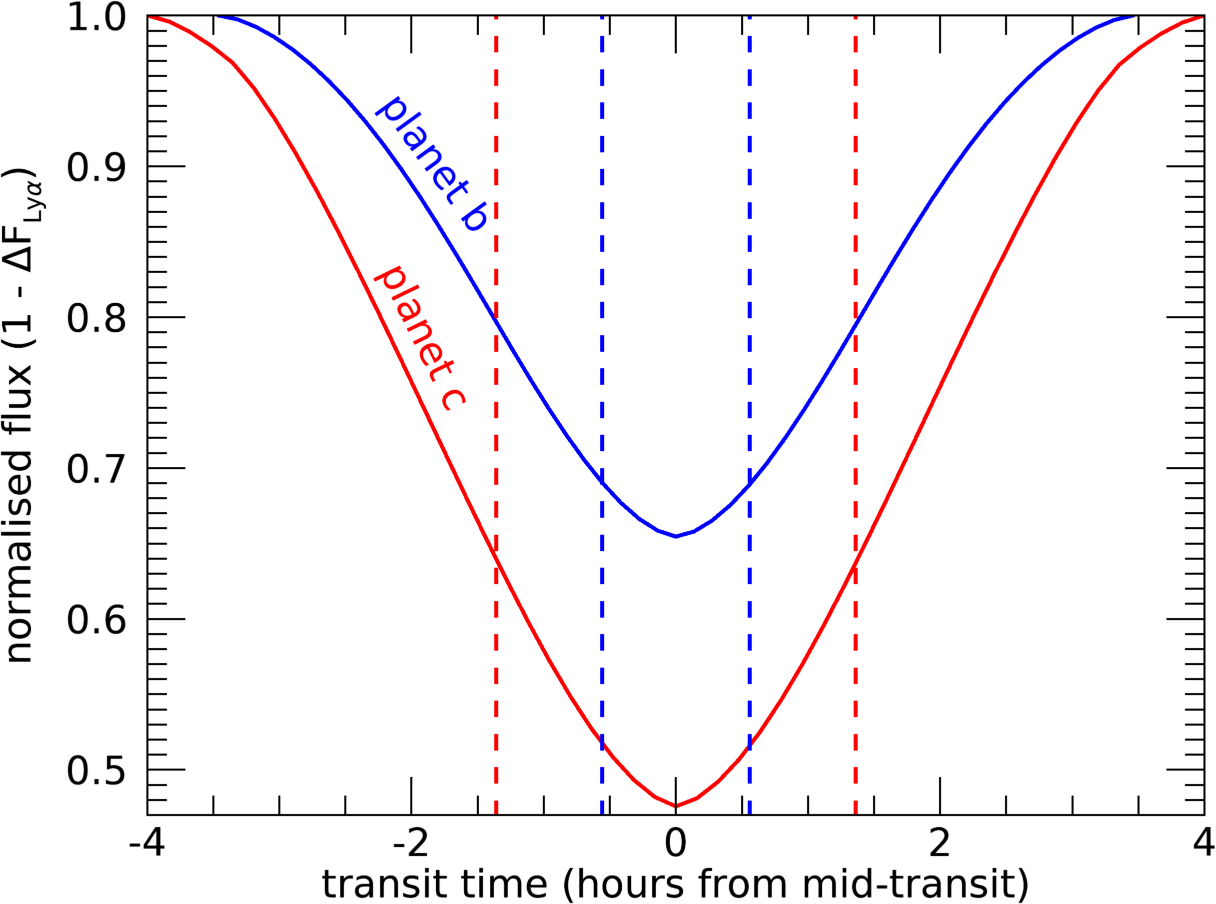}
	\caption{Predicted lightcurves for planet b (blue) and c (red) at the Ly-$\alpha$ line centre. The blue and red vertical dashed lines represent the first and fourth contact points for planet b (blue) and c (red).  
	}
	\label{fig.lyalpha}
\end{figure}

In spite of the similarities in the escape rates and velocities, we predict different Ly-$\alpha$ transit absorptions for these two planets. Figure \ref{fig.lyalpha} shows the predicted 
lightcurves at Ly-$\alpha$ line centre, 
where we see that planet b (the inner planet) shows a maximum absorption of 35\% and planet c (the outer planet) shows a maximum absorption of 53\%. These different absorptions are caused by the different ionisation fractions in each planet's atmosphere (see bottom panel of Fig. \ref{fig.profile}), with planet c showing more neutral hydrogen in its atmosphere. The lightcurves presented in Figure \ref{fig.lyalpha} are symmetric about mid-transit. This is due to the one-dimensional geometry of the model, hence of the assumption of spherically symmetric planetary atmospheres. 
However, we expect lightcurves to be asymmetric with respect to mid-transit, but these asymmetries can only be captured by 3D models that include interactions with the stellar wind \citep[e.g.,][]{Villarreal2014,Villarreal2018}, which we leave to a forthcoming paper. 

\begin{figure}[ht]
   	\centering
	\includegraphics[width=0.495\textwidth]{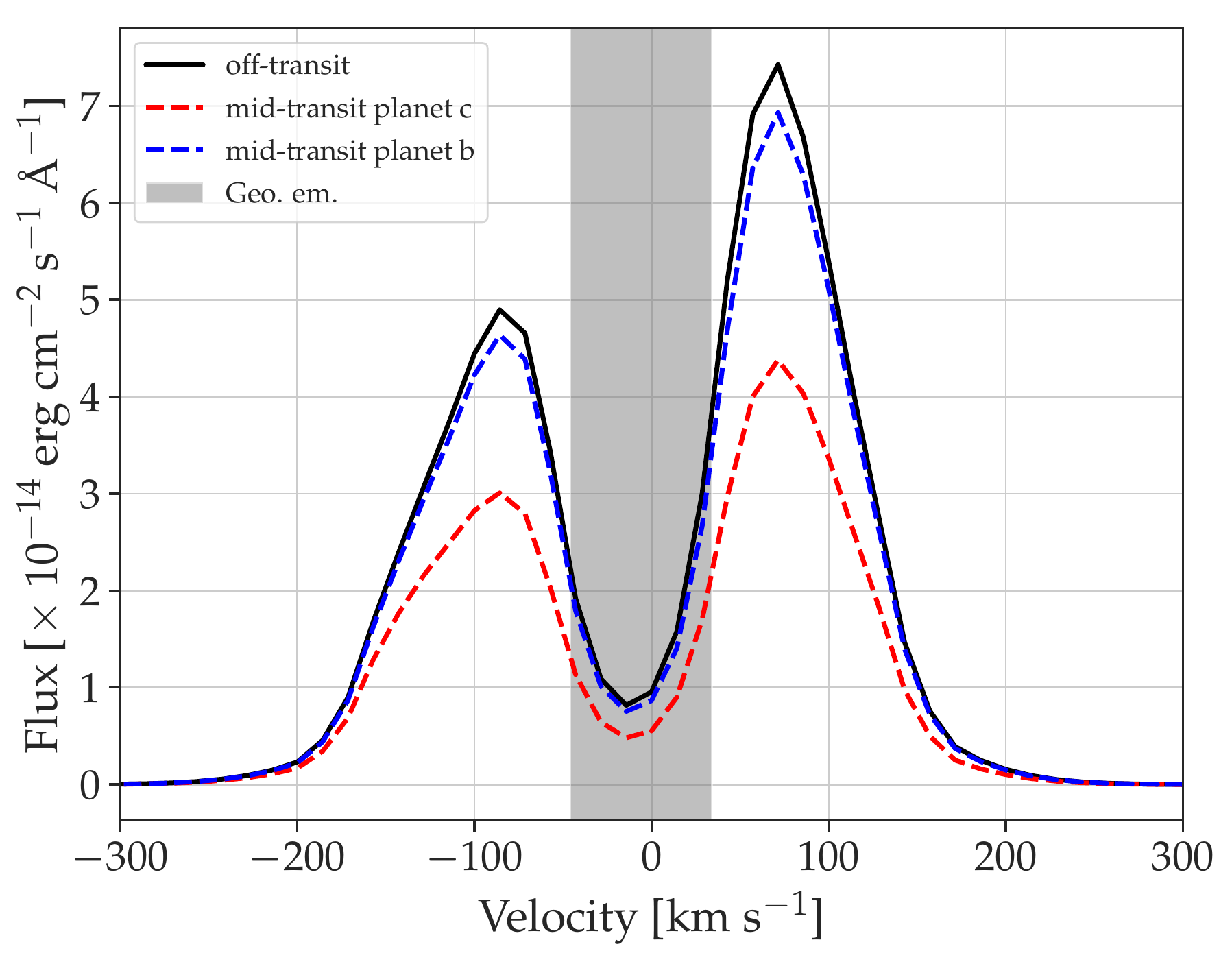}
	\caption{Predicted Ly-$\alpha$ line profile out-of-transit (black line), at mid-transit for planet b (blue-dashed line) and at mid-transit for planet c (red-dashed line) at the spectral resolution of the G140M grating of the STIS spectrograph on board HST. The gray stripe represents the part of the line expected to be contaminated by geocoronal emission.  
	}
	\label{fig.lyalpha.STIS}
\end{figure}
To encourage future observations in the UV we made use of the absorption profile obtained from the upper atmosphere simulations and computed the expected Ly-$\alpha$ profile as observed with the G140M grating of the STIS spectrograph on board HST. Figure \ref{fig.lyalpha.STIS} shows the resulting profiles at three different times reproducing an out of transit observation (black line) and an observation at mid-transit of planet b (blue-dashed line) and planet c (red-dashed line) . To compute the out of transit Ly-$\alpha$ profile of \sname, we scaled the intrinsic Ly-$\alpha$ profile of $\xi$ Boo A (G8 V, d = $6.70$ pc, $R_\star=0.78\,R_\odot$) derived by \citet{Wood2005}. We assumed that the strength of the profiles of the two stars should be similar as the stellar type and activity index for $\xi$ Boo A  (\logrhk\ $\sim - 4.40$,  \citet{Morgenthaler2012}) are close to those of \sname. 
Most of the predicted absorption will be hidden by the ISM absorption and geocoronal emission as shown in Fig. \ref{fig.lyalpha.STIS}. However, the ISM will affect the spectrum absorbing a significant fraction of its flux beyond almost 100 pc and probably most of it by 200 pc, while \sname\ is $~ 75$ pc distant. Moreover, because of the interaction of the escaping planetary upper atmosphere with the stellar wind and the resolution of the G140M grating, significant planetary Ly-$\alpha$ absorption will be visible in the line wings, making it detectable. The attenuation caused by the ISM over the intrinsic profile of \sname\ is assumed to be the same as the one estimated for $\xi$ Boo A, also computed in the work of \citet{Wood2005}. This is a crude approximation, but a more adequate estimation would require at least a near-ultraviolet observation of the ISM absorption at the position of the Mg{\sc ii}\,h\&k lines \citep{Wood2005}. 
Using HST we expect to be able to detect the absorption signal caused by planet c during transit, as this is the largest absorption modelled, in accordance to what is shown with Fig. \ref{fig.lyalpha}. 

We also compute the H-$\alpha$ transit lightcurves following the approach outlined in \citet{2019MNRAS.490.3760A}. Although we predict a large absorption in Ly-$\alpha$, no appreciable absorption is expected to be detectable in H-$\alpha$. This is because most of the hydrogen in the atmospheres of \planetb\ and \planetc\ is in the ground state -- for H-$\alpha$ to be formed, it is required some hydrogen in the first excited state. The fact that we do not predict any H-$\alpha$ absorption in our model needs to be reassessed in 3D calculations. 
Recently, Villarreal D'Angelo et al. (submitted) demonstrated that the interaction between the upper planetary atmosphere and the stellar wind, which is only captured in 3D models, could increase the atmospheric temperature in the interaction zone. As a result, this can increase the number of neutral hydrogen in the first excited state, possibly enhancing H-$\alpha$ absorption. We will further explore this in a future work.


\subsection{Planetary atmospheric evolution}\label{sec:evolution}

\begin{figure*}[ht]
   \centering
\includegraphics[width=0.99\linewidth]{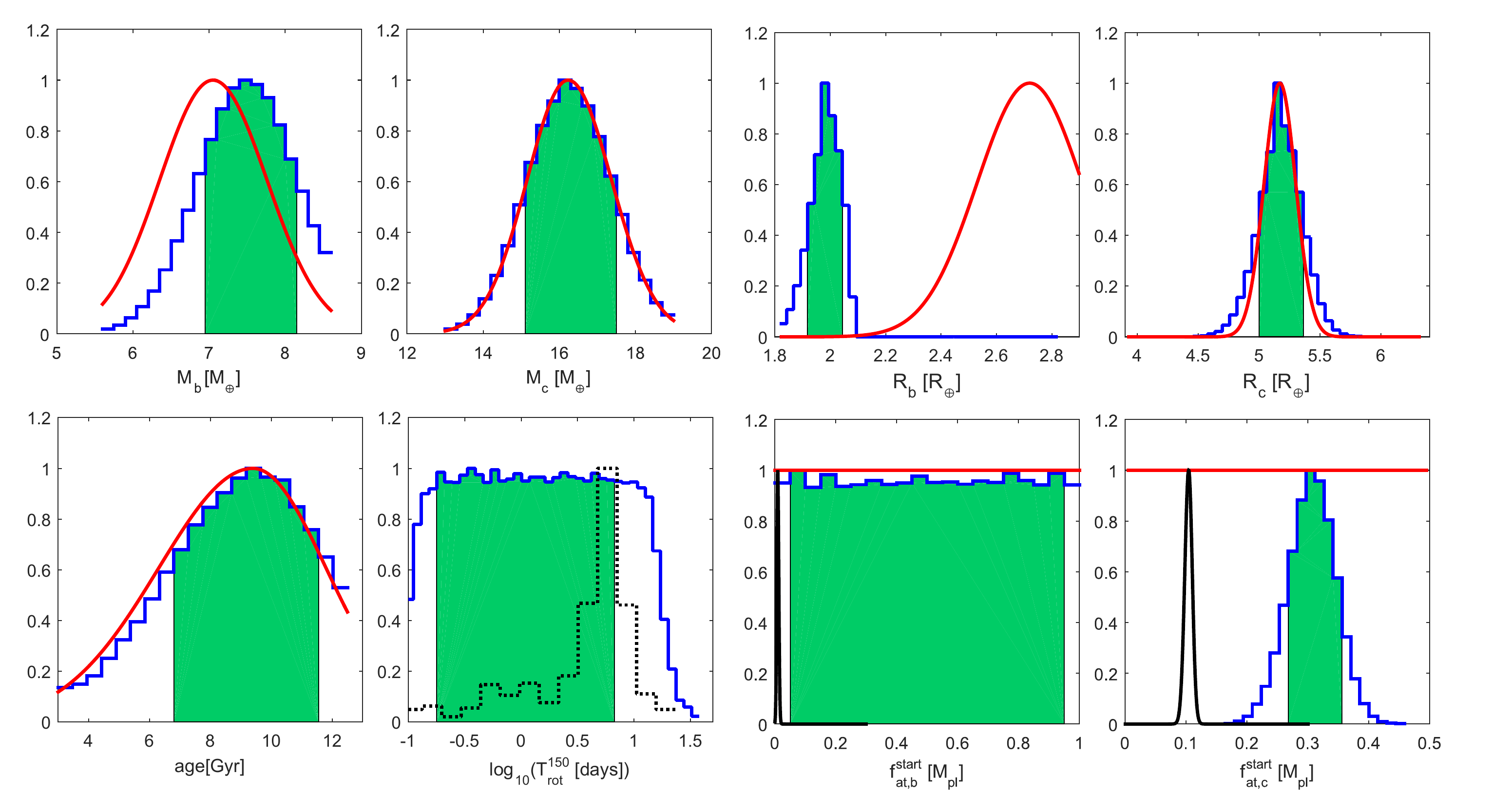}
\caption{Top, from left to right: posterior probability distribution functions for the current mass of planet b, mass of planet c, radius of planet b, radius of planet c. Bottom, from left to right: posterior probability distribution function for system's age, stellar rotation period at an age of 150\,Myr, initial atmospheric mass fraction for planet b, and initial atmospheric mass fraction for planet c. The blue solid lines indicate the posterior probabilities, the green shaded areas correspond to the 68\% HPD credible intervals, and the red solid lines are the priors. The mass and radius priors are not exactly the results of the combined RV and photometric analysis, but rather Gaussian priors with a width equal to the uncertainty on each parameter. The dashed black line in the second-from-left bottom panel shows the distribution measured for solar mass members of $\approx$150\,Myr-old open clusters \citep{johnstone2015b}. The black solid lines in the two right bottom panels illustrate the present time atmospheric mass fractions obtained for the posteriors given by MCMC.}
\label{fig:evolution}
\end{figure*}

In addition to the hydrodynamic model presented in Section \ref{sec:hyescape}, as a cross-check, we also computed the mass-loss rates employing the interpolation routine of \citet{kubyshkina2018a}, which is based on 1D hydrodynamic simulations, obtaining a value of $2.7 \times 10^{10}$g s$^{-1}$ for planet b (inner planet) and $1.6 \times 10^{10}$g s$^{-1}$ for planet c (outer planet), in agreement with the results obtained using the hydrodynamic atmospheric escape model of \citet{2019MNRAS.490.3760A}. We further notice the low $\Lambda$\footnote{The parameter $\Lambda$\,=\,$\frac{GM_{\rm p}m_{\rm H}}{k_{\rm B}T_{\rm eq}R_{\rm p}}$, where $m_{\rm H}$ is the mass of the hydrogen atom, is the restricted Jeans escape parameter and it is a measure of the thermal escape driven by the intrinsic atmospheric temperature and low planetary gravity \citep{fossati2017b}.} value ($\Lambda$\,=\,19.33) of planet b, which implies that the planetary gravity is hardly capable to hold a hydrogen-dominated atmosphere \citep{fossati2017b}. In fact, this is remarkable because other planets with similarly low $\Lambda$ values have an average density indicative of a mostly rocky composition \citep{Gandolfi2017}. This is confirmed by the mass-loss rates of planets b and c that imply that they would have lost 20\% and 5\%, respectively, of their mass over 1 Gyr. While planet c should have an atmospheric mass fraction large enough to sustain such an intense mass loss over Gyrs, for planet b the escape is probably too intense to be able to still retain a hydrogen-dominated atmosphere, as instead suggested by the average density. 

We employed the tool presented by \citet{kubyshkina2019a,kubyshkina2019b} to constrain the atmospheric evolution of both planets, their initial atmospheric mass fractions, and the evolution of the rotation rate (and therefore also of the XUV emission) of the host star. In short, the framework mixes three ingredients: a model of the stellar XUV flux evolution \citep[after][]{pizzolato2003,mamajek2008,wright2011,sanzforcada2011}, a model relating planetary parameters and atmospheric mass \citep[after][]{stokl2015,johnstone2015a}, and a model computing escape \citep[after][]{kubyshkina2018a}. The framework also accounts for the evolution of the stellar bolometric luminosity, hence planetary equilibrium temperature, using the MESA Isochrones and Stellar Tracks \citep[MIST,][]{paxton2018} grid. 

For a given core mass, the framework sets the core radius assuming an Earth-like density and the atmospheric mass is considered to be negligible \citep{owen2017}. Then, starting at 5 Myr (the assumed age of the dispersal of the protoplanetary disk), at each time step the framework extracts the mass-loss rate from the grid based on the stellar flux and system parameters, using it to update the atmospheric mass fraction and the planetary radius. This procedure is then repeated until the age of the system is reached or the planetary atmosphere is completely escaped. The framework simulates the atmospheric evolution of both planets, simultaneously. The main framework’s assumption is that the analysed planets have (or had) a hydrogen-dominated atmosphere and that the planetary orbital separation does not change after the dispersal of the protoplanetary disk.

The input parameters of the framework are planetary masses, planetary radii, orbital separations, current stellar rotation rate, and stellar mass, while the free parameters are the index of the power law describing the evolution of the stellar XUV flux and the initial planetary radius (i.e., the initial atmospheric mass fraction at the time of the dispersal of the protoplanetary disk; $f_{\rm at}$). The input parameters are set equal to the measurements with Gaussian priors having a width equal to the measurement uncertainties, while we take flat priors for the output parameters. The output parameters are constrained by implementing the atmospheric evolution algorithm in a Bayesian framework employing the Markov-chain Monte Carlo (MCMC) tool of \citet{cubillos2017a}. 

Figure~\ref{fig:evolution} shows the main results of the simulation. The posteriors on the input parameters of the host star and of the outer planet (\planetc) match the priors. The evolution of the stellar rotation rate (or XUV emission) is mostly unconstrained. For the outer planet, the analysis leads to a rather tight constraint on the initial atmospheric mass fraction of about 30\% and this result holds also when running the analysis solely on the outer planet, meaning that the anomaly found for the inner planet (see below) does not affect the other results. The outer planet could not have accreted an atmosphere much larger than $f_{\rm at}\approx30\%$ while in the disk, because otherwise the stellar XUV emission would have not been able to remove enough of it to obtain the currently observed radius, even if the star was a fast rotator. Similarly, the planet could not have accreted an atmosphere much smaller than $f_{\rm at}\approx30\%$ while in the disk, because otherwise the stellar XUV emission would have removed too much atmosphere given the observed radius, even if the star was a slow rotator.

The result obtained for the inner planet is extremely interesting. The framework is unable to find a configuration in which the planet is capable of retaining enough atmosphere to match the measured planetary mass, radius, and orbital separation. This is why the posterior of the planetary mass is slightly shifted towards higher masses compared to the prior (first left top panel of Fig.~\ref{fig:evolution}) and, moreover, the posterior of the planetary radius is significantly shifted towards smaller radii compared to the prior (third top panel of Fig.~\ref{fig:evolution}). In other words, given the system parameters, the framework finds that the inner planet always loses its hydrogen atmosphere, regardless of the evolution of the stellar XUV emission. An inspection of the atmospheric evolutionary tracks indicates that the inner planet is expected to completely lose its atmosphere within 1 Gyr, while the estimated age of the system is significantly larger. We reran the simulation looking for the planetary parameters that would enable the posteriors on mass and radius not to vary from the priors, obtaining either an orbital separation of about twice the measured one (keeping mass and radius equal to the measured values), or a planetary mass of about 16\,$M_{\oplus}$ (keeping radius and orbital separation equal to the measured values), or a planetary radius of about 2\,$R_{\oplus}$ (keeping mass and orbital separation equal to the measured values). In the last two options the planet would not host a hydrogen-dominated atmosphere.

CoRoT-24b was the first planet identified to have a low bulk density, compatible with the presence of a hydrogen-dominated atmosphere, but at the same time to be also subject to a too extreme mass loss for hosting one \citep{lammer2016}. \citet{cubillos2017b} analysed the upper atmospheric properties and high-energy irradiation of a large sample of mini-Neptunes, detected mostly by the {\it Kepler} satellite, finding that 15\% of them share this same peculiar property. There is a range of possible solutions to this puzzle. One of the main assumptions in the atmospheric evolution framework is that orbital separations do not change with time following the dispersal of the protoplanetary nebula, which may not be the case, for example, if the system had a close-enough encounter with another star in the past. It may also be that the hydrodynamic model overestimates the mass-loss rates, although past comparisons between observations and hydrodynamic models would tend to exclude orders of magnitude errors in the computed rates. One further possibility is a bias in the measured planetary parameters (mass and/or radius), maybe caused by the presence of other undetected planets in the system biasing the planetary mass measurement, but it seems unlikely, given the quantity and quality of data. The mainstream explanation for this kind of planet is the presence of high-altitude aerosols that would lead overestimation of planetary radius \citep{lammer2016,cubillos2017b,gao2020}. Future atmospheric characterisation observations, particularly those at low resolution that are more sensitive to aerosols, will be able to identify whether this is the case or not \citep[e.g.,][]{libby2020}. There is also the additional possibility that the crust released a significant amount of light gases in the atmosphere, counteracting the effect of escape \citep[e.g.,][]{kite2019}.

\subsection{Simulated \textit{HST} WFC3 retrievals}
\label{sect:Retrievals}

\begin{figure*}[ht]
   	\centering
	\includegraphics[width=0.98\textwidth]{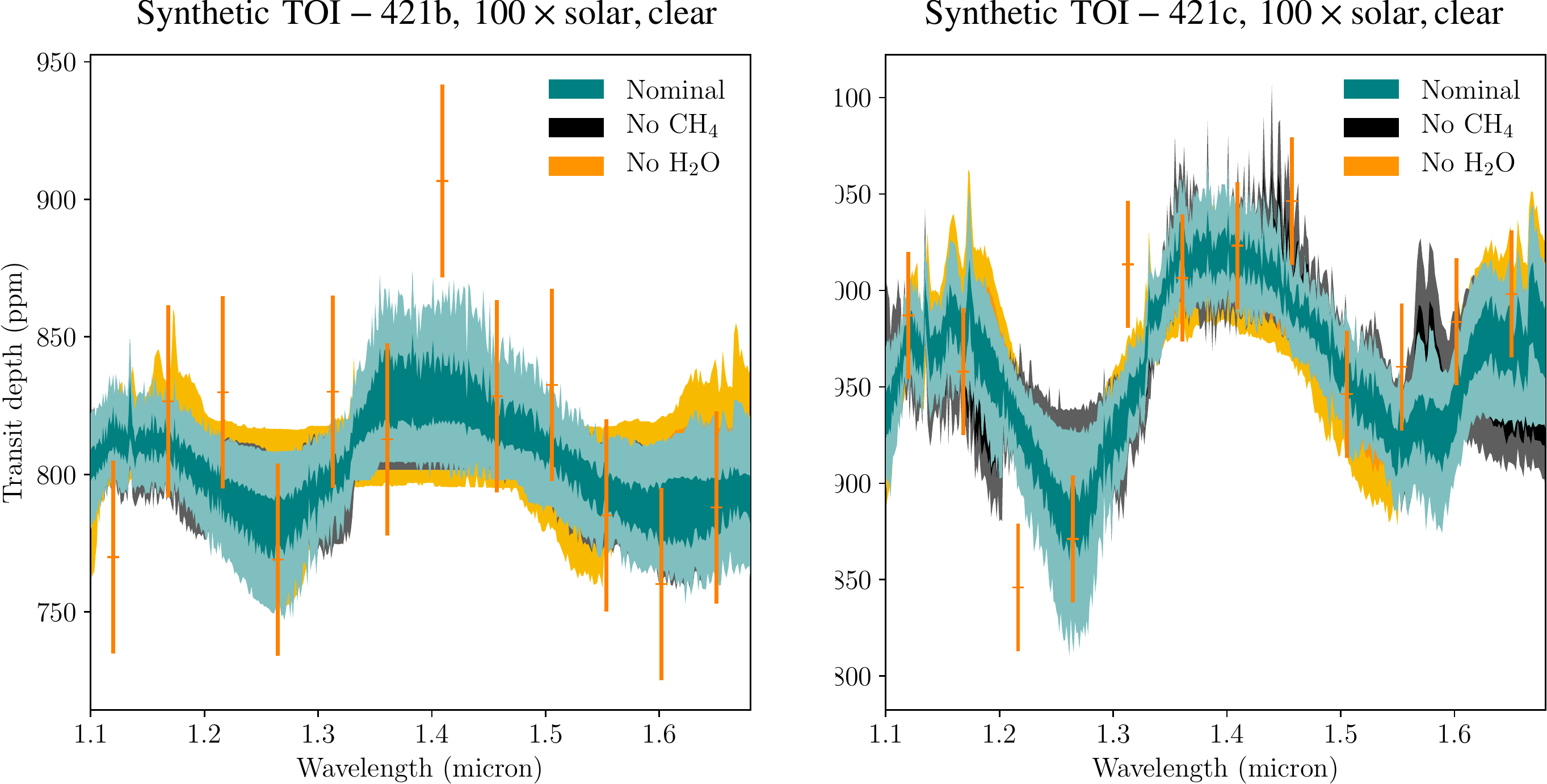}
	\caption{{\it Left panel:} synthetic HST data of \sname~b (orange), and the 1- and 2-$\sigma$ uncertainty envelopes of the retrievals with the nominal retrieval model (blue) and the model that neglected the CH$_4$ (black) or H$_2$O opacity (orange). The input data was assuming no clouds and 100 times solar enrichment. {\it Right panel:} analogous analysis plot for \sname~c.}
	\label{fig:retrieval}
\end{figure*}

With its large radius ($\sim$5~$R_\oplus$), \planetc\ represents an excellent target for atmospheric characterization. \planetb\ is somewhat more challenging; its scale height is comparable to \planetc, but because $c$ is smaller by a factor of two, so will be the transit signal corresponding to one scale height, similarly as Ly-$\alpha$ absorption. To assess how well the atmospheric properties of \planetb\ and \planetc\ could be derived from observations, we modeled the planetary atmospheres and derived transmission spectra with the open-source \texttt{petitRADTRANS} package \citep{mollierewardenier2019}. The atmospheres were set up to be isothermal, at the equilibrium temperature of the planets. The absorber abundances were obtained from assuming chemical equilibrium, calculated with the chemistry module that is part of \text{petitCODE} \citep{mollierevanboekel2017}. We assumed a solar C/O ratio of 0.55, and two different metallicity values, 3 and 100~$\times$~solar (Jupiter and Neptune-like, respectively). We also introduced a gray cloud deck, the position of which was varied between 100 and $10^{-5}$~bar, in 1~dex steps. We considered 100~bar to be the cloud-free model, as the atmosphere will become optically thick at lower pressures. The following gas opacities were included: the line opacities of H$_2$O, CH$_4$, CO, CO$_2$, Na, K, the Rayleigh scattering cross-sections of H$_2$O, CH$_4$, CO, CO$_2$, H$_2$ and He, as well as H2-H2 and H2-He collision-induced absorption.

The atmospheric models described above were then retrieved with \texttt{petitRADTRANS}, using the \texttt{PyMultiNest} package \citep{buchnergeorgakakis2014}, which makes use of the nested sampling implementation \texttt{MultiNest}, by \citet{ferozhobson2009}. We created synthetic HST observations with the WFC3 instrument, assuming 12 wavelength points spaced equidistantly from 1.12 to 1.65~$\mu$m. The error on the flux decrease during transit was assumed to be 35~ppm for \planetb\ and 33~ppm for \planetc\ per channel, which we calculated for a single transit of the two planets, using the \texttt{Pandexo\_HST} tool\footnote{\url{https://exoctk.stsci.edu/}.}. For reference, this is about 1/3 of the signal of one atmospheric scale height of \planetc, when assuming a solar composition. For every input model, we wish to characterize whether or not the molecular features of H$_2$O or CH$_4$ can be identified in the spectra. To this end we follow the technique outlined in \citet{bennekeseager2013}, that is we first retrieve the atmospheric temperature, reference pressure, cloud deck pressure and vertically constant absorber abundances freely; this is called setup (i) in the following. Then we remove the abundance parameter and opacity of either H$_2$O or CH$_4$ from the retrieval, these are setups (ii) and (iii). The Bayes factor $B$  between model (i) and (ii) will constrain how confidently the atmospheric features of H$_2$O can be detected, while the $B$ between model (i) and (iii) informs us about how reliably CH$_4$ can be detected.

It has recently been found that observational uncertainties in the range of 30 ppm can lead to significant differences in retrieved atmospheric abundances and temperatures when comparing the results of various retrieval codes \citep{barstowchangeat2020}. These discrepancies most likely arise due to differences in the opacities, either because of the use of different line lists, or the choices made when converting line lists into opacities, such as the line broadening or cutoff. Because we use the same model to make the synthetic observations and the retrievals, our results can therefore be regarded as a limiting case, where the above-mentioned issues are negligible. Moreover, because our discussion here focuses on the detection of molecular features, instead of constraining their abundances, we deem our approach acceptable for the exploratory study presented here. When running retrievals for real observations of similar data quality in order to constrain abundances and other atmospheric parameters the use of multiple retrieval codes or opacity treatments (varying line lists, the broadening description, etc.) are recommended.

\subsubsection{\planetb}
For the three times solar metallicity case we found that for cloud pressures larger than 10~mbar, substantial atmospheric features can be retrieved (we find $B>3$, see \citealt{kassraftery1995} for a definition of the $B$ thresholds). For a metallicity of 100 times solar, this transition likewise occurs for pressures larger than 10~mbar. Due to the high equilibrium temperature of the planet, no CH$_4$ can be detected in the synthetic observations. This is because at high temperatures CO is chemically favored as the main C-bearing species, instead of CH$_4$. 

As an example, the left panel of Figure \ref{fig:retrieval} shows the retrievals of the 100 times solar enrichment, clear atmosphere synthetic observation with the full model, and the models without CH$_4$ and H$_2$O. Because of the random noise instantiation H$_2$O is only weakly detected in this example ($B=2.7$), while running this test multiple times places the average $B$ in favor of including H$_2$O at a value of $B=6$.

\subsubsection{\planetc}
For the three times solar metallicity case we found that for cloud pressures larger than 1~mbar, substantial atmospheric features can be retrieved. For a metallicity of 100 times solar, this transition occurs for pressures larger than 0.1~bar. Due to the smaller temperature and the assumption of equilibrium chemistry in the atmospheric model used to generate the observations, we find that the signal of CH$_4$ can be more confidently detected in the atmosphere of the planet than that of H$_2$O.

The right panel of Figure \ref{fig:retrieval} shows the retrievals of the 100 times solar enrichment, clear atmosphere synthetic observation for the full retrieval model and the model without CH$_4$ or H$_2$O. The average $B$ value for detecting CH$_4$ or H$_2$O are $B=4$ and $B=3$, respectively.



\subsection{Prospects for detecting transit-timing variation and the Rossiter-McLaughlin effect}
\label{sec:ttvs}

The orbital periods of the two planets are close to a 3:1 commensurability (5.2 and 16.1 days) and therefore transit timing variations (TTVs) are expected. However, given the combined \tess\ and photometric follow-up observation time span of $\sim$80~days, no TTVs have been detected. We investigated possible TTVs through a 3-body simulation using the Python Tool for Transit Variations (\texttt{PyTTV}). We simulated the estimated TTVs and RVs using the stellar and planetary parameters reported in Table~\ref{tab:stellar} and \ref{tab:parstoi421}, and found an expected TTV signal with a period of $\sim$180~days and an amplitude of $\sim$4~minutes. However, two issues have prevented a TTV detection. First, the time span from the \tess\ and photometric follow-up observations covers less than half of the expected TTV period, and second the large uncertainties in the individual transit center times of $\sim$1 and 4 minutes for the outer and inner planet, respectively. \sname\ is an ideal target to compare planetary masses determined from TTVs and RVs in the future with additional transit observations.

Using the RM effect modeling and fitting code described in \cite{Espositoetal2017}, we performed simulations to assess the RV amplitude of the Rossiter-McLaughlin (RM) effect based on our determination of the relevant stellar (\vsini, \rstar, limb darkening) and planetary ($R_\mathrm{p}$, b) parameters. We found that for a sky-projected obliquity $\lambda$\,=\,0~deg (90 deg), the amplitude of the RM effect is  2.0 (4.1) \ms\ for \sname\,c. Similarly for \sname\,b, we found an amplitude of 0.3 (0.9) \ms\ for $\lambda$ = 0 deg (90 deg). We performed simulations to assess the possibility to measure the RM effect of \sname c with HARPS observations. Assuming a time series of spectra with 15 minutes exposure time covering a full transit, and a 2\,\ms\ error per RV measurement, we estimated that $\lambda$ could be measured with an uncertainty of $\lesssim$\,15 deg.

\section{Conclusions} \label{sec:concl}

We presented the discovery of a Neptune-sized planet and a sub-Neptune transiting \sname\ (BD-14~1137, TIC 94986319), a G9 dwarf star observed by \tess. The host star is the primary component of a visual binary. Our RV follow-up observations led to the confirmation of the outer Neptune-sized planet (\planetb) and the discovery of the second inner sub-Neptune (\planetc), that we also found to transit its host star. We determined both stellar and planetary parameters. We found that \sname\ is a relatively quiet star with an activity index of \logrhk\,=\,$-$4.93\,$\pm$\,0.04. Based on the analysis of the HARPS and WASP-South data, we found that the intrinsic activity of \sname\ can be explained mainly by plages.

Our TTV analysis shows that \sname\ is an ideal target to compare planetary masses determined via TTV and Doppler techniques. We aim for future additional transit observations to explore this in more detail.

\begin{figure}[t]
   	\centering
	\includegraphics[width=1.0\linewidth]{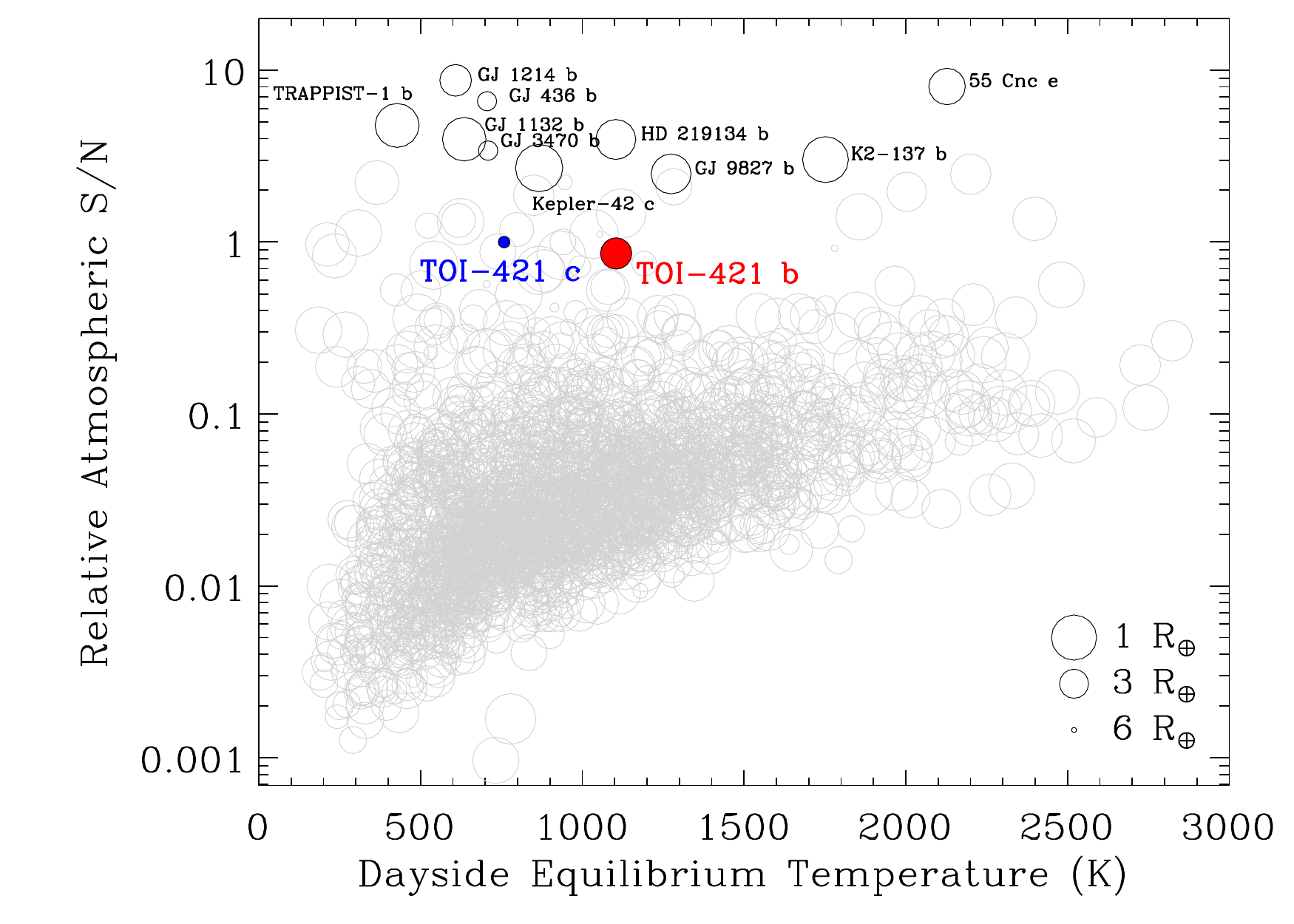}
	\caption{Predicted relative S/N of an atmospheric signal in the $J$-band for all exoplanet candidates with $R$\,$<$\,6\,$R_\oplus$. The TOI-421 planets are the filled colored symbols with TOI-421b used as the S/N reference. The top ten targets using this metric are labeled. The TOI-421 planets rank in the top 30 most favorable for atmospheric characterization from among more than 2000 exoplanets in this size range.}
	\label{fig:snrjwst}
\end{figure}

\planetb\ and \planetc\ are very appealing and suitable targets for atmospheric characterization. They are both expected to host extended atmospheres, showing significant signal in the Ly-$\alpha$ line. Moreover, the atmospheric retrievals demonstrated that we can detect CH$_4$ in the atmosphere of the outer planet (\planetc) if the atmosphere is in chemical equilibrium, and atmospheric evolution simulations showed that the inner planet (\planetb) appears to be among the small sample of peculiar super puffy mini-Neptunes, making it also more intriguing for atmospheric studies and evolution theories. 
This multi-planet system with its astonishing characteristics would be a prime target for the upcoming \textit{JWST} observations. Indeed, the two planets are among the first 30 targets with the highest expected signal-to-noise ratios, as shown in Figure~\ref{fig:snrjwst}. Using the sample of exoplanets with $R\,<\,6\,R_\oplus$, totaling more than 2000 exoplanets\footnote{\url{https://exoplanetarchive.ipac.caltech.edu}.}, \planetb\ and \planetc\ are within the top 30 most favorable targets for atmospheric characterization. This atmospheric characterization metric is based on a $J$-band, \textit{JWST} style observation, and is detailed in \citet{niraula17}. Of particular note is this metric is scaled by the frequency of transits. This is motivated by the expectation that sensitive atmospheric observations will likely require many transits to build sufficient signal, and it may be prohibitive to accumulated the needed transits for longer period exoplanets. Therefore, we used a metric that optimizes the S/N over a period of time rather than a per-transit metric. Irregardless of the nuances of the metric, the \sname's planets are highly attractive targets for characterization of both their bound and extended atmospheres.

\acknowledgments

This work was supported by the \texttt{KESPRINT} collaboration, an international consortium devoted to the characterization and research of exoplanets discovered with space-based missions.

We acknowledge the use of public \tess\ Alert data from pipelines at the \tess\ Science Office and at the \tess\ Science Processing Operations Center.

Funding for the \tess\ mission is provided by NASA’s Science Mission directorate. We acknowledge the use of public \tess\ Alert data from pipelines at the \tess\ Science Office and at the \tess\ Science Processing Operations Center. Resources supporting this work were provided by the NASA High-End Computing (HEC) Program through the NASA Advanced Supercomputing (NAS) Division at Ames Research Center for the production of the SPOC data products.

Part of this research was carried out at the Jet Propulsion Laboratory, California Institute of Technology, under a contract with the National Aeronautics and Space Administration (NASA).

We are very grateful to the NOT and ESO staff members for their unique and superb support during the observations.
A.A.V. and D.K. acknowledge funding from the European Research Council (ERC) under the European Union's Horizon 2020 research and innovation programme (grant agreement No 817540, ASTROFLOW). P.M. thanks Thomas Mikal-Evans and Laura Kreidberg for help computing predicted HST uncertainties for \sname. P.M. acknowledges support from the European Research Council under the European Union's Horizon 2020 research and innovation program under grant agreement No. 832428.

L.M.S. and D.G. gratefully acknowledge financial support from the CRT foundation under Grant No. 2018.2323 ``Gaseous or rocky? Unveiling the nature of small worlds".
R.B.\ acknowledges support from FONDECYT Post-doctoral Fellowship Project 3180246.
A.J.\ and  R.B.\ acknowledge support from project IC120009 ``Millennium Institute of Astrophysics (MAS)'' of the Millenium Science Initiative, Chilean Ministry of
Economy. A.J.\ acknowledges additional support from FONDECYT project 1171208.
C.V.D. acknowledges the funding from the Irish Research Council through the postdoctoral fellowship (Project ID: GOIPD/2018/659). C.M.P., M.F. and I.G. gratefully acknowledge the support of the Swedish National Space Agency (DNR 65/19 and 136/13). 
L.S. acknowledges financial support from the Australian Research Council (Discovery Project 170100521).
D.D. acknowledges support from NASA through Caltech/JPL grant RSA-1006130 and through the \tess\ Guest Investigator Program Grant 80NSSC19K1727.
D.H. acknowledges support from the Alfred P. Sloan Foundation, the National Aeronautics and Space Administration (80NSSC18K1585, 80NSSC19K0379), and the National Science Foundation (AST-1717000).
S.M. acknowledges support from the Spanish Ministry with the Ramon y Cajal fellowship number RYC-2015-17697. LGC thanks the support from grant FPI-SO from the Spanish Ministry of Economy and Competitiveness (MINECO) (research project SEV-2015-0548-17-2 and predoctoral contract BES-2017-082610).
A.R.G.S. acknowledges the support from NASA under Grant No. NNX17AF27G.
J.N.W. thanks the Heising-Simons foundation for support. R.A.G. acknowledges the support from PLATO and GOLF CNES grants.
M.R.D. acknowledges support of CONICYT-PFCHA/Doctorado Nacional-21140646, Chile.
KWFL, JK, SzCs, ME, SG, APH, MP and HR acknowledge support by DFG grants PA525/18-1, PA525/19-1, PA525/20-1, HA3279/12-1 and RA714/14-1 within the DFG Schwerpunkt SPP 1992, “Exploring the Diversity of Extrasolar Planets”
This research has made use of the Exoplanet Follow-up Observation Program website, which is operated by the California Institute of Technology, under contract with the National Aeronautics and Space Administration under the Exoplanet Exploration Program.

\clearpage

\begin{table}[t]
\small
\begin{center}
\caption{\label{Table:FIES}  FIES RV measurements of \sname. (a) Barycentric Julian dates are given in barycentric dynamical time; (b) S/N ratio per pixel at 550 nm.}
\begin{tabular}{lccrr}
\hline\hline
\noalign{\smallskip}
$\rm BJD_{TDB}^a$ & RV  & $\sigma$ & T$_\mathrm{exp}$ & S/N $^b$\\
-2450000          & (\kms)  & (\kms)   &      (s)         &   \\
\hline
\noalign{\smallskip}
\noalign{\smallskip}
 8517.532619 &   0.0000 & 0.0064 & 1800 & 65  \\
 8522.486513 &  -0.0070 & 0.0042 & 2400 & 88  \\
 8523.431757 &  -0.0078 & 0.0046 & 2400 & 84  \\
 8524.489067 &  -0.0126 & 0.0046 & 2400 & 82  \\
 8539.420470 &  -0.0054 & 0.0042 & 2400 & 89  \\
 8540.437340 &  -0.0146 & 0.0028 & 2700 & 120 \\
 8541.423092 &  -0.0140 & 0.0035 & 2400 & 94  \\
 8554.407973 &  -0.0064 & 0.0043 & 2400 & 88  \\
 8556.406509 &  -0.0152 & 0.0042 & 2700 & 87  \\
\noalign{\smallskip}
\hline
\end{tabular}
\end{center}
\end{table}



\begin{longtable*}{lccccccrr}
\caption{\label{Table:HARPS} HARPS RV measurements of \sname. (a) Barycentric Julian dates are given in barycentric dynamical time; (b) S/N ratio per pixel at 550 nm.}\\
\hline
\noalign{\smallskip}
$\rm BJD_{TDB}^a$ & RV      & $\sigma_\mathrm{RV}$ &    BIS  & FWHM   & \logrhk & $\sigma$~\logrhk & T$_\mathrm{exp}$ & S/N $^b$\\
-2450000          & (\kms)  & (\kms)               &  (\kms) & (\kms)  &        &                  & (s)              &   \\
\hline
\endfirsthead
\caption{Continued.} \\
\hline
$\rm BJD_{TDB}^a$ & RV  & $\sigma_\mathrm{RV}$ &    BIS  & FWHM   & \logrhk & \logrhk & T$_\mathrm{exp}$ & S/N $^b$\\
-2450000          & (\kms)  & (\kms)       &  (\kms) & (\kms)  &    &  & (s)         &   \\
\hline
\endhead
\hline
\endfoot
\noalign{\smallskip}
\noalign{\smallskip}
 8528.596454 &  79.5473 & 0.0011 & -0.0181 & 6.7449 & -4.948 & 0.019 & 1200 &  69 \\
 8529.643890 &  79.5417 & 0.0013 & -0.0198 & 6.7420 & -4.842 & 0.018 & 1200 &  61 \\
 8530.589677 &  79.5415 & 0.0011 & -0.0257 & 6.7420 & -4.978 & 0.019 & 1200 &  73 \\
 8540.601897 &  79.5363 & 0.0010 & -0.0224 & 6.7489 & -5.001 & 0.022 & 1200 &  83 \\
 8578.531222 &  79.5516 & 0.0023 & -0.0220 & 6.7481 & -5.016 & 0.061 &  900 &  41 \\
 8580.540429 &  79.5490 & 0.0014 & -0.0238 & 6.7417 & -4.942 & 0.033 &  900 &  64 \\
 8581.569558 &  79.5492 & 0.0015 & -0.0219 & 6.7370 & -5.045 & 0.049 &  900 &  62 \\
 8709.911463 &  79.5529 & 0.0007 & -0.0221 & 6.7564 & -4.879 & 0.007 & 1800 & 104 \\
 8711.894944 &  79.5463 & 0.0010 & -0.0279 & 6.7561 & -4.920 & 0.011 & 1800 &  83 \\
 8713.921013 &  79.5440 & 0.0010 & -0.0258 & 6.7547 & -4.893 & 0.011 & 1700 &  81 \\
 8714.911893 &  79.5458 & 0.0010 & -0.0191 & 6.7591 & -4.883 & 0.011 & 1800 &  82 \\
 8715.921283 &  79.5436 & 0.0012 & -0.0228 & 6.7568 & -4.870 & 0.013 & 1800 &  67 \\
 8716.915892 &  79.5401 & 0.0011 & -0.0198 & 6.7504 & -4.877 & 0.012 & 1600 &  72 \\
 8717.927169 &  79.5408 & 0.0008 & -0.0235 & 6.7513 & -4.871 & 0.008 & 1800 &  97 \\
 8718.883686 &  79.5417 & 0.0009 & -0.0262 & 6.7578 & -4.912 & 0.010 & 1800 &  87 \\
 8721.904488 &  79.5465 & 0.0010 & -0.0214 & 6.7569 & -4.911 & 0.012 & 1800 &  81 \\
 8723.878421 &  79.5460 & 0.0008 & -0.0231 & 6.7494 & -4.916 & 0.009 & 1800 &  92 \\
 8724.913234 &  79.5496 & 0.0014 & -0.0303 & 6.7434 & -4.956 & 0.021 & 1500 &  58 \\
 8725.875540 &  79.5514 & 0.0014 & -0.0309 & 6.7440 & -4.919 & 0.019 & 1800 &  58 \\
 8726.862398 &  79.5498 & 0.0012 & -0.0285 & 6.7390 & -4.970 & 0.018 & 1200 &  67 \\
 8734.857658 &  79.5470 & 0.0009 & -0.0235 & 6.7571 & -4.868 & 0.010 & 2100 &  84 \\
 8736.885722 &  79.5499 & 0.0010 & -0.0246 & 6.7596 & -4.892 & 0.011 & 2100 &  81 \\
 8737.873716 &  79.5403 & 0.0013 & -0.0185 & 6.7641 & -4.954 & 0.021 & 1800 &  61 \\
 8738.871421 &  79.5412 & 0.0011 & -0.0172 & 6.7524 & -4.897 & 0.013 & 1800 &  74 \\
 8740.862468 &  79.5464 & 0.0012 & -0.0214 & 6.7562 & -4.889 & 0.016 & 1800 &  66 \\
 8741.837012 &  79.5468 & 0.0009 & -0.0200 & 6.7482 & -4.895 & 0.011 & 1800 &  84 \\
 8744.821435 &  79.5414 & 0.0010 & -0.0254 & 6.7488 & -4.909 & 0.012 & 1800 &  79 \\
 8745.817358 &  79.5393 & 0.0016 & -0.0231 & 6.7471 & -5.004 & 0.031 & 1800 &  52 \\
 8746.804032 &  79.5414 & 0.0012 & -0.0261 & 6.7438 & -4.990 & 0.020 & 1800 &  67 \\
 8747.835235 &  79.5389 & 0.0013 & -0.0233 & 6.7452 & -4.916 & 0.018 & 1800 &  63 \\
 8748.860025 &  79.5376 & 0.0009 & -0.0264 & 6.7440 & -4.928 & 0.011 & 1800 &  84 \\
 8750.820767 &  79.5414 & 0.0010 & -0.0311 & 6.7503 & -4.924 & 0.012 & 1800 &  76 \\
 8752.867278 &  79.5418 & 0.0011 & -0.0203 & 6.7542 & -4.920 & 0.013 & 1800 &  73 \\
 8753.779744 &  79.5395 & 0.0018 & -0.0233 & 6.7559 & -5.010 & 0.034 & 1800 &  49 \\
 8754.812264 &  79.5417 & 0.0011 & -0.0278 & 6.7501 & -4.892 & 0.011 & 1800 &  75 \\
 8755.830524 &  79.5492 & 0.0008 & -0.0256 & 6.7512 & -4.908 & 0.008 & 1620 & 101 \\
 8756.882374 &  79.5524 & 0.0008 & -0.0249 & 6.7509 & -4.900 & 0.009 & 1800 &  94 \\
 8757.807038 &  79.5465 & 0.0019 & -0.0316 & 6.7564 & -4.968 & 0.036 & 1800 &  46 \\
 8760.807608 &  79.5425 & 0.0013 & -0.0232 & 6.7515 & -4.931 & 0.018 & 1800 &  62 \\
 8761.806384 &  79.5452 & 0.0009 & -0.0256 & 6.7525 & -4.884 & 0.009 & 1800 &  90 \\
 8762.847560 &  79.5450 & 0.0009 & -0.0279 & 6.7565 & -4.895 & 0.011 & 1800 &  84 \\
 8763.827086 &  79.5402 & 0.0009 & -0.0232 & 6.7547 & -4.903 & 0.010 & 1800 &  84 \\
 8766.857081 &  79.5479 & 0.0009 & -0.0195 & 6.7589 & -4.895 & 0.011 & 1800 &  85 \\
 8767.723508 &  79.5481 & 0.0011 & -0.0254 & 6.7614 & -4.889 & 0.014 & 1800 &  72 \\
 8767.743659 &  79.5491 & 0.0012 & -0.0248 & 6.7619 & -4.864 & 0.013 & 1800 &  69 \\
 8780.805190 &  79.5336 & 0.0008 & -0.0271 & 6.7448 & -4.924 & 0.010 & 1800 &  96 \\
 8781.847532 &  79.5365 & 0.0009 & -0.0299 & 6.7484 & -4.942 & 0.012 & 1800 &  90 \\
 8782.800562 &  79.5381 & 0.0017 & -0.0305 & 6.7489 & -4.925 & 0.023 & 1800 &  51 \\
 8784.840817 &  79.5395 & 0.0013 & -0.0303 & 6.7384 & -5.043 & 0.023 & 1800 &  61 \\
 8785.732151 &  79.5386 & 0.0014 & -0.0322 & 6.7431 & -4.947 & 0.019 & 1800 &  57 \\
 8785.863860 &  79.5420 & 0.0016 & -0.0373 & 6.7482 & -4.916 & 0.021 & 1800 &  50 \\
 8791.866359 &  79.5440 & 0.0010 & -0.0291 & 6.7541 & -4.908 & 0.013 & 1800 &  79 \\
 8792.789899 &  79.5466 & 0.0009 & -0.0231 & 6.7469 & -4.920 & 0.010 & 1800 &  90 \\
 8793.813558 &  79.5425 & 0.0014 & -0.0241 & 6.7486 & -4.965 & 0.019 & 1800 &  60 \\
 8794.779193 &  79.5371 & 0.0013 & -0.0228 & 6.7494 & -4.934 & 0.018 & 1800 &  61 \\
 8795.703042 &  79.5318 & 0.0014 & -0.0265 & 6.7443 & -4.943 & 0.019 & 1800 &  58 \\
 8796.819006 &  79.5386 & 0.0017 & -0.0264 & 6.7501 & -4.992 & 0.029 & 1800 &  50 \\
 8797.761303 &  79.5437 & 0.0019 & -0.0324 & 6.7361 & -4.961 & 0.029 & 1800 &  46 \\
 8798.828187 &  79.5355 & 0.0021 & -0.0316 & 6.7468 & -4.967 & 0.033 & 1800 &  42 \\
 8798.849647 &  79.5430 & 0.0020 & -0.0302 & 6.7529 & -4.982 & 0.033 & 1800 &  45 \\
 8799.804743 &  79.5415 & 0.0016 & -0.0213 & 6.7485 & -4.935 & 0.022 & 1800 &  52 \\
 8802.684679 &  79.5488 & 0.0007 & -0.0260 & 6.7498 & -4.880 & 0.006 & 1800 & 111 \\
 8803.723755 &  79.5537 & 0.0008 & -0.0257 & 6.7537 & -4.873 & 0.007 & 1800 & 101 \\
 8805.801094 &  79.5514 & 0.0009 & -0.0266 & 6.7587 & -4.890 & 0.011 & 1800 &  87 \\
 8806.812098 &  79.5532 & 0.0009 & -0.0205 & 6.7557 & -4.897 & 0.010 & 1800 &  92 \\
 8820.826561 &  79.5478 & 0.0008 & -0.0262 & 6.7460 & -4.954 & 0.012 & 1800 & 102 \\
 8821.721745 &  79.5424 & 0.0007 & -0.0243 & 6.7479 & -4.964 & 0.010 & 1800 & 110 \\
 8823.814612 &  79.5494 & 0.0008 & -0.0267 & 6.7466 & -5.000 & 0.016 & 1800 & 100 \\
 8824.842097 &  79.5463 & 0.0010 & -0.0259 & 6.7576 & -4.948 & 0.015 & 1800 &  86 \\
 8826.728971 &  79.5352 & 0.0007 & -0.0247 & 6.7491 & -4.936 & 0.008 & 1800 & 123 \\
 8830.721095 &  79.5404 & 0.0008 & -0.0259 & 6.7451 & -4.924 & 0.010 & 1800 &  99 \\
 8831.736359 &  79.5368 & 0.0009 & -0.0244 & 6.7450 & -4.943 & 0.012 & 1800 &  90 \\
 8832.741251 &  79.5387 & 0.0011 & -0.0260 & 6.7508 & -4.894 & 0.014 & 1800 &  74 \\
 8833.766142 &  79.5419 & 0.0009 & -0.0248 & 6.7506 & -4.923 & 0.011 & 1800 &  95 \\
 8834.754203 &  79.5460 & 0.0008 & -0.0280 & 6.7481 & -4.896 & 0.010 & 1800 &  97 \\
 8835.678212 &  79.5481 & 0.0009 & -0.0247 & 6.7447 & -4.920 & 0.012 & 1800 &  87 \\
 8835.804299 &  79.5456 & 0.0007 & -0.0245 & 6.7512 & -4.914 & 0.009 & 1800 & 119 \\
 8841.709874 &  79.5455 & 0.0017 & -0.0244 & 6.7541 & -4.941 & 0.026 & 1500 &  50 \\
 8843.694063 &  79.5398 & 0.0013 & -0.0252 & 6.7475 & -4.906 & 0.018 & 1500 &  62 \\
 8845.701891 &  79.5434 & 0.0009 & -0.0212 & 6.7444 & -4.916 & 0.013 & 1500 &  90 \\
 8847.726300 &  79.5392 & 0.0008 & -0.0224 & 6.7489 & -4.923 & 0.012 & 1500 &  97 \\
 8856.746648 &  79.5476 & 0.0012 & -0.0297 & 6.7456 & -4.967 & 0.022 & 1800 &  70 \\
 8856.768244 &  79.5443 & 0.0014 & -0.0242 & 6.7371 & -4.969 & 0.027 & 1800 &  64 \\
 8857.687817 &  79.5417 & 0.0008 & -0.0245 & 6.7430 & -4.947 & 0.011 & 1800 & 107 \\
 8857.709216 &  79.5415 & 0.0007 & -0.0254 & 6.7424 & -4.958 & 0.011 & 1800 & 110 \\
 8858.570410 &  79.5403 & 0.0010 & -0.0234 & 6.7414 & -4.956 & 0.014 & 1800 &  77 \\
 8858.689630 &  79.5424 & 0.0009 & -0.0248 & 6.7450 & -4.933 & 0.013 & 1800 &  93 \\
 8859.582685 &  79.5422 & 0.0018 & -0.0185 & 6.7456 & -4.962 & 0.033 & 1800 &  49 \\
 8859.675147 &  79.5433 & 0.0011 & -0.0281 & 6.7448 & -4.966 & 0.016 & 1800 &  72 \\
 8860.603211 &  79.5408 & 0.0010 & -0.0237 & 6.7412 & -4.948 & 0.013 & 1800 &  85 \\
 8860.695071 &  79.5420 & 0.0010 & -0.0277 & 6.7480 & -4.979 & 0.017 & 1800 &  80 \\
 8861.558044 &  79.5435 & 0.0011 & -0.0266 & 6.7424 & -4.961 & 0.016 & 1800 &  73 \\
 8861.720386 &  79.5431 & 0.0009 & -0.0250 & 6.7457 & -4.939 & 0.014 & 1800 &  90 \\
 8862.575813 &  79.5388 & 0.0008 & -0.0257 & 6.7449 & -4.955 & 0.011 & 1800 &  94 \\
 8862.688991 &  79.5341 & 0.0009 & -0.0294 & 6.7392 & -4.953 & 0.013 & 1800 &  94 \\
 8863.592677 &  79.5362 & 0.0010 & -0.0253 & 6.7433 & -4.959 & 0.015 & 1800 &  87 \\
 8863.684930 &  79.5385 & 0.0008 & -0.0243 & 6.7416 & -4.936 & 0.012 & 1800 & 109 \\
 8864.587945 &  79.5413 & 0.0010 & -0.0257 & 6.7473 & -4.931 & 0.013 & 1800 &  80 \\
 8864.681911 &  79.5418 & 0.0009 & -0.0235 & 6.7408 & -4.926 & 0.012 & 1800 &  90 \\
 8865.614228 &  79.5448 & 0.0010 & -0.0248 & 6.7429 & -4.928 & 0.013 & 1800 &  81 \\
 8865.717198 &  79.5444 & 0.0010 & -0.0180 & 6.7384 & -4.954 & 0.015 & 1800 &  86 \\
 8866.667986 &  79.5441 & 0.0008 & -0.0222 & 6.7457 & -4.935 & 0.012 & 1800 &  100 \\
 8868.665966 &  79.5436 & 0.0009 & -0.0225 & 6.7426 & -4.949 & 0.013 & 1650 &  93 \\
 8869.699467 &  79.5441 & 0.0007 & -0.0246 & 6.7449 & -4.990 & 0.012 & 1660 & 109 \\
 8871.661192 &  79.5486 & 0.0010 & -0.0245 & 6.7466 & -4.967 & 0.016 & 1800 &  78 \\
\hline
\end{longtable*}


\clearpage

\begin{table}[t]
\small
\begin{center}
\caption{\label{Table:HIRES} HIRES RV measurements of \sname. (a) Barycentric Julian dates are given in barycentric dynamical time; (b) S/N ratio per pixel at 550 nm.}
\begin{tabular}{lcccccr}
\hline\hline
\noalign{\smallskip}
$\rm BJD_{TDB}^a$ & RV  & $\sigma$ & T$_\mathrm{exp}$ & S/N $^b$\\
-2450000          & (\kms)  & (\kms)   &      (s)         &   \\
\hline
\noalign{\smallskip}
\noalign{\smallskip}
 8744.056124 &  -0.0047 & 0.0011 & 770 & 218 \\
 8777.027485 &   0.0047 & 0.0011 & 592 & 218 \\
 8788.071425 &   0.0061 & 0.0011 & 680 & 221 \\
 8794.976150 &  -0.0076 & 0.0009 & 716 & 220 \\
 8796.022921 &  -0.0090 & 0.0011 & 774 & 221 \\
 8797.052515 &  -0.0061 & 0.0010 & 900 & 212 \\
 8798.104180 &  -0.0030 & 0.0010 & 706 & 218 \\
 8798.917940 &  -0.0062 & 0.0013 & 900 & 186 \\
 8802.899539 &   0.0038 & 0.0010 & 537 & 219 \\
 8809.050906 &   0.0086 & 0.0012 & 650 & 218 \\
 8815.908118 &  -0.0055 & 0.0012 & 808 & 216 \\
 8819.967016 &   0.0069 & 0.0010 & 900 & 214 \\
 8827.942590 &  -0.0138 & 0.0011 & 595 & 218 \\
 8832.964946 &  -0.0009 & 0.0012 & 877 & 221 \\
 8833.929016 &   0.0011 & 0.0012 & 582 & 220 \\
 8844.885674 &  -0.0048 & 0.0011 & 634 & 219 \\
 8845.918812 &  -0.0050 & 0.0011 & 651 & 217 \\
 8852.868230 &   0.0035 & 0.0012 & 633 & 220 \\
 8855.804440 &   0.0072 & 0.0012 & 900 & 197 \\
 8856.831220 &   0.0043 & 0.0011 & 900 & 182 \\
 8857.840778 &  -0.0011 & 0.0011 & 824 & 220 \\
 8869.855111 &   0.0026 & 0.0011 & 774 & 219 \\
 8870.880081 &   0.0055 & 0.0011 & 762 & 219 \\
 8878.833553 &  -0.0047 & 0.0011 & 567 & 219 \\
 8879.783275 &  -0.0060 & 0.0011 & 865 & 220 \\
 8880.810664 &  -0.0012 & 0.0011 & 582 & 219 \\
 8884.813904 &   0.0038 & 0.0011 & 512 & 219 \\
 8885.935404 &   0.0021 & 0.0015 & 899 & 164 \\
 8903.828998  &  0.0020 & 0.0012 & 900 & 174 \\
 8905.785063  & -0.0007 & 0.0010 & 631 & 218 \\
 8906.756423  & -0.0017 & 0.0011 & 670 & 218 \\ 
 8907.751496  & -0.0039 & 0.0012 & 900 & 187 \\
 8911.799720  & -0.0028 & 0.0010 & 668 & 216 \\
\noalign{\smallskip}
\hline
\end{tabular}
\end{center}
\end{table}

\begin{table}[t]
\small
\begin{center}
\caption{\label{Table:PFS} PFS RV measurements of \sname. (a) Barycentric Julian dates are given in barycentric dynamical time; (b) S/N ratio per pixel at 550 nm.}
\begin{tabular}{lccrr}
\hline\hline
\noalign{\smallskip}
$\rm BJD_{TDB}^a$ & RV  & $\sigma$ & T$_\mathrm{exp}$ & S/N $^b$\\
-2450000          & (\kms)  & (\kms)   &      (s)         &   \\
\hline
\noalign{\smallskip}
\noalign{\smallskip}
 8592.508995 &  -0.0006 & 0.0009 & 2400 & 73\\
 8708.926166 &   0.0059 & 0.0009 & 1200 & 61\\
 8717.901009 &  -0.0017 & 0.0010 & 1200 & 50\\
 8738.854892 &  -0.0002 & 0.0012 & 1200 & 51\\
 8739.858020 &  -0.0007 & 0.0010 & 1200 & 57\\
 8763.842388 &  -0.0023 & 0.0008 & 1200 & 66\\
 8764.859431 &   0.0011 & 0.0007 & 1200 & 71\\
 8767.887884 &   0.0057 & 0.0007 & 1200 & 77\\
 8768.852571 &   0.0042 & 0.0009 &  900 & 55\\
\noalign{\smallskip}
\hline
\end{tabular}
\end{center}
\end{table}
\clearpage

\bibliography{biblio}{}

\begin{thebibliography}{}
\expandafter\ifx\csname natexlab\endcsname\relax\def\natexlab#1{#1}\fi
\providecommand{\url}[1]{\href{#1}{#1}}
\providecommand{\dodoi}[1]{doi:~\href{http://doi.org/#1}{\nolinkurl{#1}}}
\providecommand{\doeprint}[1]{\href{http://ascl.net/#1}{\nolinkurl{http://ascl.net/#1}}}
\providecommand{\doarXiv}[1]{\href{https://arxiv.org/abs/#1}{\nolinkurl{https://arxiv.org/abs/#1}}}

\bibitem[{{Allan} \& {Vidotto}(2019)}]{2019MNRAS.490.3760A}
{Allan}, A., \& {Vidotto}, A.~A. 2019, \mnras, 490, 3760,
  \dodoi{10.1093/mnras/stz2842}

\bibitem[{{Allard} {et~al.}(2012){Allard}, {Homeier}, \&
  {Freytag}}]{Allard2012}
{Allard}, F., {Homeier}, D., \& {Freytag}, B. 2012, Philosophical Transactions
  of the Royal Society of London Series A, 370, 2765,
  \dodoi{10.1098/rsta.2011.0269}

\bibitem[{{Awiphan} {et~al.}(2016){Awiphan}, {Kerins}, {Pichadee},
  {Komonjinda}, {Dhillon}, {Rujopakarn}, {Poshyachinda}, {Marsh}, {Reichart},
  {Ivarsen}, \& {Haislip}}]{Awiphanetal2016}
{Awiphan}, S., {Kerins}, E., {Pichadee}, S., {et~al.} 2016, \mnras, 463, 2574,
  \dodoi{10.1093/mnras/stw2148}

\bibitem[{{Baranne} {et~al.}(1996){Baranne}, {Queloz}, {Mayor}, {Adrianzyk},
  {Knispel}, {Kohler}, {Lacroix}, {Meunier}, {Rimbaud}, \& {Vin}}]{Baranne1996}
{Baranne}, A., {Queloz}, D., {Mayor}, M., {et~al.} 1996, \aaps, 119, 373

\bibitem[{{Barclay} {et~al.}(2018){Barclay}, {Pepper}, \&
  {Quintana}}]{Barclayetal2018}
{Barclay}, T., {Pepper}, J., \& {Quintana}, E.~V. 2018, \apjs, 239, 2,
  \dodoi{10.3847/1538-4365/aae3e9}

\bibitem[{{Barrag{\'a}n} {et~al.}(2019){Barrag{\'a}n}, {Gandolfi}, \&
  {Antoniciello}}]{pyaneti}
{Barrag{\'a}n}, O., {Gandolfi}, D., \& {Antoniciello}, G. 2019, \mnras, 482,
  1017, \dodoi{10.1093/mnras/sty2472}

\bibitem[{{Barrag{\'a}n} {et~al.}(2018){Barrag{\'a}n}, {Gandolfi}, {Dai},
  {Livingston}, {Persson}, {Hirano}, {Narita}, {Csizmadia}, {Winn}, {Nespral},
  {Prieto-Arranz}, {Smith}, {Nowak}, {Albrecht}, {Antoniciello}, {Bo Justesen},
  {Cabrera}, {Cochran}, {Deeg}, {Eigmuller}, {Endl}, {Erikson}, {Fridlund},
  {Fukui}, {Grziwa}, {Guenther}, {Hatzes}, {Hidalgo}, {Johnson}, {Korth},
  {Palle}, {Patzold}, {Rauer}, {Tanaka}, \& {Van Eylen}}]{Barragan2018}
{Barrag{\'a}n}, O., {Gandolfi}, D., {Dai}, F., {et~al.} 2018, \aap, 612, A95,
  \dodoi{10.1051/0004-6361/201732217}

\bibitem[{{Barstow} {et~al.}(2020){Barstow}, {Changeat}, {Garland}, {Line},
  {Rocchetto}, \& {Waldmann}}]{barstowchangeat2020}
{Barstow}, J.~K., {Changeat}, Q., {Garland}, R., {et~al.} 2020, \mnras, 493,
  4884, \dodoi{10.1093/mnras/staa548}

\bibitem[{{Benneke} \& {Seager}(2013)}]{bennekeseager2013}
{Benneke}, B., \& {Seager}, S. 2013, \apj, 778, 153,
  \dodoi{10.1088/0004-637X/778/2/153}

\bibitem[{{Benneke} {et~al.}(2019){Benneke}, {Wong}, {Piaulet}, {Knutson},
  {Lothringer}, {Morley}, {Crossfield}, {Gao}, {Greene}, {Dressing},
  {Dragomir}, {Howard}, {McCullough}, {Kempton}, {Fortney}, \&
  {Fraine}}]{Bennekeetal2019}
{Benneke}, B., {Wong}, I., {Piaulet}, C., {et~al.} 2019, \apjl, 887, L14,
  \dodoi{10.3847/2041-8213/ab59dc}

\bibitem[{{Bourrier} {et~al.}(2018){Bourrier}, {Lecavelier des Etangs},
  {Ehrenreich}, {Sanz-Forcada}, {Allart}, {Ballester}, {Buchhave}, {Cohen},
  {Deming}, {Evans}, {Garc{\'\i}a Mu{\~n}oz}, {Henry}, {Kataria}, {Lavvas},
  {Lewis}, {L{\'o}pez-Morales}, {Marley}, {Sing}, \& {Wakeford}}]{Bourrier2018}
{Bourrier}, V., {Lecavelier des Etangs}, A., {Ehrenreich}, D., {et~al.} 2018,
  \aap, 620, A147, \dodoi{10.1051/0004-6361/201833675}

\bibitem[{{Brahm} {et~al.}(2019{\natexlab{a}}){Brahm}, {Espinoza},
  {Jord{\'a}n}, {Henning}, {Sarkis}, {Jones}, {D{\'\i}az}, {Jenkins}, {Vanzi},
  {Zapata}, {Petrovich}, {Kossakowski}, {Rabus}, {Rojas}, \&
  {Torres}}]{Brahmetal2019}
{Brahm}, R., {Espinoza}, N., {Jord{\'a}n}, A., {et~al.} 2019{\natexlab{a}},
  \aj, 158, 45, \dodoi{10.3847/1538-3881/ab279a}

\bibitem[{{Brahm} {et~al.}(2019{\natexlab{b}}){Brahm}, {Espinoza},
  {Jord{\'a}n}, {Henning}, {Sarkis}, {Jones}, {D{\'\i}az}, {Jenkins}, {Vanzi},
  {Zapata}, {Petrovich}, {Kossakowski}, {Rabus}, {Rojas}, \&
  {Torres}}]{Brahm2019}
---. 2019{\natexlab{b}}, \aj, 158, 45, \dodoi{10.3847/1538-3881/ab279a}

\bibitem[{{Bressan} {et~al.}(2012){Bressan}, {Marigo}, {Girardi}, {Salasnich},
  {Dal Cero}, {Rubele}, \& {Nanni}}]{2012MNRAS.427..127B}
{Bressan}, A., {Marigo}, P., {Girardi}, L., {et~al.} 2012, \mnras, 427, 127,
  \dodoi{10.1111/j.1365-2966.2012.21948.x}

\bibitem[{{Brown} {et~al.}(2013){Brown}, {Baliber}, {Bianco}, {Bowman},
  {Burleson}, {Conway}, {Crellin}, {Depagne}, {De Vera}, {Dilday}, {Dragomir},
  {Dubberley}, {Eastman}, {Elphick}, {Falarski}, {Foale}, {Ford}, {Fulton},
  {Garza}, {Gomez}, {Graham}, {Greene}, {Haldeman}, {Hawkins}, {Haworth},
  {Haynes}, {Hidas}, {Hjelstrom}, {Howell}, {Hygelund}, {Lister}, {Lobdill},
  {Martinez}, {Mullins}, {Norbury}, {Parrent}, {Paulson}, {Petry}, {Pickles},
  {Posner}, {Rosing}, {Ross}, {Sand}, {Saunders}, {Shobbrook}, {Shporer},
  {Street}, {Thomas}, {Tsapras}, {Tufts}, {Valenti}, {Vander Horst}, {Walker},
  {White}, \& {Willis}}]{Brown:2013}
{Brown}, T.~M., {Baliber}, N., {Bianco}, F.~B., {et~al.} 2013, Publications of
  the Astronomical Society of the Pacific, 125, 1031, \dodoi{10.1086/673168}

\bibitem[{{Bruntt} {et~al.}(2010){Bruntt}, {Bedding}, {Quirion}, {Lo Curto},
  {Carrier}, {Smalley}, {Dall}, {Arentoft}, {Bazot}, \& {Butler}}]{Bruntt2010b}
{Bruntt}, H., {Bedding}, T.~R., {Quirion}, P.-O., {et~al.} 2010, \mnras, 405,
  1907, \dodoi{10.1111/j.1365-2966.2010.16575.x}

\bibitem[{{Buchhave} {et~al.}(2010){Buchhave}, {Bakos}, {Hartman}, {Torres},
  {Kov{\'a}cs}, {Latham}, {Noyes}, {Esquerdo}, {Everett}, {Howard}, {Marcy},
  {Fischer}, {Johnson}, {Andersen}, {F{\H{u}}r{\'e}sz}, {Perumpilly},
  {Sasselov}, {Stefanik}, {B{\'e}ky}, {L{\'a}z{\'a}r}, {Papp}, \&
  {S{\'a}ri}}]{Buchhave2010}
{Buchhave}, L.~A., {Bakos}, G.~{\'A}., {Hartman}, J.~D., {et~al.} 2010, \apj,
  720, 1118, \dodoi{10.1088/0004-637X/720/2/1118}

\bibitem[{{Buchner} {et~al.}(2014){Buchner}, {Georgakakis}, {Nandra}, {Hsu},
  {Rangel}, {Brightman}, {Merloni}, {Salvato}, {Donley}, \&
  {Kocevski}}]{buchnergeorgakakis2014}
{Buchner}, J., {Georgakakis}, A., {Nandra}, K., {et~al.} 2014, \aap, 564, A125,
  \dodoi{10.1051/0004-6361/201322971}

\bibitem[{Burnham \& Anderson(2004)}]{Burnham2014}
Burnham, K.~P., \& Anderson, D.~R. 2004, Sociological Methods \& Research, 33,
  261, \dodoi{10.1177/0049124104268644}

\bibitem[{{Butler} {et~al.}(1996){Butler}, {Marcy}, {Williams}, {McCarthy},
  {Dosanjh}, \& {Vogt}}]{Butler1996}
{Butler}, R.~P., {Marcy}, G.~W., {Williams}, E., {et~al.} 1996, \pasp, 108,
  500, \dodoi{10.1086/133755}

\bibitem[{{Cabrera} {et~al.}(2012){Cabrera}, {Csizmadia}, {Erikson}, {Rauer},
  \& {Kirste}}]{Cabrera2012}
{Cabrera}, J., {Csizmadia}, S., {Erikson}, A., {Rauer}, H., \& {Kirste}, S.
  2012, \aap, 548, A44, \dodoi{10.1051/0004-6361/201219337}

\bibitem[{{Cardelli} {et~al.}(1989){Cardelli}, {Clayton}, \&
  {Mathis}}]{Cardelli1998}
{Cardelli}, J.~A., {Clayton}, G.~C., \& {Mathis}, J.~S. 1989, \apj, 345, 245,
  \dodoi{10.1086/167900}

\bibitem[{{Carleo} {et~al.}(2020){Carleo}, {Malavolta}, {Lanza}, {Damasso},
  {Desidera}, {Borsa}, {Mallonn}, {Pinamonti}, {Gratton}, {Alei}, {Benatti},
  {Mancini}, {Maldonado}, {Biazzo}, {Esposito}, {Frustagli},
  {Gonz{\'a}lez-{\'A}lvarez}, {Micela}, {Scandariato}, {Sozzetti}, {Affer},
  {Bignamini}, {Bonomo}, {Claudi}, {Cosentino}, {Covino}, {Fiorenzano},
  {Giacobbe}, {Harutyunyan}, {Leto}, {Maggio}, {Molinari}, {Nascimbeni},
  {Pagano}, {Pedani}, {Piotto}, {Poretti}, {Rainer}, {Redfield}, {Baffa},
  {Baruffolo}, {Buchschacher}, {Billotti}, {Cecconi}, {Falcini}, {Fantinel},
  {Fini}, {Galli}, {Ghedina}, {Ghinassi}, {Giani}, {Gonzalez}, {Gonzalez},
  {Guerra}, {Hernandez Diaz}, {Hernandez}, {Iuzzolino}, {Lodi}, {Oliva},
  {Origlia}, {Perez Ventura}, {Puglisi}, {Riverol}, {Riverol}, {San Juan},
  {Sanna}, {Scuderi}, {Seemann}, {Sozzi}, \& {Tozzi}}]{carleo2020}
{Carleo}, I., {Malavolta}, L., {Lanza}, A.~F., {et~al.} 2020, arXiv e-prints,
  arXiv:2002.10562.
\newblock \doarXiv{2002.10562}

\bibitem[{{Ceillier} {et~al.}(2017){Ceillier}, {Tayar}, {Mathur}, {Salabert},
  {Garc{\'\i}a}, {Stello}, {Pinsonneault}, {van Saders}, {Beck}, \&
  {Bloemen}}]{Ceillieretal2017}
{Ceillier}, T., {Tayar}, J., {Mathur}, S., {et~al.} 2017, \aap, 605, A111,
  \dodoi{10.1051/0004-6361/201629884}

\bibitem[{{Choi} {et~al.}(2016){Choi}, {Dotter}, {Conroy}, {Cantiello},
  {Paxton}, \& {Johnson}}]{2016ApJ...823..102C}
{Choi}, J., {Dotter}, A., {Conroy}, C., {et~al.} 2016, \apj, 823, 102,
  \dodoi{10.3847/0004-637X/823/2/102}

\bibitem[{{Claret} \& {Bloemen}(2011)}]{Claret1}
{Claret}, A., \& {Bloemen}, S. 2011, \aap, 529, A75,
  \dodoi{10.1051/0004-6361/201116451}

\bibitem[{{Collins} {et~al.}(2017){Collins}, {Kielkopf}, {Stassun}, \&
  {Hessman}}]{Collins:2017}
{Collins}, K.~A., {Kielkopf}, J.~F., {Stassun}, K.~G., \& {Hessman}, F.~V.
  2017, \aj, 153, 77, \dodoi{10.3847/1538-3881/153/2/77}

\bibitem[{{Crane} {et~al.}(2006){Crane}, {Shectman}, \&
  {Butler}}]{craneetal2006}
{Crane}, J.~D., {Shectman}, S.~A., \& {Butler}, R.~P. 2006, Society of
  Photo-Optical Instrumentation Engineers (SPIE) Conference Series, Vol. 6269,
  {The Carnegie Planet Finder Spectrograph}, 626931, \dodoi{10.1117/12.672339}

\bibitem[{{Crane} {et~al.}(2010){Crane}, {Shectman}, {Butler}, {Thompson},
  {Birk}, {Jones}, \& {Burley}}]{Craneetal2010}
{Crane}, J.~D., {Shectman}, S.~A., {Butler}, R.~P., {et~al.} 2010, Society of
  Photo-Optical Instrumentation Engineers (SPIE) Conference Series, Vol. 7735,
  {The Carnegie Planet Finder Spectrograph: integration and commissioning},
  773553, \dodoi{10.1117/12.857792}

\bibitem[{{Crane} {et~al.}(2008){Crane}, {Shectman}, {Butler}, {Thompson}, \&
  {Burley}}]{craneetal2008}
{Crane}, J.~D., {Shectman}, S.~A., {Butler}, R.~P., {Thompson}, I.~B., \&
  {Burley}, G.~S. 2008, Society of Photo-Optical Instrumentation Engineers
  (SPIE) Conference Series, Vol. 7014, {The Carnegie Planet Finder
  Spectrograph: a status report}, 701479, \dodoi{10.1117/12.789637}

\bibitem[{{Cubillos} {et~al.}(2017{\natexlab{a}}){Cubillos}, {Harrington},
  {Loredo}, {Lust}, {Blecic}, \& {Stemm}}]{cubillos2017a}
{Cubillos}, P., {Harrington}, J., {Loredo}, T.~J., {et~al.} 2017{\natexlab{a}},
  \aj, 153, 3, \dodoi{10.3847/1538-3881/153/1/3}

\bibitem[{{Cubillos} {et~al.}(2017{\natexlab{b}}){Cubillos}, {Erkaev}, {Juvan},
  {Fossati}, {Johnstone}, {Lammer}, {Lendl}, {Odert}, \&
  {Kislyakova}}]{cubillos2017b}
{Cubillos}, P., {Erkaev}, N.~V., {Juvan}, I., {et~al.} 2017{\natexlab{b}},
  \mnras, 466, 1868, \dodoi{10.1093/mnras/stw3103}

\bibitem[{{Cutri} \& {et al.}(2013)}]{cutri2013}
{Cutri}, R.~M., \& {et al.} 2013, VizieR Online Data Catalog, II/328

\bibitem[{{Cutri} {et~al.}(2003){Cutri}, {Skrutskie}, {van Dyk}, {Beichman},
  {Carpenter}, {Chester}, {Cambresy}, {Evans}, {Fowler}, {Gizis}, {Howard},
  {Huchra}, {Jarrett}, {Kopan}, {Kirkpatrick}, {Light}, {Marsh}, {McCallon},
  {Schneider}, {Stiening}, {Sykes}, {Weinberg}, {Wheaton}, {Wheelock}, \&
  {Zacarias}}]{cutri2003}
{Cutri}, R.~M., {Skrutskie}, M.~F., {van Dyk}, S., {et~al.} 2003, {2MASS All
  Sky Catalog of point sources.}

\bibitem[{{da Silva} {et~al.}(2006){da Silva}, {Girardi}, {Pasquini},
  {Setiawan}, {von der L{\"u}he}, {de Medeiros}, {Hatzes}, {D{\"o}llinger}, \&
  {Weiss}}]{daSilva2006}
{da Silva}, L., {Girardi}, L., {Pasquini}, L., {et~al.} 2006, \aap, 458, 609,
  \dodoi{10.1051/0004-6361:20065105}

\bibitem[{{Dai} {et~al.}(2017){Dai}, {Winn}, {Gandolfi}, {Wang}, {Teske},
  {Burt}, {Albrecht}, {Barrag{\'a}n}, {Cochran}, {Endl}, {Fridlund}, {Hatzes},
  {Hirano}, {Hirsch}, {Johnson}, {Justesen}, {Livingston}, {Persson},
  {Prieto-Arranz}, {Vanderburg}, {Alonso}, {Antoniciello}, {Arriagada},
  {Butler}, {Cabrera}, {Crane}, {Cusano}, {Csizmadia}, {Deeg}, {Dieterich},
  {Eigm{\"u}ller}, {Erikson}, {Everett}, {Fukui}, {Grziwa}, {Guenther},
  {Henry}, {Howell}, {Johnson}, {Korth}, {Kuzuhara}, {Narita}, {Nespral},
  {Nowak}, {Palle}, {P{\"a}tzold}, {Rauer}, {Monta{\~n}{\'e}s Rodr{\'\i}guez},
  {Shectman}, {Smith}, {Thompson}, {Van Eylen}, {Williamson}, \&
  {Wittenmyer}}]{2017AJ....154..226D}
{Dai}, F., {Winn}, J.~N., {Gandolfi}, D., {et~al.} 2017, \aj, 154, 226,
  \dodoi{10.3847/1538-3881/aa9065}

\bibitem[{{D{\'\i}az} {et~al.}(2020){D{\'\i}az}, {Jenkins}, {Gand olfi},
  {Lopez}, {Soto}, {Cort{\'e}s-Zuleta}, {Berdi{\~n}as}, {Stassun}, {Collins},
  {Vines}, {Ziegler}, {Fridlund}, {Jensen}, {Murgas}, {Santerne}, {Wilson},
  {Esposito}, {Hatzes}, {Johnson}, {Lam}, {Livingston}, {Van Eylen}, {Narita},
  {Brice{\~n}o}, {Collins}, {Csizmadia}, {Fausnaugh}, {Gan}, {Garc{\'\i}a},
  {Georgieva}, {Glidden}, {Gonz{\'a}lez-Cuesta}, {Jenkins}, {Latham}, {Law},
  {Mann}, {Mathur}, {Mireles}, {Morris}, {Pall{\'e}}, {Persson}, {Ricker},
  {Rinehart}, {Rose}, {Seager}, {Smith}, {Tan}, {Tokovinin}, {Vanderburg},
  {Vanderspek}, {Winn}, \& {Yahalomi}}]{diazetal2020}
{D{\'\i}az}, M.~R., {Jenkins}, J.~S., {Gand olfi}, D., {et~al.} 2020, \mnras,
  \dodoi{10.1093/mnras/staa277}

\bibitem[{{Donati} \& {Landstreet}(2009)}]{Donati09}
{Donati}, J.~F., \& {Landstreet}, J.~D. 2009, \araa, 47, 333,
  \dodoi{10.1146/annurev-astro-082708-101833}

\bibitem[{{dos Santos} {et~al.}(2020){dos Santos}, {Ehrenreich}, {Bourrier},
  {Astudillo-Defru}, {Bonfils}, {Forget}, {Lovis}, {Pepe}, \&
  {Udry}}]{dosSantos2020}
{dos Santos}, L.~A., {Ehrenreich}, D., {Bourrier}, V., {et~al.} 2020, \aap,
  634, L4, \dodoi{10.1051/0004-6361/201937327}

\bibitem[{{Doyle} {et~al.}(2014){Doyle}, {Davies}, {Smalley}, {Chaplin}, \&
  {Elsworth}}]{Doyle2014}
{Doyle}, A.~P., {Davies}, G.~R., {Smalley}, B., {Chaplin}, W.~J., \&
  {Elsworth}, Y. 2014, \mnras, 444, 3592, \dodoi{10.1093/mnras/stu1692}

\bibitem[{{Dragomir} {et~al.}(2019){Dragomir}, {Teske}, {G{\"u}nther},
  {S{\'e}gransan}, {Burt}, {Huang}, {Vanderburg}, {Matthews}, {Dumusque},
  {Stassun}, {Pepper}, {Ricker}, {Vanderspek}, {Latham}, {Seager}, {Winn},
  {Jenkins}, {Beatty}, {Bouchy}, {Brown}, {Butler}, {Ciardi}, {Crane},
  {Eastman}, {Fossati}, {Francis}, {Fulton}, {Gaudi}, {Goeke}, {James},
  {Klaus}, {Kuhn}, {Lovis}, {Lund}, {McDermott}, {Paegert}, {Pepe},
  {Rodriguez}, {Sha}, {Shectman}, {Shporer}, {Siverd}, {Garcia Soto},
  {Stevens}, {Twicken}, {Udry}, {Villanueva}, {Wang}, {Wohler}, {Yao}, \&
  {Zhan}}]{Dragomiretal2019}
{Dragomir}, D., {Teske}, J., {G{\"u}nther}, M.~N., {et~al.} 2019, \apjl, 875,
  L7, \dodoi{10.3847/2041-8213/ab12ed}

\bibitem[{{Dumusque} {et~al.}(2014){Dumusque}, {Boisse}, \&
  {Santos}}]{Dumusque}
{Dumusque}, X., {Boisse}, I., \& {Santos}, N.~C. 2014, \apj, 796, 132,
  \dodoi{10.1088/0004-637X/796/2/132}

\bibitem[{{Esposito} {et~al.}(2017){Esposito}, {Covino}, {Desidera}, {Mancini},
  {Nascimbeni}, {Zanmar Sanchez}, {Biazzo}, {Lanza}, {Leto}, {Southworth},
  {Bonomo}, {Su{\'a}rez Mascare{\~n}o}, {Boccato}, {Cosentino}, {Claudi},
  {Gratton}, {Maggio}, {Micela}, {Molinari}, {Pagano}, {Piotto}, {Poretti},
  {Smareglia}, {Sozzetti}, {Affer}, {Anderson}, {Andreuzzi}, {Benatti},
  {Bignamini}, {Borsa}, {Borsato}, {Ciceri}, {Damasso}, {di Fabrizio},
  {Giacobbe}, {Granata}, {Harutyunyan}, {Henning}, {Malavolta}, {Maldonado},
  {Martinez Fiorenzano}, {Masiero}, {Molaro}, {Molinaro}, {Pedani}, {Rainer},
  {Scandariato}, \& {Turner}}]{Espositoetal2017}
{Esposito}, M., {Covino}, E., {Desidera}, S., {et~al.} 2017, \aap, 601, A53,
  \dodoi{10.1051/0004-6361/201629720}

\bibitem[{{Esposito} {et~al.}(2019){Esposito}, {Armstrong}, {Gandolfi},
  {Adibekyan}, {Fridlund}, {Santos}, {Livingston}, {Delgado Mena}, {Fossati},
  {Lillo-Box}, {Barrag{\'a}n}, {Barrado}, {Cubillos}, {Cooke}, {Justesen},
  {Meru}, {D{\'\i}az}, {Dai}, {Nielsen}, {Persson}, {Wheatley}, {Hatzes}, {Van
  Eylen}, {Musso}, {Alonso}, {Beck}, {Barros}, {Bayliss}, {Bonomo}, {Bouchy},
  {Brown}, {Bryant}, {Cabrera}, {Cochran}, {Csizmadia}, {Deeg}, {Demangeon},
  {Deleuil}, {Dumusque}, {Eigm{\"u}ller}, {Endl}, {Erikson}, {Faedi},
  {Figueira}, {Fukui}, {Grziwa}, {Guenther}, {Hidalgo}, {Hjorth}, {Hirano},
  {Hojjatpanah}, {Knudstrup}, {Korth}, {Lam}, {de Leon}, {Lund}, {Luque},
  {Mathur}, {Monta{\~n}{\'e}s Rodr{\'\i}guez}, {Narita}, {Nespral}, {Niraula},
  {Nowak}, {Osborn}, {Pall{\'e}}, {P{\"a}tzold}, {Pollacco}, {Prieto-Arranz},
  {Rauer}, {Redfield}, {Ribas}, {Sousa}, {Smith}, {Tala-Pinto}, {Udry}, \&
  {Winn}}]{Espositoetal2019}
{Esposito}, M., {Armstrong}, D.~J., {Gandolfi}, D., {et~al.} 2019, \aap, 623,
  A165, \dodoi{10.1051/0004-6361/201834853}

\bibitem[{{Feroz} {et~al.}(2009){Feroz}, {Hobson}, \&
  {Bridges}}]{ferozhobson2009}
{Feroz}, F., {Hobson}, M.~P., \& {Bridges}, M. 2009, \mnras, 398, 1601,
  \dodoi{10.1111/j.1365-2966.2009.14548.x}

\bibitem[{{Feroz} {et~al.}(2013){Feroz}, {Hobson}, {Cameron}, \&
  {Pettitt}}]{2013arXiv1306.2144F}
{Feroz}, F., {Hobson}, M.~P., {Cameron}, E., \& {Pettitt}, A.~N. 2013, ArXiv
  e-prints.
\newblock \doarXiv{1306.2144}

\bibitem[{{Fossati} {et~al.}(2017{\natexlab{a}}){Fossati}, {Marcelja}, {Staab},
  {Cubillos}, {France}, {Haswell}, {Ingrassia}, {Jenkins}, {Koskinen}, {Lanza},
  {Redfield}, {Youngblood}, \& {Pelzmann}}]{fossati2017a}
{Fossati}, L., {Marcelja}, S.~E., {Staab}, D., {et~al.} 2017{\natexlab{a}},
  \aap, 601, A104, \dodoi{10.1051/0004-6361/201630339}

\bibitem[{{Fossati} {et~al.}(2017{\natexlab{b}}){Fossati}, {Erkaev}, {Lammer},
  {Cubillos}, {Odert}, {Juvan}, {Kislyakova}, {Lendl}, {Kubyshkina}, \&
  {Bauer}}]{fossati2017b}
{Fossati}, L., {Erkaev}, N.~V., {Lammer}, H., {et~al.} 2017{\natexlab{b}},
  \aap, 598, A90, \dodoi{10.1051/0004-6361/201629716}

\bibitem[{{Frandsen} \& {Lindberg}(1999)}]{Frandsen1999}
{Frandsen}, S., \& {Lindberg}, B. 1999, in Astrophysics with the NOT, ed.
  H.~{Karttunen} \& V.~{Piirola}, 71

\bibitem[{{Fridlund} {et~al.}(2017){Fridlund}, {Gaidos}, {Barrag{\'a}n},
  {Persson}, {Gandolfi}, {Cabrera}, {Hirano}, {Kuzuhara}, {Csizmadia}, {Nowak},
  {Endl}, {Grziwa}, {Korth}, {Pfaff}, {Bitsch}, {Johansen}, {Mustill},
  {Davies}, {Deeg}, {Palle}, {Cochran}, {Eigm{\"u}ller}, {Erikson}, {Guenther},
  {Hatzes}, {Kiilerich}, {Kudo}, {MacQueen}, {Narita}, {Nespral},
  {P{\"a}tzold}, {Prieto-Arranz}, {Rauer}, \& {Van
  Eylen}}]{2017A&A...604A..16F}
{Fridlund}, M., {Gaidos}, E., {Barrag{\'a}n}, O., {et~al.} 2017, \aap, 604,
  A16, \dodoi{10.1051/0004-6361/201730822}

\bibitem[{{Fulton} {et~al.}(2018){Fulton}, {Petigura}, {Blunt}, \&
  {Sinukoff}}]{2018PASP..130d4504F}
{Fulton}, B.~J., {Petigura}, E.~A., {Blunt}, S., \& {Sinukoff}, E. 2018, \pasp,
  130, 044504, \dodoi{10.1088/1538-3873/aaaaa8}

\bibitem[{{Furlan} {et~al.}(2017){Furlan}, {Ciardi}, {Everett}, {Saylors},
  {Teske}, {Horch}, {Howell}, {van Belle}, {Hirsch}, {Gautier}, {Adams},
  {Barrado}, {Cartier}, {Dressing}, {Dupree}, {Gilliland}, {Lillo-Box},
  {Lucas}, \& {Wang}}]{Furlanetal2017}
{Furlan}, E., {Ciardi}, D.~R., {Everett}, M.~E., {et~al.} 2017, \aj, 153, 71,
  \dodoi{10.3847/1538-3881/153/2/71}

\bibitem[{{Gaia Collaboration} {et~al.}(2016){Gaia Collaboration}, {Prusti},
  {de Bruijne}, {Brown}, {Vallenari}, {Babusiaux}, {Bailer-Jones}, {Bastian},
  {Biermann}, {Evans}, \& et~al.}]{2016A&A...595A...1G}
{Gaia Collaboration}, {Prusti}, T., {de Bruijne}, J.~H.~J., {et~al.} 2016,
  \aap, 595, A1, \dodoi{10.1051/0004-6361/201629272}

\bibitem[{{Gaia Collaboration} {et~al.}(2018{\natexlab{a}}){Gaia
  Collaboration}, {Brown}, {Vallenari}, {Prusti}, {de Bruijne}, {Babusiaux},
  {Bailer-Jones}, {Biermann}, {Evans}, {Eyer}, \& et~al.}]{GaiaDR2}
{Gaia Collaboration}, {Brown}, A.~G.~A., {Vallenari}, A., {et~al.}
  2018{\natexlab{a}}, \aap, 616, A1, \dodoi{10.1051/0004-6361/201833051}

\bibitem[{{Gaia Collaboration} {et~al.}(2018{\natexlab{b}}){Gaia
  Collaboration}, {Brown}, {Vallenari}, {Prusti}, {de Bruijne}, {Babusiaux},
  {Bailer-Jones}, {Biermann}, {Evans}, {Eyer}, \& et~al.}]{2018A&A...616A...1G}
---. 2018{\natexlab{b}}, \aap, 616, A1, \dodoi{10.1051/0004-6361/201833051}

\bibitem[{{Gandolfi} {et~al.}(2008){Gandolfi}, {Alcal{\'a}}, {Leccia},
  {Frasca}, {Spezzi}, {Covino}, {Testi}, {Marilli}, \&
  {Kainulainen}}]{Gandolfi2008}
{Gandolfi}, D., {Alcal{\'a}}, J.~M., {Leccia}, S., {et~al.} 2008, \apj, 687,
  1303, \dodoi{10.1086/591729}

\bibitem[{{Gandolfi} {et~al.}(2013){Gandolfi}, {Parviainen}, {Fridlund},
  {Hatzes}, {Deeg}, {Frasca}, {Lanza}, {Prada Moroni}, {Tognelli}, {McQuillan},
  {Aigrain}, {Alonso}, {Antoci}, {Cabrera}, {Carone}, {Csizmadia}, {Djupvik},
  {Guenther}, {Jessen-Hansen}, {Ofir}, \& {Telting}}]{Gandolfi2013}
{Gandolfi}, D., {Parviainen}, H., {Fridlund}, M., {et~al.} 2013, \aap, 557,
  A74, \dodoi{10.1051/0004-6361/201321901}

\bibitem[{{Gandolfi} {et~al.}(2017){Gandolfi}, {Barrag{\'a}n}, {Hatzes},
  {Fridlund}, {Fossati}, {Donati}, {Johnson}, {Nowak}, {Prieto-Arranz},
  {Albrecht}, {Dai}, {Deeg}, {Endl}, {Grziwa}, {Hjorth}, {Korth}, {Nespral},
  {Saario}, {Smith}, {Antoniciello}, {Alarcon}, {Bedell}, {Blay}, {Brems},
  {Cabrera}, {Csizmadia}, {Cusano}, {Cochran}, {Eigm{\"u}ller}, {Erikson},
  {Gonz{\'a}lez Hern{\'a}ndez}, {Guenther}, {Hirano}, {Su{\'a}rez
  Mascare{\~n}o}, {Narita}, {Palle}, {Parviainen}, {P{\"a}tzold}, {Persson},
  {Rauer}, {Saviane}, {Schmidtobreick}, {Van Eylen}, {Winn}, \&
  {Zakhozhay}}]{Gandolfi2017}
{Gandolfi}, D., {Barrag{\'a}n}, O., {Hatzes}, A.~P., {et~al.} 2017, \aj, 154,
  123, \dodoi{10.3847/1538-3881/aa832a}

\bibitem[{{Gandolfi} {et~al.}(2018){Gandolfi}, {Barrag{\'a}n}, {Livingston},
  {Fridlund}, {Justesen}, {Redfield}, {Fossati}, {Mathur}, {Grziwa}, {Cabrera},
  {Garc{\'\i}a}, {Persson}, {Van Eylen}, {Hatzes}, {Hidalgo}, {Albrecht},
  {Bugnet}, {Cochran}, {Csizmadia}, {Deeg}, {Eigm{\"u}ller}, {Endl}, {Erikson},
  {Esposito}, {Guenther}, {Korth}, {Luque}, {Monta{\~n}es Rodr{\'\i}guez},
  {Nespral}, {Nowak}, {P{\"a}tzold}, \& {Prieto-Arranz}}]{Gandolfietal2018}
{Gandolfi}, D., {Barrag{\'a}n}, O., {Livingston}, J.~H., {et~al.} 2018, \aap,
  619, L10, \dodoi{10.1051/0004-6361/201834289}

\bibitem[{{Gandolfi} {et~al.}(2019){Gandolfi}, {Fossati}, {Livingston},
  {Stassun}, {Grziwa}, {Barrag{\'a}n}, {Fridlund}, {Kubyshkina}, {Persson},
  {Dai}, {Lam}, {Albrecht}, {Batalha}, {Beck}, {Justesen}, {Cabrera},
  {Cartwright}, {Cochran}, {Csizmadia}, {Davies}, {Deeg}, {Eigm{\"u}ller},
  {Endl}, {Erikson}, {Esposito}, {Garc{\'\i}a}, {Goeke}, {Gonz{\'a}lez-Cuesta},
  {Guenther}, {Hatzes}, {Hidalgo}, {Hirano}, {Hjorth}, {Kabath}, {Knudstrup},
  {Korth}, {Li}, {Luque}, {Mathur}, {Monta{\~n}es Rodr{\'\i}guez}, {Narita},
  {Nespral}, {Niraula}, {Nowak}, {Palle}, {P{\"a}tzold}, {Prieto-Arranz},
  {Rauer}, {Redfield}, {Ribas}, {Skarka}, {Smith}, {Rowden}, {Torres}, {Van
  Eylen}, \& {Vezie}}]{Gandolfietal2019}
{Gandolfi}, D., {Fossati}, L., {Livingston}, J.~H., {et~al.} 2019, \apjl, 876,
  L24, \dodoi{10.3847/2041-8213/ab17d9}

\bibitem[{{Gao} \& {Zhang}(2020)}]{gao2020}
{Gao}, P., \& {Zhang}, X. 2020, \apj, 890, 93, \dodoi{10.3847/1538-4357/ab6a9b}

\bibitem[{{Garc{\'\i}a} \& {Ballot}(2019)}]{garciaballot2019}
{Garc{\'\i}a}, R.~A., \& {Ballot}, J. 2019, Living Reviews in Solar Physics,
  16, 4, \dodoi{10.1007/s41116-019-0020-1}

\bibitem[{{Garc{\'\i}a} {et~al.}(2011){Garc{\'\i}a}, {Hekker}, {Stello},
  {Guti{\'e}rrez-Soto}, {Handberg}, {Huber}, {Karoff}, {Uytterhoeven},
  {Appourchaux}, {Chaplin}, {Elsworth}, {Mathur}, {Ballot},
  {Christensen-Dalsgaard}, {Gilliland}, {Houdek}, {Jenkins}, {Kjeldsen},
  {McCauliff}, {Metcalfe}, {Middour}, {Molenda-Zakowicz}, {Monteiro}, {Smith},
  \& {Thompson}}]{garciaetal2011}
{Garc{\'\i}a}, R.~A., {Hekker}, S., {Stello}, D., {et~al.} 2011, \mnras, 414,
  L6, \dodoi{10.1111/j.1745-3933.2011.01042.x}

\bibitem[{{Garc{\'\i}a} {et~al.}(2014{\natexlab{a}}){Garc{\'\i}a}, {Mathur},
  {Pires}, {R{\'e}gulo}, {Bellamy}, {Pall{\'e}}, {Ballot}, {Barcel{\'o}
  Forteza}, {Beck}, {Bedding}, {Ceillier}, {Roca Cort{\'e}s}, {Salabert}, \&
  {Stello}}]{garciaetal2014a}
{Garc{\'\i}a}, R.~A., {Mathur}, S., {Pires}, S., {et~al.} 2014{\natexlab{a}},
  \aap, 568, A10, \dodoi{10.1051/0004-6361/201323326}

\bibitem[{{Garc{\'\i}a} {et~al.}(2014{\natexlab{b}}){Garc{\'\i}a}, {Ceillier},
  {Salabert}, {Mathur}, {van Saders}, {Pinsonneault}, {Ballot}, {Beck},
  {Bloemen}, {Campante}, {Davies}, {do Nascimento}, {Mathis}, {Metcalfe},
  {Nielsen}, {Su{\'a}rez}, {Chaplin}, {Jim{\'e}nez}, \&
  {Karoff}}]{garciaetal2014b}
{Garc{\'\i}a}, R.~A., {Ceillier}, T., {Salabert}, D., {et~al.}
  2014{\natexlab{b}}, \aap, 572, A34, \dodoi{10.1051/0004-6361/201423888}

\bibitem[{{Gelman} \& {Rubin}(1992)}]{Gelman1992}
{Gelman}, A., \& {Rubin}, D.~B. 1992, Statistical Science, 7, 457,
  \dodoi{10.1214/ss/1177011136}

\bibitem[{{Grunblatt} {et~al.}(2015){Grunblatt}, {Howard}, \&
  {Haywood}}]{2015ApJ...808..127G}
{Grunblatt}, S.~K., {Howard}, A.~W., \& {Haywood}, R.~D. 2015, \apj, 808, 127,
  \dodoi{10.1088/0004-637X/808/2/127}

\bibitem[{{Grziwa} \& {P{\"a}tzold}(2016)}]{Grziwa2016}
{Grziwa}, S., \& {P{\"a}tzold}, M. 2016, arXiv e-prints, arXiv:1607.08417.
\newblock \doarXiv{1607.08417}

\bibitem[{{Grziwa} {et~al.}(2016){Grziwa}, {Gandolfi}, {Csizmadia}, {Fridlund},
  {Parviainen}, {Deeg}, {Cabrera}, {Djupvik}, {Albrecht}, {Palle},
  {P{\"a}tzold}, {B{\'e}jar}, {Prieto-Arranz}, {Eigm{\"u}ller}, {Erikson},
  {Fynbo}, {Guenther}, {Hatzes}, {Kiilerich}, {Korth}, {Kuutma},
  {Monta{\~n}{\'e}s-Rodr{\'\i}guez}, {Nespral}, {Nowak}, {Rauer}, {Saario},
  {Sebastian}, \& {Slumstrup}}]{Grziwa2016b}
{Grziwa}, S., {Gandolfi}, D., {Csizmadia}, S., {et~al.} 2016, \aj, 152, 132,
  \dodoi{10.3847/0004-6256/152/5/132}

\bibitem[{{G{\"u}nther} {et~al.}(2019){G{\"u}nther}, {Pozuelos}, {Dittmann},
  {Dragomir}, {Kane}, {Daylan}, {Feinstein}, {Huang}, {Morton}, {Bonfanti},
  {Bouma}, {Burt}, {Collins}, {Lissauer}, {Matthews}, {Montet}, {Vand erburg},
  {Wang}, {Winters}, {Ricker}, {Vanderspek}, {Latham}, {Seager}, {Winn},
  {Jenkins}, {Armstrong}, {Barkaoui}, {Batalha}, {Bean}, {Caldwell}, {Ciardi},
  {Collins}, {Crossfield}, {Fausnaugh}, {Furesz}, {Gan}, {Gillon}, {Guerrero},
  {Horne}, {Howell}, {Ireland }, {Isopi}, {Jehin}, {Kielkopf}, {Lepine},
  {Mallia}, {Matson}, {Myers}, {Palle}, {Quinn}, {Relles}, {Rojas-Ayala},
  {Schlieder}, {Sefako}, {Shporer}, {Su{\'a}rez}, {Tan}, {Ting}, {Twicken}, \&
  {Waite}}]{Gunteretal2019}
{G{\"u}nther}, M.~N., {Pozuelos}, F.~J., {Dittmann}, J.~A., {et~al.} 2019,
  Nature Astronomy, 3, 1099, \dodoi{10.1038/s41550-019-0845-5}

\bibitem[{{Hatzes}(2016)}]{Hatzes2016}
{Hatzes}, A.~P. 2016, Astrophysics and Space Science Library, Vol. 428, {The
  Radial Velocity Method for the Detection of Exoplanets}, ed. V.~{Bozza},
  L.~{Mancini}, \& A.~{Sozzetti}, 3, \dodoi{10.1007/978-3-319-27458-4_1}

\bibitem[{{Haywood} {et~al.}(2014){Haywood}, {Collier Cameron}, {Queloz},
  {Barros}, {Deleuil}, {Fares}, {Gillon}, {Lanza}, {Lovis}, {Moutou}, {Pepe},
  {Pollacco}, {Santerne}, {S{\'e}gransan}, \& {Unruh}}]{2014MNRAS.443.2517H}
{Haywood}, R.~D., {Collier Cameron}, A., {Queloz}, D., {et~al.} 2014, \mnras,
  443, 2517, \dodoi{10.1093/mnras/stu1320}

\bibitem[{{Henden} {et~al.}(2015){Henden}, {Levine}, {Terrell}, \&
  {Welch}}]{Henden2015}
{Henden}, A.~A., {Levine}, S., {Terrell}, D., \& {Welch}, D.~L. 2015, in
  American Astronomical Society Meeting Abstracts, Vol. 225, American
  Astronomical Society Meeting Abstracts \#225, 336.16

\bibitem[{{Hirano} {et~al.}(2018){Hirano}, {Dai}, {Gandolfi}, {Fukui},
  {Livingston}, {Miyakawa}, {Endl}, {Cochran}, {Alonso-Floriano}, {Kuzuhara},
  {Montes}, {Ryu}, {Albrecht}, {Barragan}, {Cabrera}, {Csizmadia}, {Deeg},
  {Eigm{\"u}ller}, {Erikson}, {Fridlund}, {Grziwa}, {Guenther}, {Hatzes},
  {Korth}, {Kudo}, {Kusakabe}, {Narita}, {Nespral}, {Nowak}, {P{\"a}tzold},
  {Palle}, {Persson}, {Prieto-Arranz}, {Rauer}, {Ribas}, {Sato}, {Smith},
  {Tamura}, {Tanaka}, {Van Eylen}, \& {Winn}}]{Hirano2018}
{Hirano}, T., {Dai}, F., {Gandolfi}, D., {et~al.} 2018, \aj, 155, 127,
  \dodoi{10.3847/1538-3881/aaa9c1}

\bibitem[{{Howard} {et~al.}(2010){Howard}, {Johnson}, {Marcy}, {Fischer},
  {Wright}, {Bernat}, {Henry}, {Peek}, {Isaacson}, {Apps}, {Endl}, {Cochran},
  {Valenti}, {Anderson}, \& {Piskunov}}]{Howard2010}
{Howard}, A.~W., {Johnson}, J.~A., {Marcy}, G.~W., {et~al.} 2010, \apj, 721,
  1467, \dodoi{10.1088/0004-637X/721/2/1467}

\bibitem[{{Huang} {et~al.}(2018{\natexlab{a}}){Huang}, {Shporer}, {Dragomir},
  {Fausnaugh}, {Levine}, {Morgan}, {Nguyen}, {Ricker}, {Wall}, {Woods}, \&
  {Vanderspek}}]{Huangetal2018a}
{Huang}, C.~X., {Shporer}, A., {Dragomir}, D., {et~al.} 2018{\natexlab{a}},
  arXiv e-prints, arXiv:1807.11129.
\newblock \doarXiv{1807.11129}

\bibitem[{{Huang} {et~al.}(2018{\natexlab{b}}){Huang}, {Burt}, {Vanderburg},
  {G{\"u}nther}, {Shporer}, {Dittmann}, {Winn}, {Wittenmyer}, {Sha}, {Kane},
  {Ricker}, {Vand erspek}, {Latham}, {Seager}, {Jenkins}, {Caldwell},
  {Collins}, {Guerrero}, {Smith}, {Quinn}, {Udry}, {Pepe}, {Bouchy},
  {S{\'e}gransan}, {Lovis}, {Ehrenreich}, {Marmier}, {Mayor}, {Wohler},
  {Haworth}, {Morgan}, {Fausnaugh}, {Ciardi}, {Christiansen}, {Charbonneau},
  {Dragomir}, {Deming}, {Glidden}, {Levine}, {McCullough}, {Yu}, {Narita},
  {Nguyen}, {Morton}, {Pepper}, {P{\'a}l}, {Rodriguez}, {Stassun}, {Torres},
  {Sozzetti}, {Doty}, {Christensen-Dalsgaard}, {Laughlin}, {Clampin}, {Bean},
  {Buchhave}, {Bakos}, {Sato}, {Ida}, {Kaltenegger}, {Palle}, {Sasselov},
  {Butler}, {Lissauer}, {Ge}, \& {Rinehart}}]{Huangetal2018b}
{Huang}, C.~X., {Burt}, J., {Vanderburg}, A., {et~al.} 2018{\natexlab{b}},
  \apjl, 868, L39, \dodoi{10.3847/2041-8213/aaef91}

\bibitem[{Husser {et~al.}(2013)Husser, {Wende-von Berg}, Dreizler, Homeier,
  Reiners, Barman, \& Hauschildt}]{Husser2013}
Husser, T.-O., {Wende-von Berg}, S., Dreizler, S., {et~al.} 2013, A{\&}A, 553,
  A6, \dodoi{10.1051/0004-6361/201219058}

\bibitem[{{Jenkins} {et~al.}(2016){Jenkins}, {Twicken}, {McCauliff},
  {Campbell}, {Sanderfer}, {Lung}, {Mansouri-Samani}, {Girouard}, {Tenenbaum},
  {Klaus}, {Smith}, {Caldwell}, {Chacon}, {Henze}, {Heiges}, {Latham},
  {Morgan}, {Swade}, {Rinehart}, \& {Vanderspek}}]{Jenkins2016}
{Jenkins}, J.~M., {Twicken}, J.~D., {McCauliff}, S., {et~al.} 2016, Society of
  Photo-Optical Instrumentation Engineers (SPIE) Conference Series, Vol. 9913,
  {The TESS science processing operations center}, 99133E,
  \dodoi{10.1117/12.2233418}

\bibitem[{{Jensen}(2013)}]{Jensen:2013}
{Jensen}, E. 2013, {Tapir: A web interface for transit/eclipse observability},
  Astrophysics Source Code Library.
\newblock \doeprint{1306.007}

\bibitem[{{Johnstone} {et~al.}(2015{\natexlab{a}}){Johnstone}, {G{\"u}del},
  {Brott}, \& {L{\"u}ftinger}}]{johnstone2015b}
{Johnstone}, C.~P., {G{\"u}del}, M., {Brott}, I., \& {L{\"u}ftinger}, T.
  2015{\natexlab{a}}, \aap, 577, A28, \dodoi{10.1051/0004-6361/201425301}

\bibitem[{{Johnstone} {et~al.}(2015{\natexlab{b}}){Johnstone}, {G{\"u}del},
  {St{\"o}kl}, {Lammer}, {Tu}, {Kislyakova}, {L{\"u}ftinger}, {Odert},
  {Erkaev}, \& {Dorfi}}]{johnstone2015a}
{Johnstone}, C.~P., {G{\"u}del}, M., {St{\"o}kl}, A., {et~al.}
  2015{\natexlab{b}}, \apjl, 815, L12, \dodoi{10.1088/2041-8205/815/1/L12}

\bibitem[{{Jord{\'a}n} {et~al.}(2019){Jord{\'a}n}, {Brahm}, {Espinoza},
  {Henning}, {Jones}, {Kossakowski}, {Sarkis}, {Trifonov}, {Rojas}, {Torres},
  {Drass}, {Nandakumar}, {Barbieri}, {Davis}, {Wang}, {Bayliss}, {Bouma},
  {Dragomir}, {Eastman}, {Daylan}, {Guerrero}, {Barclay}, {Ting}, {Henze},
  {Ricker}, {Vanderspek}, {Latham}, {Seager}, {Winn}, {Jenkins}, {Wittenmyer},
  {Bowler}, {Crossfield}, {Horner}, {Kane}, {Kielkopf}, {Morton}, {Plavchan},
  {Tinney}, {Addison}, {Mengel}, {Okumura}, {Shahaf}, {Mazeh}, {Rabus},
  {Shporer}, {Ziegler}, {Mann}, \& {Hart}}]{Jordan2019}
{Jord{\'a}n}, A., {Brahm}, R., {Espinoza}, N., {et~al.} 2019, arXiv e-prints,
  arXiv:1911.05574.
\newblock \doarXiv{1911.05574}

\bibitem[{Kass \& Raftery(1995)}]{kassraftery1995}
Kass, R.~E., \& Raftery, A.~E. 1995, Journal of the American Statistical
  Association, 90, 773.
\newblock \url{http://www.jstor.org/stable/2291091}

\bibitem[{{Kipping}(2013)}]{Kipping2013}
{Kipping}, D.~M. 2013, \mnras, 435, 2152, \dodoi{10.1093/mnras/stt1435}

\bibitem[{{Kite} {et~al.}(2019){Kite}, {Fegley}, {Schaefer}, \&
  {Ford}}]{kite2019}
{Kite}, E.~S., {Fegley}, Bruce, J., {Schaefer}, L., \& {Ford}, E.~B. 2019,
  \apjl, 887, L33, \dodoi{10.3847/2041-8213/ab59d9}

\bibitem[{{Korth} {et~al.}(2019){Korth}, {Csizmadia}, {Gandolfi}, {Fridlund},
  {P{\"a}tzold}, {Hirano}, {Livingston}, {Persson}, {Deeg}, {Justesen},
  {Barrag{\'a}n}, {Grziwa}, {Endl}, {Tronsgaard}, {Dai}, {Cochran}, {Albrecht},
  {Alonso}, {Cabrera}, {Cauley}, {Cusano}, {Eigm{\"u}ller}, {Erikson},
  {Esposito}, {Guenther}, {Hatzes}, {Hidalgo}, {Kuzuhara}, {Monta{\~n}es},
  {Napolitano}, {Narita}, {Niraula}, {Nespral}, {Nowak}, {Palle}, {Petrillo},
  {Redfield}, {Prieto-Arranz}, {Rauer}, {Smith}, {Tortora}, {Van Eylen}, \&
  {Winn}}]{Korth2019}
{Korth}, J., {Csizmadia}, S., {Gandolfi}, D., {et~al.} 2019, \mnras, 482, 1807,
  \dodoi{10.1093/mnras/sty2760}

\bibitem[{{Kov{\'a}cs} {et~al.}(2002){Kov{\'a}cs}, {Zucker}, \&
  {Mazeh}}]{Kovacs2002}
{Kov{\'a}cs}, G., {Zucker}, S., \& {Mazeh}, T. 2002, \aap, 391, 369,
  \dodoi{10.1051/0004-6361:20020802}

\bibitem[{{Kubyshkina} {et~al.}(2018){Kubyshkina}, {Fossati}, {Erkaev},
  {Johnstone}, {Cubillos}, {Kislyakova}, {Lammer}, {Lendl}, \&
  {Odert}}]{kubyshkina2018a}
{Kubyshkina}, D., {Fossati}, L., {Erkaev}, N.~V., {et~al.} 2018, \aap, 619,
  A151, \dodoi{10.1051/0004-6361/201833737}

\bibitem[{{Kubyshkina} {et~al.}(2019{\natexlab{a}}){Kubyshkina}, {Cubillos},
  {Fossati}, {Erkaev}, {Johnstone}, {Kislyakova}, {Lammer}, {Lendl}, {Odert},
  \& {G{\"u}del}}]{kubyshkina2019a}
{Kubyshkina}, D., {Cubillos}, P.~E., {Fossati}, L., {et~al.}
  2019{\natexlab{a}}, \apj, 879, 26, \dodoi{10.3847/1538-4357/ab1e42}

\bibitem[{{Kubyshkina} {et~al.}(2019{\natexlab{b}}){Kubyshkina}, {Fossati},
  {Mustill}, {Cubillos}, {Davies}, {Erkaev}, {Johnstone}, {Kislyakova},
  {Lammer}, {Lendl}, \& {Odert}}]{kubyshkina2019b}
{Kubyshkina}, D., {Fossati}, L., {Mustill}, A.~J., {et~al.} 2019{\natexlab{b}},
  \aap, 632, A65, \dodoi{10.1051/0004-6361/201936581}

\bibitem[{{Kurucz}(2013)}]{Kurucz2013}
{Kurucz}, R.~L. 2013, {ATLAS12: Opacity sampling model atmosphere program},
  Astrophysics Source Code Library.
\newblock \doeprint{1303.024}

\bibitem[{{Lammer} {et~al.}(2016){Lammer}, {Erkaev}, {Fossati}, {Juvan},
  {Odert}, {Cubillos}, {Guenther}, {Kislyakova}, {Johnstone}, {L{\"u}ftinger},
  \& {G{\"u}del}}]{lammer2016}
{Lammer}, H., {Erkaev}, N.~V., {Fossati}, L., {et~al.} 2016, \mnras, 461, L62,
  \dodoi{10.1093/mnrasl/slw095}

\bibitem[{{Latham} {et~al.}(2011){Latham}, {Rowe}, {Quinn}, {Batalha},
  {Borucki}, {Brown}, {Bryson}, {Buchhave}, {Caldwell}, {Carter},
  {Christiansen}, {Ciardi}, {Cochran}, {Dunham}, {Fabrycky}, {Ford}, {Gautier},
  {Gilliland}, {Holman}, {Howell}, {Ibrahim}, {Isaacson}, {Jenkins}, {Koch},
  {Lissauer}, {Marcy}, {Quintana}, {Ragozzine}, {Sasselov}, {Shporer},
  {Steffen}, {Welsh}, \& {Wohler}}]{lathametal2011}
{Latham}, D.~W., {Rowe}, J.~F., {Quinn}, S.~N., {et~al.} 2011, \apjl, 732, L24,
  \dodoi{10.1088/2041-8205/732/2/L24}

\bibitem[{{Lendl} {et~al.}(2020){Lendl}, {Bouchy}, {Gill}, {Nielsen}, {Turner},
  {Stassun}, {Acton}, {Anderson}, {Armstrong}, {Bayliss}, {Belardi}, {Bryant},
  {Burleigh}, {Chaushev}, {Casewell}, {Cooke}, {Eigm{\"u}ller}, {Gillen},
  {Goad}, {G{\"u}nther}, {Hagelberg}, {Jenkins}, {Louden}, {Marmier},
  {McCormac}, {Moyano}, {Pollacco}, {Raynard}, {Tilbrook}, {Udry}, {Vines},
  {West}, {Wheatley}, {Ricker}, {Vanderspek}, {Latham}, {Seager}, {Winn},
  {Jenkins}, {Addison}, {Brice{\~n}o}, {Brahm}, {Caldwell}, {Doty}, {Espinoza},
  {Goeke}, {Henning}, {Jord{\'a}n}, {Krishnamurthy}, {Law}, {Morris},
  {Okumura}, {Mann}, {Rodriguez}, {Sarkis}, {Schlieder}, {Twicken},
  {Villanueva}, {Wittenmyer}, {Wright}, \& {Ziegler}}]{Lendletal2020}
{Lendl}, M., {Bouchy}, F., {Gill}, S., {et~al.} 2020, \mnras, 492, 1761,
  \dodoi{10.1093/mnras/stz3545}

\bibitem[{{Libby-Roberts} {et~al.}(2020){Libby-Roberts}, {Berta-Thompson},
  {D{\'e}sert}, {Masuda}, {Morley}, {Lopez}, {Deck}, {Fabrycky}, {Fortney},
  {Line}, {Sanchis-Ojeda}, \& {Winn}}]{libby2020}
{Libby-Roberts}, J.~E., {Berta-Thompson}, Z.~K., {D{\'e}sert}, J.-M., {et~al.}
  2020, \aj, 159, 57, \dodoi{10.3847/1538-3881/ab5d36}

\bibitem[{{Linsky} {et~al.}(2014){Linsky}, {Fontenla}, \&
  {France}}]{linsky2014}
{Linsky}, J.~L., {Fontenla}, J., \& {France}, K. 2014, \apj, 780, 61,
  \dodoi{10.1088/0004-637X/780/1/61}

\bibitem[{{Linsky} {et~al.}(2013){Linsky}, {France}, \& {Ayres}}]{linsky2013}
{Linsky}, J.~L., {France}, K., \& {Ayres}, T. 2013, \apj, 766, 69,
  \dodoi{10.1088/0004-637X/766/2/69}

\bibitem[{{Lissauer} {et~al.}(2011){Lissauer}, {Ragozzine}, {Fabrycky},
  {Steffen}, {Ford}, {Jenkins}, {Shporer}, {Holman}, {Rowe}, {Quintana},
  {Batalha}, {Borucki}, {Bryson}, {Caldwell}, {Carter}, {Ciardi}, {Dunham},
  {Fortney}, {Gautier}, {Howell}, {Koch}, {Latham}, {Marcy}, {Morehead}, \&
  {Sasselov}}]{Lissaueretal2011}
{Lissauer}, J.~J., {Ragozzine}, D., {Fabrycky}, D.~C., {et~al.} 2011, \apjs,
  197, 8, \dodoi{10.1088/0067-0049/197/1/8}

\bibitem[{{Lissauer} {et~al.}(2014){Lissauer}, {Marcy}, {Bryson}, {Rowe},
  {Jontof-Hutter}, {Agol}, {Borucki}, {Carter}, {Ford}, {Gilliland}, {Kolbl},
  {Star}, {Steffen}, \& {Torres}}]{Lissaueretal2014}
{Lissauer}, J.~J., {Marcy}, G.~W., {Bryson}, S.~T., {et~al.} 2014, \apj, 784,
  44, \dodoi{10.1088/0004-637X/784/1/44}

\bibitem[{{Lovis} \& {Pepe}(2007)}]{Lovis2007}
{Lovis}, C., \& {Pepe}, F. 2007, \aap, 468, 1115,
  \dodoi{10.1051/0004-6361:20077249}

\bibitem[{{Lovis} {et~al.}(2006){Lovis}, {Pepe}, {Bouchy}, {Lo Curto}, {Mayor},
  {Pasquini}, {Queloz}, {Rupprecht}, {Udry}, \& {Zucker}}]{Lovis2006}
{Lovis}, C., {Pepe}, F., {Bouchy}, F., {et~al.} 2006, Society of Photo-Optical
  Instrumentation Engineers (SPIE) Conference Series, Vol. 6269, {The exoplanet
  hunter HARPS: unequalled accuracy and perspectives toward 1 cm s $^{-1}$
  precision}, 62690P, \dodoi{10.1117/12.669991}

\bibitem[{{Luri} {et~al.}(2018){Luri}, {Brown}, {Sarro}, {Arenou},
  {Bailer-Jones}, {Castro-Ginard}, {de Bruijne}, {Prusti}, {Babusiaux}, \&
  {Delgado}}]{2018A&A...616A...9L}
{Luri}, X., {Brown}, A.~G.~A., {Sarro}, L.~M., {et~al.} 2018, \aap, 616, A9,
  \dodoi{10.1051/0004-6361/201832964}

\bibitem[{{Mamajek} \& {Hillenbrand}(2008)}]{mamajek2008}
{Mamajek}, E.~E., \& {Hillenbrand}, L.~A. 2008, \apj, 687, 1264,
  \dodoi{10.1086/591785}

\bibitem[{{Mandel} \& {Agol}(2002)}]{Mandel2002}
{Mandel}, K., \& {Agol}, E. 2002, \apjl, 580, L171, \dodoi{10.1086/345520}

\bibitem[{{Mathur} {et~al.}(2010){Mathur}, {Garc{\'\i}a}, {R{\'e}gulo},
  {Creevey}, {Ballot}, {Salabert}, {Arentoft}, {Quirion}, {Chaplin}, \&
  {Kjeldsen}}]{Mathuretal2010}
{Mathur}, S., {Garc{\'\i}a}, R.~A., {R{\'e}gulo}, C., {et~al.} 2010, \aap, 511,
  A46, \dodoi{10.1051/0004-6361/200913266}

\bibitem[{{Mathur} {et~al.}(2014){Mathur}, {Garc{\'\i}a}, {Ballot}, {Ceillier},
  {Salabert}, {Metcalfe}, {R{\'e}gulo}, {Jim{\'e}nez}, \&
  {Bloemen}}]{Mathuretal2014}
{Mathur}, S., {Garc{\'\i}a}, R.~A., {Ballot}, J., {et~al.} 2014, \aap, 562,
  A124, \dodoi{10.1051/0004-6361/201322707}

\bibitem[{{Maxted} {et~al.}(2011){Maxted}, {Anderson}, {Collier Cameron},
  {Hellier}, {Queloz}, {Smalley}, {Street}, {Triaud}, {West}, {Gillon},
  {Lister}, {Pepe}, {Pollacco}, {S{\'e}gransan}, {Smith}, \&
  {Udry}}]{2011PASP..123..547M}
{Maxted}, P.~F.~L., {Anderson}, D.~R., {Collier Cameron}, A., {et~al.} 2011,
  \pasp, 123, 547, \dodoi{10.1086/660007}

\bibitem[{{Mayor} {et~al.}(2003){Mayor}, {Pepe}, {Queloz}, {Bouchy},
  {Rupprecht}, {Lo Curto}, {Avila}, {Benz}, {Bertaux}, {Bonfils}, {Dall},
  {Dekker}, {Delabre}, {Eckert}, {Fleury}, {Gilliotte}, {Gojak}, {Guzman},
  {Kohler}, {Lizon}, {Longinotti}, {Lovis}, {Megevand}, {Pasquini}, {Reyes},
  {Sivan}, {Sosnowska}, {Soto}, {Udry}, {van Kesteren}, {Weber}, \&
  {Weilenmann}}]{Mayor2003}
{Mayor}, M., {Pepe}, F., {Queloz}, D., {et~al.} 2003, The Messenger, 114, 20

\bibitem[{{McCully} {et~al.}(2018){McCully}, {Volgenau}, {Harbeck}, {Lister},
  {Saunders}, {Turner}, {Siiverd}, \& {Bowman}}]{McCully:2018}
{McCully}, C., {Volgenau}, N.~H., {Harbeck}, D.-R., {et~al.} 2018, in Society
  of Photo-Optical Instrumentation Engineers (SPIE) Conference Series, Vol.
  10707, \procspie, 107070K, \dodoi{10.1117/12.2314340}

\bibitem[{{Meunier} {et~al.}(2010){Meunier}, {Desort}, \& {Lagrange}}]{Meunier}
{Meunier}, N., {Desort}, M., \& {Lagrange}, A.-M. 2010, \aap, 512, A39,
  \dodoi{10.1051/0004-6361/200913551}

\bibitem[{{Milbourne} {et~al.}(2019){Milbourne}, {Haywood}, {Phillips}, {Saar},
  {Cegla}, {Cameron}, {Costes}, {Dumusque}, {Langellier}, {Latham},
  {Maldonado}, {Malavolta}, {Mortier}, {Palumbo}, {Thompson}, {Watson},
  {Bouchy}, {Buchschacher}, {Cecconi}, {Charbonneau}, {Cosentino}, {Ghedina},
  {Glenday}, {Gonzalez}, {Li}, {Lodi}, {L{\'o}pez-Morales}, {Lovis}, {Mayor},
  {Micela}, {Molinari}, {Pepe}, {Piotto}, {Rice}, {Sasselov}, {S{\'e}gransan},
  {Sozzetti}, {Szentgyorgyi}, {Udry}, \& {Walsworth}}]{2019ApJ...874..107M}
{Milbourne}, T.~W., {Haywood}, R.~D., {Phillips}, D.~F., {et~al.} 2019, \apj,
  874, 107, \dodoi{10.3847/1538-4357/ab064a}

\bibitem[{{Molli{\`e}re} {et~al.}(2017){Molli{\`e}re}, {van Boekel}, {Bouwman},
  {Henning}, {Lagage}, \& {Min}}]{mollierevanboekel2017}
{Molli{\`e}re}, P., {van Boekel}, R., {Bouwman}, J., {et~al.} 2017, \aap, 600,
  A10, \dodoi{10.1051/0004-6361/201629800}

\bibitem[{{Molli{\`e}re} {et~al.}(2019){Molli{\`e}re}, {Wardenier}, {van
  Boekel}, {Henning}, {Molaverdikhani}, \& {Snellen}}]{mollierewardenier2019}
{Molli{\`e}re}, P., {Wardenier}, J.~P., {van Boekel}, R., {et~al.} 2019, \aap,
  627, A67, \dodoi{10.1051/0004-6361/201935470}

\bibitem[{{Morgenthaler} {et~al.}(2012){Morgenthaler}, {Petit}, {Saar},
  {Solanki}, {Morin}, {Marsden}, {Auri{\`e}re}, {Dintrans}, {Fares}, {Gastine},
  {Lanoux}, {Ligni{\`e}res}, {Paletou}, {Ram{\'\i}rez V{\'e}lez}, {Th{\'e}ado},
  \& {Van Grootel}}]{Morgenthaler2012}
{Morgenthaler}, A., {Petit}, P., {Saar}, S., {et~al.} 2012, \aap, 540, A138,
  \dodoi{10.1051/0004-6361/201118139}

\bibitem[{{Morris} {et~al.}(2017){Morris}, {Twicken}, {Smith}, {Clarke},
  {Jenkins}, {Bryson}, {Girouard}, \& {Klaus}}]{morrisetal2017}
{Morris}, R.~L., {Twicken}, J.~D., {Smith}, J.~C., {et~al.} 2017, {Kepler Data
  Processing Handbook: Photometric Analysis}, Kepler Science Document

\bibitem[{{Morton}(2015)}]{2015ascl.soft03010M}
{Morton}, T.~D. 2015, {isochrones: Stellar model grid package}, Astrophysics
  Source Code Library.
\newblock \doeprint{1503.010}

\bibitem[{{Murdoch} {et~al.}(1993){Murdoch}, {Hearnshaw}, \&
  {Clark}}]{Murdoch1993}
{Murdoch}, K.~A., {Hearnshaw}, J.~B., \& {Clark}, M. 1993, \apj, 413, 349,
  \dodoi{10.1086/173003}

\bibitem[{{Murray-Clay} {et~al.}(2009){Murray-Clay}, {Chiang}, \&
  {Murray}}]{2009ApJ...693...23M}
{Murray-Clay}, R.~A., {Chiang}, E.~I., \& {Murray}, N. 2009, \apj, 693, 23,
  \dodoi{10.1088/0004-637X/693/1/23}

\bibitem[{{Niraula} {et~al.}(2017){Niraula}, {Redfield}, {Dai}, {Barrag{\'a}n},
  {Gandolfi}, {Cauley}, {Hirano}, {Korth}, {Smith}, {Prieto-Arranz}, {Grziwa},
  {Fridlund}, {Persson}, {Justesen}, {Winn}, {Albrecht}, {Cochran},
  {Csizmadia}, {Duvvuri}, {Endl}, {Hatzes}, {Livingston}, {Narita}, {Nespral},
  {Nowak}, {P{\"a}tzold}, {Palle}, \& {Van Eylen}}]{niraula17}
{Niraula}, P., {Redfield}, S., {Dai}, F., {et~al.} 2017, \aj, 154, 266,
  \dodoi{10.3847/1538-3881/aa957c}

\bibitem[{{Nowak} {et~al.}(2017){Nowak}, {Palle}, {Gandolfi}, {Dai}, {Lanza},
  {Hirano}, {Barrag{\'a}n}, {Fukui}, {Bruntt}, {Endl}, {Cochran}, {Prada
  Moroni}, {Prieto-Arranz}, {Kiilerich}, {Nespral}, {Hatzes}, {Albrecht},
  {Deeg}, {Winn}, {Yu}, {Kuzuhara}, {Grziwa}, {Smith}, {Guenther}, {Van Eylen},
  {Csizmadia}, {Fridlund}, {Cabrera}, {Eigm{\"u}ller}, {Erikson}, {Korth},
  {Narita}, {P{\"a}tzold}, {Rauer}, \& {Ribas}}]{Nowak2017}
{Nowak}, G., {Palle}, E., {Gandolfi}, D., {et~al.} 2017, \aj, 153, 131,
  \dodoi{10.3847/1538-3881/aa5cb6}

\bibitem[{{Oklop{\v{c}}i{\'c}} \& {Hirata}(2018)}]{oklopcicetal2018}
{Oklop{\v{c}}i{\'c}}, A., \& {Hirata}, C.~M. 2018, \apjl, 855, L11,
  \dodoi{10.3847/2041-8213/aaada9}

\bibitem[{{Owen} \& {Wu}(2017)}]{owen2017}
{Owen}, J.~E., \& {Wu}, Y. 2017, \apj, 847, 29,
  \dodoi{10.3847/1538-4357/aa890a}

\bibitem[{Parviainen \& Aigrain(2015)}]{Parviainen2015}
Parviainen, H., \& Aigrain, S. 2015, MNRAS, 453, 3821,
  \dodoi{10.1093/mnras/stv1857}

\bibitem[{{Paxton} {et~al.}(2018){Paxton}, {Schwab}, {Bauer}, {Bildsten},
  {Blinnikov}, {Duffell}, {Farmer}, {Goldberg}, {Marchant}, {Sorokina},
  {Thoul}, {Townsend}, \& {Timmes}}]{paxton2018}
{Paxton}, B., {Schwab}, J., {Bauer}, E.~B., {et~al.} 2018, \apjs, 234, 34,
  \dodoi{10.3847/1538-4365/aaa5a8}

\bibitem[{{Pecaut} \& {Mamajek}(2013)}]{Pecaut2013}
{Pecaut}, M.~J., \& {Mamajek}, E.~E. 2013, \apjs, 208, 9,
  \dodoi{10.1088/0067-0049/208/1/9}

\bibitem[{{Pepe} {et~al.}(2002){Pepe}, {Mayor}, {Galland}, {Naef}, {Queloz},
  {Santos}, {Udry}, \& {Burnet}}]{Pepe2002a}
{Pepe}, F., {Mayor}, M., {Galland}, F., {et~al.} 2002, \aap, 388, 632,
  \dodoi{10.1051/0004-6361:20020433}

\bibitem[{{Persson} {et~al.}(2018){Persson}, {Fridlund}, {Barrag{\'a}n}, {Dai},
  {Gandolfi}, {Hatzes}, {Hirano}, {Grziwa}, {Korth}, {Prieto-Arranz},
  {Fossati}, {Van Eylen}, {Justesen}, {Livingston}, {Kubyshkina}, {Deeg},
  {Guenther}, {Nowak}, {Cabrera}, {Eigm{\"u}ller}, {Csizmadia}, {Smith},
  {Erikson}, {Albrecht}, {Sobrino}, {Cochran}, {Endl}, {Esposito}, {Fukui},
  {Heeren}, {Hidalgo}, {Hjorth}, {Kuzuhara}, {Narita}, {Nespral}, {Palle},
  {P{\"a}tzold}, {Rauer}, {Rodler}, \& {Winn}}]{2018A&A...618A..33P}
{Persson}, C.~M., {Fridlund}, M., {Barrag{\'a}n}, O., {et~al.} 2018, \aap, 618,
  A33, \dodoi{10.1051/0004-6361/201832867}

\bibitem[{{Persson} {et~al.}(2019){Persson}, {Csizmadia}, {Mustill},
  {Fridlund}, {Hatzes}, {Nowak}, {Georgieva}, {Gandolfi}, {Davies},
  {Livingston}, {Palle}, {Monta{\~n}es Rodr{\'\i}guez}, {Endl}, {Hirano},
  {Prieto-Arranz}, {Korth}, {Grziwa}, {Esposito}, {Albrecht}, {Johnson},
  {Barrag{\'a}n}, {Parviainen}, {Van Eylen}, {Alonso Sobrino}, {Beck},
  {Cabrera}, {Carleo}, {Cochran}, {Dai}, {Deeg}, {de Leon}, {Eigm{\"u}ller},
  {Erikson}, {Fukui}, {Gonz{\'a}lez-Cuesta}, {Guenther}, {Hidalgo}, {Hjorth},
  {Kabath}, {Knudstrup}, {Kusakabe}, {Lam}, {Lund}, {Luque}, {Mathur},
  {Murgas}, {Narita}, {Nespral}, {Niraula}, {Olofsson}, {P{\"a}tzold}, {Rauer},
  {Redfield}, {Ribas}, {Skarka}, {Smith}, {Subjak}, \&
  {Tamura}}]{Perssonetal2019}
{Persson}, C.~M., {Csizmadia}, S., {Mustill}, A. e.~J., {et~al.} 2019, \aap,
  628, A64, \dodoi{10.1051/0004-6361/201935505}

\bibitem[{{Piskunov} \& {Valenti}(2017)}]{pv2017}
{Piskunov}, N., \& {Valenti}, J.~A. 2017, \aap, 597, A16,
  \dodoi{10.1051/0004-6361/201629124}

\bibitem[{{Pizzolato} {et~al.}(2003){Pizzolato}, {Maggio}, {Micela},
  {Sciortino}, \& {Ventura}}]{pizzolato2003}
{Pizzolato}, N., {Maggio}, A., {Micela}, G., {Sciortino}, S., \& {Ventura}, P.
  2003, \aap, 397, 147, \dodoi{10.1051/0004-6361:20021560}

\bibitem[{{Pollacco} {et~al.}(2006){Pollacco}, {Skillen}, {Collier Cameron},
  {Christian}, {Hellier}, {Irwin}, {Lister}, {Street}, {West}, {Anderson},
  {Clarkson}, {Deeg}, {Enoch}, {Evans}, {Fitzsimmons}, {Haswell}, {Hodgkin},
  {Horne}, {Kane}, {Keenan}, {Maxted}, {Norton}, {Osborne}, {Parley}, {Ryans},
  {Smalley}, {Wheatley}, \& {Wilson}}]{2006PASP..118.1407P}
{Pollacco}, D.~L., {Skillen}, I., {Collier Cameron}, A., {et~al.} 2006, \pasp,
  118, 1407, \dodoi{10.1086/508556}

\bibitem[{{Press} {et~al.}(2002){Press}, {Teukolsky}, {Vetterling}, \&
  {Flannery}}]{Press2002}
{Press}, W.~H., {Teukolsky}, S.~A., {Vetterling}, W.~T., \& {Flannery}, B.~P.
  2002, {Numerical recipes in C++ : the art of scientific computing}

\bibitem[{{Queloz} {et~al.}(2001){Queloz}, {Mayor}, {Udry}, {Burnet},
  {Carrier}, {Eggenberger}, {Naef}, {Santos}, {Pepe}, {Rupprecht}, {Avila},
  {Baeza}, {Benz}, {Bertaux}, {Bouchy}, {Cavadore}, {Delabre}, {Eckert},
  {Fischer}, {Fleury}, {Gilliotte}, {Goyak}, {Guzman}, {Kohler}, {Lacroix},
  {Lizon}, {Megevand}, {Sivan}, {Sosnowska}, \& {Weilenmann}}]{CORALIE}
{Queloz}, D., {Mayor}, M., {Udry}, S., {et~al.} 2001, The Messenger, 105, 1

\bibitem[{{Quinn} {et~al.}(2019){Quinn}, {Becker}, {Rodriguez}, {Hadden},
  {Huang}, {Morton}, {Adams}, {Armstrong}, {Eastman}, {Horner}, {Kane},
  {Lissauer}, {Twicken}, {Vanderburg}, {Wittenmyer}, {Ricker}, {Vanderspek},
  {Latham}, {Seager}, {Winn}, {Jenkins}, {Agol}, {Barkaoui}, {Beichman},
  {Bouchy}, {Bouma}, {Burdanov}, {Campbell}, {Carlino}, {Cartwright},
  {Charbonneau}, {Christiansen}, {Ciardi}, {Collins}, {Collins}, {Conti},
  {Crossfield}, {Daylan}, {Dittmann}, {Doty}, {Dragomir}, {Ducrot}, {Gillon},
  {Glidden}, {Goeke}, {Gonzales}, {He{\l}miniak}, {Horch}, {Howell}, {Jehin},
  {Jensen}, {Kielkopf}, {Kristiansen}, {Law}, {Mann}, {Marmier}, {Matson},
  {Matthews}, {Mazeh}, {Mori}, {Murgas}, {Murray}, {Narita}, {Nielsen},
  {Ottoni}, {Palle}, {Paw{\l}aszek}, {Pepe}, {Pitogo de Leon}, {Pozuelos},
  {Relles}, {Schlieder}, {Sebastian}, {S{\'e}gransan}, {Shporer}, {Stassun},
  {Tamura}, {Udry}, {Waite}, {Winters}, \& {Ziegler}}]{Quinnetal2019}
{Quinn}, S.~N., {Becker}, J.~C., {Rodriguez}, J.~E., {et~al.} 2019, \aj, 158,
  177, \dodoi{10.3847/1538-3881/ab3f2b}

\bibitem[{{Ricker} {et~al.}(2014){Ricker}, {Winn}, {Vanderspek}, {Latham},
  {Bakos}, {Bean}, {Berta-Thompson}, {Brown}, {Buchhave}, {Butler}, {Butler},
  {Chaplin}, {Charbonneau}, {Christensen-Dalsgaard}, {Clampin}, {Deming},
  {Doty}, {De Lee}, {Dressing}, {Dunham}, {Endl}, {Fressin}, {Ge}, {Henning},
  {Holman}, {Howard}, {Ida}, {Jenkins}, {Jernigan}, {Johnson}, {Kaltenegger},
  {Kawai}, {Kjeldsen}, {Laughlin}, {Levine}, {Lin}, {Lissauer}, {MacQueen},
  {Marcy}, {McCullough}, {Morton}, {Narita}, {Paegert}, {Palle}, {Pepe},
  {Pepper}, {Quirrenbach}, {Rinehart}, {Sasselov}, {Sato}, {Seager},
  {Sozzetti}, {Stassun}, {Sullivan}, {Szentgyorgyi}, {Torres}, {Udry}, \&
  {Villasenor}}]{Rickeretal2014}
{Ricker}, G.~R., {Winn}, J.~N., {Vanderspek}, R., {et~al.} 2014, Society of
  Photo-Optical Instrumentation Engineers (SPIE) Conference Series, Vol. 9143,
  {Transiting Exoplanet Survey Satellite (TESS)}, 914320,
  \dodoi{10.1117/12.2063489}

\bibitem[{{Rowe} {et~al.}(2014){Rowe}, {Bryson}, {Marcy}, {Lissauer},
  {Jontof-Hutter}, {Mullally}, {Gilliland}, {Issacson}, {Ford}, {Howell},
  {Borucki}, {Haas}, {Huber}, {Steffen}, {Thompson}, {Quintana}, {Barclay},
  {Still}, {Fortney}, {Gautier}, {Hunter}, {Caldwell}, {Ciardi}, {Devore},
  {Cochran}, {Jenkins}, {Agol}, {Carter}, \& {Geary}}]{Roweetal2014}
{Rowe}, J.~F., {Bryson}, S.~T., {Marcy}, G.~W., {et~al.} 2014, \apj, 784, 45,
  \dodoi{10.1088/0004-637X/784/1/45}

\bibitem[{{Ryabchikova} {et~al.}(2015){Ryabchikova}, {Piskunov}, {Kurucz},
  {Stempels}, {Heiter}, {Pakhomov}, \& {Barklem}}]{Ryabchikova2015}
{Ryabchikova}, T., {Piskunov}, N., {Kurucz}, R.~L., {et~al.} 2015, \physscr,
  90, 054005, \dodoi{10.1088/0031-8949/90/5/054005}

\bibitem[{{Santos} {et~al.}(2019){Santos}, {Garc{\'\i}a}, {Mathur}, {Bugnet},
  {van Saders}, {Metcalfe}, {Simonian}, \& {Pinsonneault}}]{Santosetal2019}
{Santos}, A.~R.~G., {Garc{\'\i}a}, R.~A., {Mathur}, S., {et~al.} 2019, \apjs,
  244, 21, \dodoi{10.3847/1538-4365/ab3b56}

\bibitem[{{Sanz-Forcada} {et~al.}(2011){Sanz-Forcada}, {Micela}, {Ribas},
  {Pollock}, {Eiroa}, {Velasco}, {Solano}, \&
  {Garc{\'\i}a-{\'A}lvarez}}]{sanzforcada2011}
{Sanz-Forcada}, J., {Micela}, G., {Ribas}, I., {et~al.} 2011, \aap, 532, A6,
  \dodoi{10.1051/0004-6361/201116594}

\bibitem[{{Savitzky} \& {Golay}(1964)}]{Savitzky1964}
{Savitzky}, A., \& {Golay}, M.~J.~E. 1964, Analytical Chemistry, 36, 1627

\bibitem[{{Schofield} {et~al.}(2019){Schofield}, {Chaplin}, {Huber},
  {Campante}, {Davies}, {Miglio}, {Ball}, {Appourchaux}, {Basu}, {Bedding},
  {Christensen-Dalsgaard}, {Creevey}, {Garc{\'\i}a}, {Handberg}, {Kawaler},
  {Kjeldsen}, {Latham}, {Lund}, {Metcalfe}, {Ricker}, {Serenelli}, {Silva
  Aguirre}, {Stello}, \& {Vanderspek}}]{Schofieldetal2019}
{Schofield}, M., {Chaplin}, W.~J., {Huber}, D., {et~al.} 2019, \apjs, 241, 12,
  \dodoi{10.3847/1538-4365/ab04f5}

\bibitem[{{Shapiro} {et~al.}(2016){Shapiro}, {Solanki}, {Krivova}, {Yeo}, \&
  {Schmutz}}]{Shapiroetal2016}
{Shapiro}, A.~I., {Solanki}, S.~K., {Krivova}, N.~A., {Yeo}, K.~L., \&
  {Schmutz}, W.~K. 2016, \aap, 589, A46, \dodoi{10.1051/0004-6361/201527527}

\bibitem[{{Skrutskie} {et~al.}(2006){Skrutskie}, {Cutri}, {Stiening},
  {Weinberg}, {Schneider}, {Carpenter}, {Beichman}, {Capps}, {Chester},
  {Elias}, {Huchra}, {Liebert}, {Lonsdale}, {Monet}, {Price}, {Seitzer},
  {Jarrett}, {Kirkpatrick}, {Gizis}, {Howard}, {Evans}, {Fowler}, {Fullmer},
  {Hurt}, {Light}, {Kopan}, {Marsh}, {McCallon}, {Tam}, {Van Dyk}, \&
  {Wheelock}}]{2006AJ....131.1163S}
{Skrutskie}, M.~F., {Cutri}, R.~M., {Stiening}, R., {et~al.} 2006, \aj, 131,
  1163, \dodoi{10.1086/498708}

\bibitem[{{Smith} {et~al.}(2012){Smith}, {Stumpe}, {Van Cleve}, {Jenkins},
  {Barclay}, {Fanelli}, {Girouard}, {Kolodziejczak}, {McCauliff}, {Morris}, \&
  {Twicken}}]{Smith2012}
{Smith}, J.~C., {Stumpe}, M.~C., {Van Cleve}, J.~E., {et~al.} 2012, \pasp, 124,
  1000, \dodoi{10.1086/667697}

\bibitem[{{Southworth}(2011)}]{tepcat}
{Southworth}, J. 2011, \mnras, 417, 2166,
  \dodoi{10.1111/j.1365-2966.2011.19399.x}

\bibitem[{{Stassun} {et~al.}(2018){Stassun}, {Oelkers}, {Pepper}, {Paegert},
  {De Lee}, {Torres}, {Latham}, {Charpinet}, {Dressing}, {Huber}, {Kane},
  {L{\'e}pine}, {Mann}, {Muirhead}, {Rojas-Ayala}, {Silvotti}, {Fleming},
  {Levine}, \& {Plavchan}}]{Stassun2018}
{Stassun}, K.~G., {Oelkers}, R.~J., {Pepper}, J., {et~al.} 2018, \aj, 156, 102,
  \dodoi{10.3847/1538-3881/aad050}

\bibitem[{{St{\"o}kl} {et~al.}(2015){St{\"o}kl}, {Dorfi}, \&
  {Lammer}}]{stokl2015}
{St{\"o}kl}, A., {Dorfi}, E., \& {Lammer}, H. 2015, \aap, 576, A87,
  \dodoi{10.1051/0004-6361/201423638}

\bibitem[{{Strassmeier}(2009)}]{Strassmeier09}
{Strassmeier}, K.~G. 2009, \aapr, 17, 251, \dodoi{10.1007/s00159-009-0020-6}

\bibitem[{{Stumpe} {et~al.}(2014){Stumpe}, {Smith}, {Catanzarite}, {Van Cleve},
  {Jenkins}, {Twicken}, \& {Girouard}}]{Stumpeetal2014}
{Stumpe}, M.~C., {Smith}, J.~C., {Catanzarite}, J.~H., {et~al.} 2014, \pasp,
  126, 100, \dodoi{10.1086/674989}

\bibitem[{{Stumpe} {et~al.}(2012){Stumpe}, {Smith}, {Van Cleve}, {Twicken},
  {Barclay}, {Fanelli}, {Girouard}, {Jenkins}, {Kolodziejczak}, {McCauliff}, \&
  {Morris}}]{Stumpe2012}
{Stumpe}, M.~C., {Smith}, J.~C., {Van Cleve}, J.~E., {et~al.} 2012, \pasp, 124,
  985, \dodoi{10.1086/667698}

\bibitem[{{St{\"u}rmer} {et~al.}(2018){St{\"u}rmer}, {Seifahrt}, {Schwab},
  {Gandolfi}, {Monta{\~n}es-Rodriguez}, \& {Buchhave}}]{Stuermer2018}
{St{\"u}rmer}, J., {Seifahrt}, A., {Schwab}, C., {et~al.} 2018, in Society of
  Photo-Optical Instrumentation Engineers (SPIE) Conference Series, Vol. 10702,
  \procspie, 107022S, \dodoi{10.1117/12.2313052}

\bibitem[{{Sullivan} {et~al.}(2015){Sullivan}, {Winn}, {Berta-Thompson},
  {Charbonneau}, {Deming}, {Dressing}, {Latham}, {Levine}, {McCullough},
  {Morton}, {Ricker}, {Vanderspek}, \& {Woods}}]{Sullivanetal2015}
{Sullivan}, P.~W., {Winn}, J.~N., {Berta-Thompson}, Z.~K., {et~al.} 2015, \apj,
  809, 77, \dodoi{10.1088/0004-637X/809/1/77}

\bibitem[{{Telting} {et~al.}(2014){Telting}, {Avila}, {Buchhave}, {Frandsen},
  {Gandolfi}, {Lindberg}, {Stempels}, {Prins}, \& {NOT staff}}]{Telting2014}
{Telting}, J.~H., {Avila}, G., {Buchhave}, L., {et~al.} 2014, Astronomische
  Nachrichten, 335, 41, \dodoi{10.1002/asna.201312007}

\bibitem[{{Twicken} {et~al.}(2010){Twicken}, {Clarke}, {Bryson}, {Tenenbaum},
  {Wu}, {Jenkins}, {Girouard}, \& {Klaus}}]{Twickenetal2010}
{Twicken}, J.~D., {Clarke}, B.~D., {Bryson}, S.~T., {et~al.} 2010, Society of
  Photo-Optical Instrumentation Engineers (SPIE) Conference Series, Vol. 7740,
  {Photometric analysis in the Kepler Science Operations Center pipeline},
  774023, \dodoi{10.1117/12.856790}

\bibitem[{{Twicken} {et~al.}(2018){Twicken}, {Catanzarite}, {Clarke},
  {Girouard}, {Jenkins}, {Klaus}, {Li}, {McCauliff}, {Seader}, {Tenenbaum},
  {Wohler}, {Bryson}, {Burke}, {Caldwell}, {Haas}, {Henze}, \&
  {Sanderfer}}]{Twicken2018}
{Twicken}, J.~D., {Catanzarite}, J.~H., {Clarke}, B.~D., {et~al.} 2018, \pasp,
  130, 064502, \dodoi{10.1088/1538-3873/aab694}

\bibitem[{{Valenti} \& {Piskunov}(1996)}]{vp96}
{Valenti}, J.~A., \& {Piskunov}, N. 1996, \aaps, 118, 595

\bibitem[{{Van Eylen} \& {Albrecht}(2015)}]{vaneylen2015}
{Van Eylen}, V., \& {Albrecht}, S. 2015, \apj, 808, 126,
  \dodoi{10.1088/0004-637X/808/2/126}

\bibitem[{{Van Eylen} {et~al.}(2016){Van Eylen}, {Nowak}, {Albrecht}, {Palle},
  {Ribas}, {Bruntt}, {Perger}, {Gandolfi}, {Hirano}, {Sanchis-Ojeda},
  {Kiilerich}, {Prieto-Arranz}, {Badenas}, {Dai}, {Deeg}, {Guenther},
  {Monta{\~n}{\'e}s-Rodr{\'\i}guez}, {Narita}, {Rogers}, {B{\'e}jar},
  {Shrotriya}, {Winn}, \& {Sebastian}}]{VanEylenetal2016}
{Van Eylen}, V., {Nowak}, G., {Albrecht}, S., {et~al.} 2016, \apj, 820, 56,
  \dodoi{10.3847/0004-637X/820/1/56}

\bibitem[{{Villarreal D'Angelo} {et~al.}(2018){Villarreal D'Angelo},
  {Esquivel}, {Schneiter}, \& {Sgr{\'o}}}]{Villarreal2018}
{Villarreal D'Angelo}, C., {Esquivel}, A., {Schneiter}, M., \& {Sgr{\'o}},
  M.~A. 2018, \mnras, 479, 3115, \dodoi{10.1093/mnras/sty1544}

\bibitem[{{Villarreal D'Angelo} {et~al.}(2014){Villarreal D'Angelo},
  {Schneiter}, {Costa}, {Vel{\'a}zquez}, {Raga}, \&
  {Esquivel}}]{Villarreal2014}
{Villarreal D'Angelo}, C., {Schneiter}, M., {Costa}, A., {et~al.} 2014, \mnras,
  438, 1654, \dodoi{10.1093/mnras/stt2303}

\bibitem[{{Vogt} {et~al.}(1994){Vogt}, {Allen}, {Bigelow}, {Bresee}, {Brown},
  {Cantrall}, {Conrad}, {Couture}, {Delaney}, {Epps}, {Hilyard}, {Hilyard},
  {Horn}, {Jern}, {Kanto}, {Keane}, {Kibrick}, {Lewis}, {Osborne},
  {Pardeilhan}, {Pfister}, {Ricketts}, {Robinson}, {Stover}, {Tucker}, {Ward},
  \& {Wei}}]{Vogt1994}
{Vogt}, S.~S., {Allen}, S.~L., {Bigelow}, B.~C., {et~al.} 1994, Society of
  Photo-Optical Instrumentation Engineers (SPIE) Conference Series, Vol. 2198,
  {HIRES: the high-resolution echelle spectrometer on the Keck 10-m Telescope},
  ed. D.~L. {Crawford} \& E.~R. {Craine}, 362, \dodoi{10.1117/12.176725}

\bibitem[{{Winn}(2010)}]{Winn2010}
{Winn}, J.~N. 2010, arXiv e-prints, arXiv:1001.2010.
\newblock \doarXiv{1001.2010}

\bibitem[{{Wood} {et~al.}(2005){Wood}, {Redfield}, {Linsky}, {M{\"u}ller}, \&
  {Zank}}]{Wood2005}
{Wood}, B.~E., {Redfield}, S., {Linsky}, J.~L., {M{\"u}ller}, H.-R., \& {Zank},
  G.~P. 2005, \apjs, 159, 118, \dodoi{10.1086/430523}

\bibitem[{{Wright} {et~al.}(2011){Wright}, {Drake}, {Mamajek}, \&
  {Henry}}]{wright2011}
{Wright}, N.~J., {Drake}, J.~J., {Mamajek}, E.~E., \& {Henry}, G.~W. 2011,
  \apj, 743, 48, \dodoi{10.1088/0004-637X/743/1/48}

\bibitem[{{Yee} {et~al.}(2017){Yee}, {Petigura}, \& {von Braun}}]{Yee2017}
{Yee}, S.~W., {Petigura}, E.~A., \& {von Braun}, K. 2017, \apj, 836, 77,
  \dodoi{10.3847/1538-4357/836/1/77}

\bibitem[{{Zechmeister} \& {K{\"u}rster}(2009)}]{Zechmeister2009}
{Zechmeister}, M., \& {K{\"u}rster}, M. 2009, \aap, 496, 577,
  \dodoi{10.1051/0004-6361:200811296}

\bibitem[{{Zeng} {et~al.}(2016){Zeng}, {Sasselov}, \& {Jacobsen}}]{Zeng2016}
{Zeng}, L., {Sasselov}, D.~D., \& {Jacobsen}, S.~B. 2016, \apj, 819, 127,
  \dodoi{10.3847/0004-637X/819/2/127}

\end{thebibliography}
\bibliographystyle{aasjournal}



\end{document}